\documentclass[12pt]{amsart}
\usepackage{amsmath,amssymb,url}
\usepackage{geometry} 
\usepackage{graphicx}

\usepackage{hyperref}

\newcommand{\pder}[2]{\ensuremath{\frac{ \partial #1}{\partial #2}}}

\newcommand{\R}{\ensuremath{\mathbb{R}}}

\newtheorem{thm}{Theorem}[section]

\theoremstyle{remark}
\newtheorem{rmk}[thm]{Remark}

\DeclareMathOperator{\SDiff}{SDiff}

\DeclareMathOperator{\SO}{SO}

\DeclareMathOperator{\Orb}{Orb}

\title[Multipole vortex blobs]{Multipole vortex blobs (MVB):\\ Symplectic geometry and dynamics}
\author[D.D. Holm \& H.O. Jacobs]{Darryl D. Holm \& Henry O. Jacobs\\ \\
Department of Mathematics\\
Imperial College, London SW7 2AZ, UK}
\date{16 August 2015}

\begin{document}

\begin{abstract}
Vortex blob methods are typically characterized by a regularization length scale, below which the dynamics are trivial for isolated blobs.
In this article we observe that the dynamics need not be trivial if one is willing to consider distributional derivatives of Dirac delta functionals as valid vorticity distributions.
More specifically, a new singular vortex theory is presented for regularized Euler fluid equations of ideal incompressible flow in the plane.
We determine the conditions under which such regularized Euler fluid equations may admit vorticity singularities which are stronger than delta functions, e.g., derivatives of delta functions.
We also describe the symplectic geometry associated to these augmented vortex structures and we characterize the dynamics as Hamiltonian.
Applications to the design of numerical methods similar to vortex blob methods are also discussed.
Such findings illuminate the rich dynamics which occur below the regularization length scale and
enlighten our perspective on the potential for regularized fluid models to capture multiscale phenomena.
\end{abstract}

\maketitle

\tableofcontents

\section{Introduction}
\label{sec:intro}
Vortices are important in hydrodynamics because they are the sources for the incompressible flow field. The vorticity distribution at any instant of time determines both the current state of the flow and its future evolution, for given boundary conditions. This property holds for any Hamiltonian system, and it can indeed be shown that the dynamics of vortices can be usefully expressed in Hamiltonian form. In the vorticity and stream function formulation of an ideal incompressible planar fluid, the evolution of the vorticity distribution $\omega(x,y,t)$ is given by
\begin{align}
  \partial_t \omega - \{\omega,\psi\} \equiv 
  \partial_t \omega - \partial_x  \omega\,  \partial_y \psi
  + \partial_y \omega \, \partial_x \psi = 0\,, \label{eq:vorticity}
\end{align}
where $\omega = - \Delta \psi$ is the vorticity, $\psi$ is the stream function, and $\Delta = \partial_{xx} + \partial_{yy}$ is the Laplace operator.
The corresponding $(x,y)$ components of the Eulerian velocity field are given by
\begin{align*}
	 (u,v) = (\partial_y \psi , - \partial_x \psi).
\end{align*}

If one is willing to view the vorticity $\omega$ as a distribution, one can consider point vortex solutions.
In particular, point vortices are obtained if one considers the vorticity solution ansatz 
\[\omega(z,t) = \sum_{i} \Gamma_i(t) \delta_{z_i(t)}\,,\]
where $\Gamma_i(t) \in \mathbb{R}$, $z=(x,y) \in \mathbb{R}^2$ and $\delta_{z_i(t)}$ is the Dirac delta distribution centered at the point $z_i(t) = (x_i(t),y_i(t)) \in \mathbb{R}^2$
at a given time $t \in \mathbb{R}$.
Substitution of this ansatz into \eqref{eq:vorticity} yields the following well known finite dimensional system in the form  of Hamilton's canonical equations,
\begin{align}
\begin{split}
	\frac{d \Gamma_i}{dt} &= 0 \,,\quad  \quad \psi(z,t) = \sum_i  \Gamma_i(t) G(z-z_i(t)) \,, \\
	\frac{dx_i}{dt} &= \partial_y \psi(z_i) \,,\quad  \quad \frac{dy_i}{dt} = - \partial_x \psi(z_i)\,,
\end{split} \label{eq:point_vortex}
\end{align}
where $G(z) =  - (2\pi)^{-1} \ln( \| z \|)$ is the Green's function for the planar Laplacian. 

A point-vortex approximation to a continuous distribution of vorticity for Euler's fluid equations is problematic, though, because a point vortex induces a flow velocity which becomes unbounded. However, when the point vortex is made smooth and bounded (regularized) the approximation becomes reasonable  \cite{Chorin1973}.

For example, one may consider the regularized form of the vorticity equation given by
choosing a translationally and rotationally invariant smoothing kernel $K_\delta$ of width $\delta > 0$ and defining the regularized vorticity as 
$K_\delta* \omega = -\Delta \psi$ while continuing to use \eqref{eq:vorticity} to evolve $\omega$ in time.
For example, $K_\delta (z) = \exp( - \| z\|^2 / \delta^2)$ is considered in \cite{BealeMajda1985}.
In this case the point vortex ansatz yields \eqref{eq:point_vortex} again, except that the singular Green's function $G$ is replaced by the smooth kernel
\begin{align}
	G_\delta(z) := K_\delta*G(z) = \frac{1}{4\pi} \left( {\rm Ei}(- \| z \|^2 / \delta^2) - 2\ln ( \| z \|) \right), \label{eq:kernel}
\end{align}
where ${\rm Ei}( \cdot )$ denotes the exponential integral function.
The vorticity kernel $G_\delta$ has no singularity at the origin for $\delta > 0$, and is known as a \textit{vortex blob}.
This system is the starting point for the vortex blob method, introduced in \cite{Chorin1973} (albeit with a different regularization).

The economy of the vortex blob method derives from the property that Dirac delta distributions are hyper-local (i.e. parametrized by position), and the property that the vorticity equation  \eqref{eq:vorticity} admits Dirac delta distributions as solutions.
However, there are many distributions which are localized to a similar degree (e.g. derivatives of delta functions, $\partial_x \delta_{z_i}$).

In this paper, we study the more general vorticity solution ansatz,
\[\omega(z,t) = \sum_{i,m,n} \Gamma_i^{mn}(t) \partial_x^m \partial_y^n \delta_{z_i}\,.\]
We find that this ansatz yields a closed finite dimensional system which generalizes vortex blobs. We call these new carriers of vorticity \emph{multipole vortex blobs} or \emph{MVBs}.

\subsection{Main contributions}
\begin{enumerate}
        \item Section \ref{sec:background} briefly reviews the background for vortex methods in fluid modeling.
        \item Section \ref{sec:regularized} reviews the relationship between regularized fluids and vortex blob methods.
        \item Section \ref{sec:EOM} derives the equations of motion for point vortices and MVBs as exact solutions of a regularized vorticity equation.
        \item Section \ref{sec:conserved} derives the conservation laws for these equations, such as energy, linear momentum, and angular momentum, and circulation.
        The derivation of these conserved quantities as symplectic momentum maps can be found in Appendix \ref{sec:symmetries}.
        \item Section \ref{sec:moments} explains the relationship between the dynamical systems for MVBs and an implicitly defined \emph{closed} dynamical system which governs the spatial moments of the vorticity distribution.
        \item Section \ref{sec:numerics} discusses numerical aspects of using MVBs to model fluid dynamics, such as approximations of initial conditions (subsection \ref{sec:approximation}), and grouping of computational nodes (subsection \ref{sec:grouping}).
        \item  Section \ref{sec:numerical_experiments} presents the results of several numerical experiments involving small numbers of vortices, for $N=1,2$, and $3$.
        	\item MVB dynamics is Hamiltonian. We present the symplectic and Hamiltonian structure of MVB dynamics in Section \ref{sec:symplectic}.
\end{enumerate}

\section{Background}
\label{sec:background}
Vortex methods for fluid modeling predate the computer age
and references to them can be found in the work of Helmholtz \cite[see the introductory section]{Smith2011}
For example, the use of point vortices as idealized solutions can already be found in a 1931 paper concerning a ``line of discontinuity'' in planar fluid flow \cite{Rosenhead1931}.
At the beginning of their development, the infinite velocities (and energies) associated to point vortices caused great difficulties, both numerically and theoretically.
In fact, the point vortex approach did not produce a competitive numerical method until the 1970s, when the problems related to singularities were overcome by regularizing the singular vortex kernel to form a \emph{vortex blob}.
Stochastic perturbations were further included to model viscosity \cite{Chorin1973}.
These adjustments to the classical point vortex method yielded the \emph{vortex blob method}, which quickly became of practical use for realistic fluid flow modeling.
In particular, the regularized system proved more amenable to error analysis.
It was shown that the solutions of the vortex blob method converge to solutions of the Navier-Stokes equations in \cite{Hald1979}.
Later, stronger convergence rates were achieved by judicious choice of vortex kernels.
By convolving the singular vortex kernel with sums of Gaussian smoothing kernels, a sequence of vortex blob kernels with faster convergence rates was found.
Specifically, the convergence rate of the $m$th kernel was found to be of order $h^{mq}$ for any $q \in (0,1)$ where $h = \delta^q$
is a grid-spacing parameter and $\delta > 0$ is a length scale associated to the smoothing kernel \cite{BealeMajda1982,BealeMajda1985}.

Simultaneously, the symplectic geometry of point vortices was clarified in \cite{MarsdenWeinstein1983}
by invoking Arnold's interpretation of ideal fluids \cite{Arnold1966}.
The findings of \cite{MarsdenWeinstein1983} were developed further in \cite{GayBalmazVizman2012} to handle fluid flow on manifolds with nontrivial homology.
While this theoretical development clarified the geometry of point vortices, vortex blobs were sometimes thought to be a numerical ``trick" which violated the geometric interpretation.
However, this thought was banished with the invention of the Euler-$\alpha$ model, a regularized model of ideal fluids with a parameter $\alpha$ representing the typical correlation length of fluctuations away from the mean of a Lagrangian fluid path \cite{FoiasHolmTiti2001}. In particular, vortex blob solutions associated to a specific kernel serve as \emph{exact} solutions to the Euler-$\alpha$ model \cite{OliverShkoller2001}.
The Euler-$\alpha$ kernel is different from the kernels used in \cite{Chorin1973} and \cite{BealeMajda1985}.
A comparison of the Euler-$\alpha$ kernel to the $m=1$ kernel of \cite{BealeMajda1985} is given in \cite{HolmNitschePutkaradze2006} for vortex filament and vortex sheet motion.

While vortex blobs performed well, they did not capture all of the qualitative richness observed in fluid vorticity dynamics.
In particular, blobs of vorticity in real ideal fluids are known to change shape and deviate from initially circular distributions.
A numerical method is proposed in \cite{Rossi1997,Rossi2005} to capture these shape dynamics by adding
basis functions with non-trivial moments in the study of vortex merger (see for example \cite{MelanderZabuskyMcWilliams1998,DizesVerga2002,MeunierDizesLeweke2005}).
Another distinct model obtained by projection onto a Hermite basis is described in \cite{NagemSandriUminskyWayne2009}.
This projection yielded a finite-dimensional systems which modeled the (truncated) moments of the vorticity of an ideal incompressible fluid.
The derivation of simplified combinatorial formulas invoked by the dynamics of this model were discovered in \cite{UminskyWayneBarbaro2010}
and these formulas have made the method numerically tractable for a large number of moments.

A dual approach to the moment based methods of the previous paragraph \cite{Rossi1997,Rossi2005,NagemSandriUminskyWayne2009} is to consider multipole based methods.
This is the approach proposed in \cite{Nicolaides1986}, where an initial vortex ansatz consisting of sums of distributional derivatives of dirac delta distributions is considered.
Such an idea has occured intermittently in various forms in the literature, over many years.
For example, a regularized vortex blob model, in the spirit of \cite{BealeMajda1982,BealeMajda1985}, which considered vorticity distributions of the form $\omega = \sum \Gamma_i \delta_{z_i} + \Gamma_i^x \partial_x \delta_{z_i} + \Gamma_i^y \partial_y \delta_{z_i}$ was investigated in \cite{ChiuNivolaides1988}.
Here it was proven that this augmentation of the traditional vortex method will yield faster spectral convergence than that of a traditional vortex blob method.
The current article considers higher order derivatives and can be seen as a natural follow up to \cite{ChiuNivolaides1988}.
More recently, dynamics have been derived for interactions of pure vortices and pure dipoles.
These come from vorticity distributions of the form $\omega = \sum_{i} \Gamma_i \delta_{z_{v,i}} +  \sum_{j} \left(\Gamma_i^x \partial_x \delta_{z_{d,i}} + \Gamma_i^y \partial_y \delta_{z_{d,i}}\right)$
with the assumption that the locations of the dipoles and the vortices never overlap and that their self-interaction terms may be neglected \cite{Yanovsky2009,TurYanovskyKonstantin2011}.
In a different approach, approximations of dipoles are created by holonomically constraining vortices of opposite strength to be a fixed distance from one another, \cite{TchieuKansoNewton2012}. The question remains, however, to what extent the dynamics of \cite{TchieuKansoNewton2012} approximates those of \cite{Yanovsky2009,TurYanovskyKonstantin2011} after self-interaction terms have been neglected.
In summary, the removal of self-interaction terms is one of the primary obstacles to obtaining a multipole based generalization of the point vortex method \cite{Smith2011}.
Moreover, the spectral error decay rates found in \cite{Hald1979,BealeMajda1982,BealeMajda1985,ChiuNivolaides1988} arise from the use of vortex blobs in place of (singular) point vortices.
In this article we will follow this regularization based approach.

\section{Vortex blobs and regularized fluid models}
\label{sec:regularized}
In this section we review a class of regularized fluid models and their relationship with vortex blob methods (for a more detailed discussion see~\cite{HolmNitschePutkaradze2006}).
The sort of fluid models we consider take the form
\begin{align*}
	\partial_{t} \omega + \vec{u} \cdot \nabla \omega = 0 \\
	\omega = {\rm curl}( L_{\alpha} \cdot u ).
\end{align*}
Where $L_{\alpha}$ is a $\operatorname{SE}(2)$ invariant linear psuedo-differential operator with a length-scale parameter $\alpha > 0$ such that $\lim_{\alpha \to 0} Q_{op} = 1$.
When $L_{\alpha}$ is just the identity, the above ``model'' is Euler's equations of motion for an ideal fluid.
When $L_{\alpha} = (1- \alpha^{-2} \Delta)$ where $\Delta$ is the Laplace operator, then we obtain the the Euler-$\alpha$ model, the solutions of 
which will converge to solutions of Euler's ideal fluid equations as $\alpha >0$ vanishes \cite{FoiasHolmTiti2001}.

We may replace $u$ with its stream function, $\psi$, in order to rewrite the above equations as
\begin{align}
	\partial_{t} \omega + \{ \psi , \omega \} = 0 \label{eq:omega eom} \\
	\omega = \Delta( L_{\alpha} \cdot \psi) \label{eq:psi to omega}
\end{align}
This allows us to represent planar fluid dynamics in terms of scalar functions and distributions rather than vector-fields.

The relationship between these regularized models and vortex blobs methods comes from first considering the point-vortex ansatz
\begin{align*}
	\omega (z ; t) = \sum_{i} \Gamma_{i} \delta(z - z_{i}(t) ).
\end{align*}
If the operator, $\Delta \circ L_\alpha$, has a non-singular Green's function, $G_\alpha$, then substituting the ansatz into \eqref{eq:psi to omega} implies that
\begin{align}
	\psi(z;t) = \sum_{i} \Gamma_{i} G_{\alpha}(z - z_{i}(t)) \label{eq:stream}
\end{align}
We should note that when $L_\alpha$ is the identity (i.e. for an Euler fluid), then $G_\alpha$ is singular, and an extra argument (perhaps a physical one) must be presented in order to allow the resulting singular velocity fields.
In this paper no such issues with singularity arise because we are modelling an Euler fluid with a regularized fluid where $L_{\alpha}$ has a non-singular Green's function.

Substitution of $\psi$ into \eqref{eq:omega eom} then implies the following equations of motion
for the vortex cores $z_{i} = (x_{i},y_{i})$ and the strengths $\Gamma_{i}(t)$:
\begin{align}
	\frac{dx_{i}}{dt} = \partial_{y}\psi(z_{i}(t);t) \quad, \quad \frac{dy_{i}}{dt} = - \partial_{x} \psi(z_{i}(t);t ) \quad ,\quad \frac{d\Gamma_{i}}{dt} = 0. \label{eq:ode}
\end{align}
When $\alpha = 0$ and $L_{\alpha} = 1$ this is nothing but the point-vortex method.
When $\alpha > 0$ it is possible for $\psi$ to be much more regular, and we obtain various vortex-blob methods.
In particular, we obtain the smooth vortex blobs of~\cite{BealeMajda1985}.

It is notable that \eqref{eq:stream} and \eqref{eq:ode} together form a finite dimensional ODE.
The solutions of this ODE are \emph{exact} solutions to the regularized fluid model.
Again, this is valuable because the solutions of many regularized fluid models have been shown to converge to solutions of the ideal fluid equations as $\alpha$ vanishes.
This paper seeks to generalize these point-like solutions to regularized fluid models to obtain a richer class of solutions with richer conservation properties.

\section{Equations of motion}
\label{sec:EOM}
In this section we derive the equations of motion for the time-dependent parameters which specify multipole vortex blobs (MVBs).
The zeroth order MVBs are just standard vortex blobs and the resulting equations of motion are those of the standard (non-stochastic) vortex blob algorithm \cite{Chorin1973}.
The first order MVBs are regularized dipoles and the equations of motion are those of \cite{ChiuNivolaides1988}.
Here we will derive the equations of motion for $N$th order MVBs following the approach of \cite{ChiuNivolaides1988}.

\begin{figure}[p] 
\centering
	\includegraphics[clip,trim=0.6in 0.5in 0.6in 0.6in,width=0.3\textwidth]{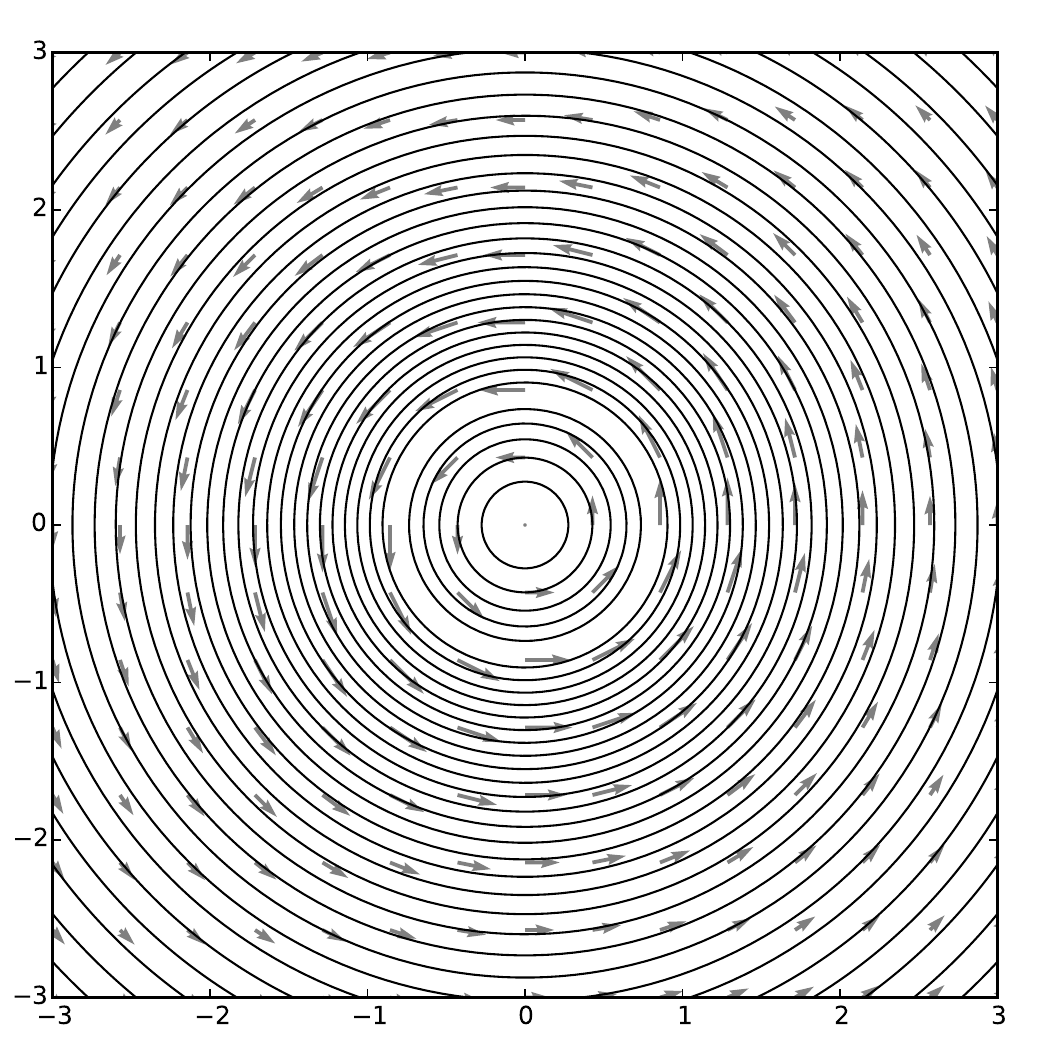} 
	\caption{
		A 0th order MVB with $z = 0$ and $\Gamma=1$, using the kernel $G_\delta$ of equation \eqref{eq:kernel}.
		This form of the kernel produces one of the vortex blobs presented in\cite{BealeMajda1985}
		and the resulting numerical method is identical.}
	\label{fig:zero}
\end{figure}

\begin{figure}
	\centering
	\includegraphics[clip,trim=0.6in 0.5in 0.6in 0.6in,width=0.4\textwidth]{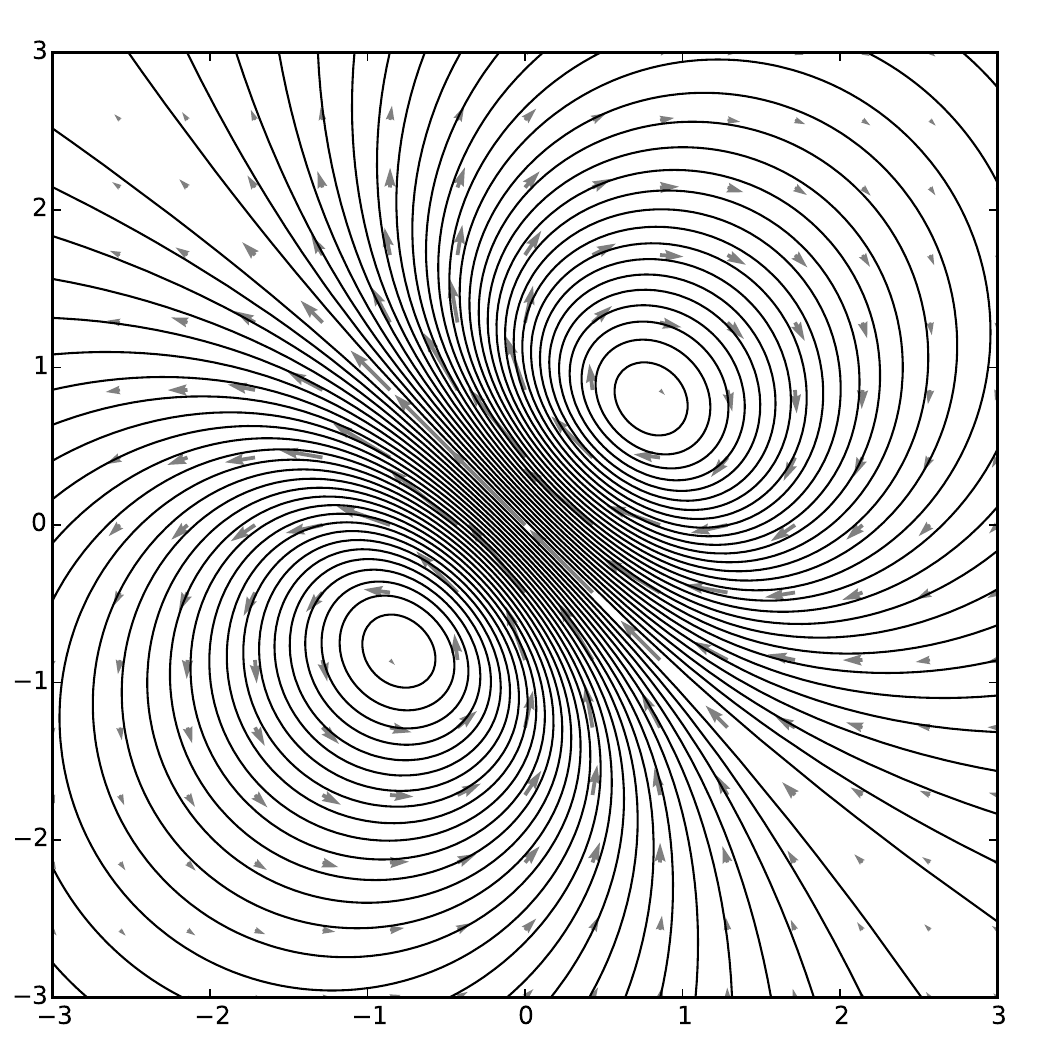} 
	\caption{The flow field around a 1st order MVB with $\Gamma = 0,\Gamma^x = 1,\Gamma^y=1$ is that of a regularized dipole.}
	\label{fig:one}
\end{figure}

\begin{figure}
	\includegraphics[clip,trim =1in 1in 1in 1in,width=0.3\textwidth]{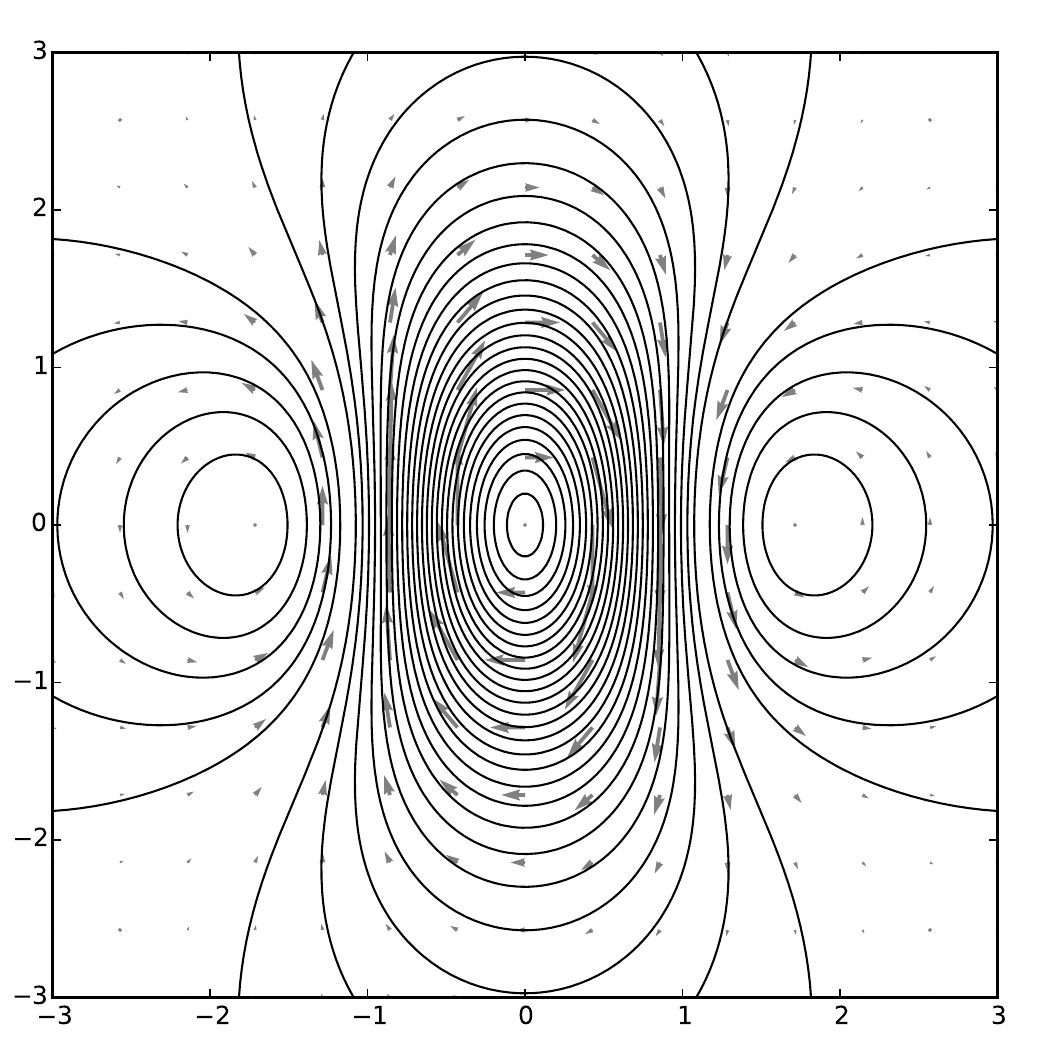} 
	\includegraphics[clip,trim =1in 1in 1in 1in,width=0.3\textwidth]{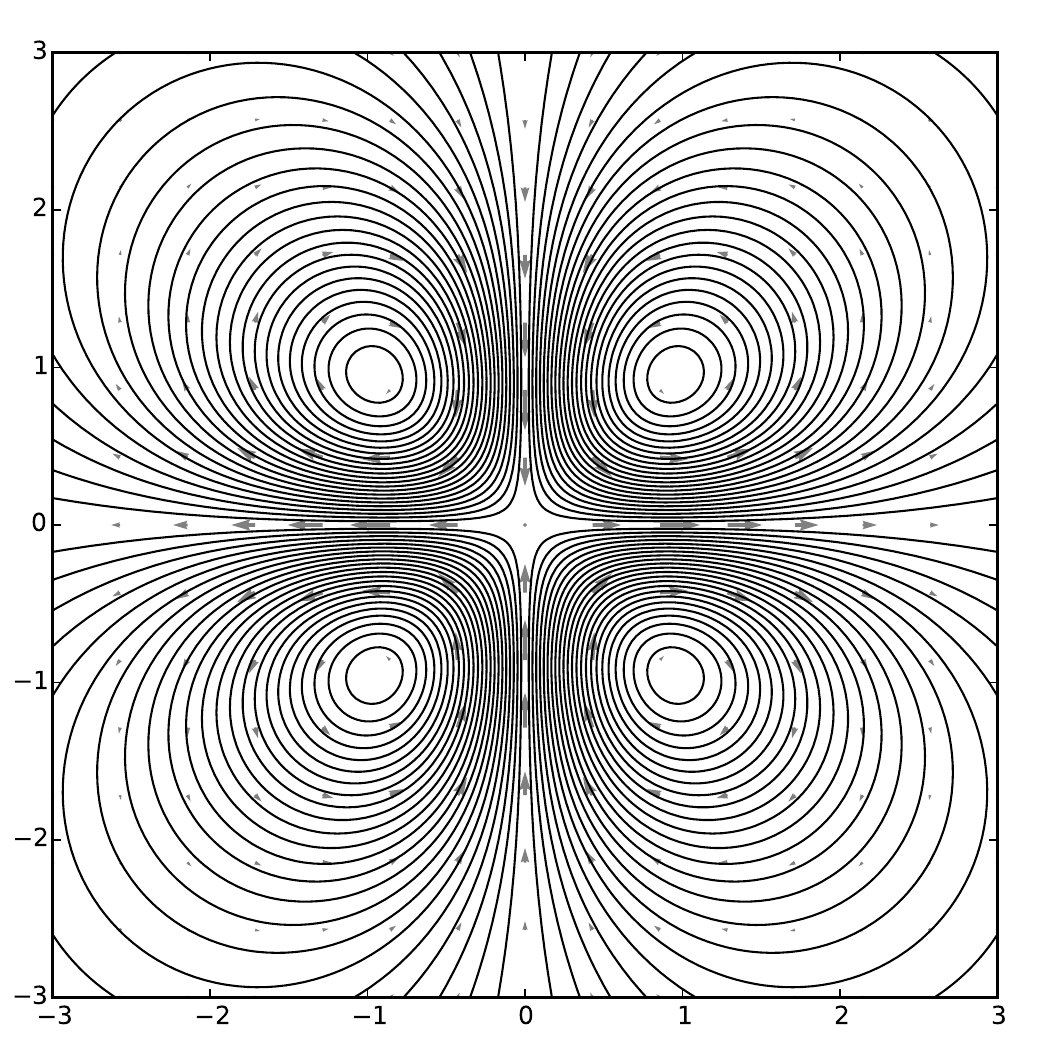} 
	\includegraphics[clip,trim =1in 1in 1in 1in,width=0.3\textwidth]{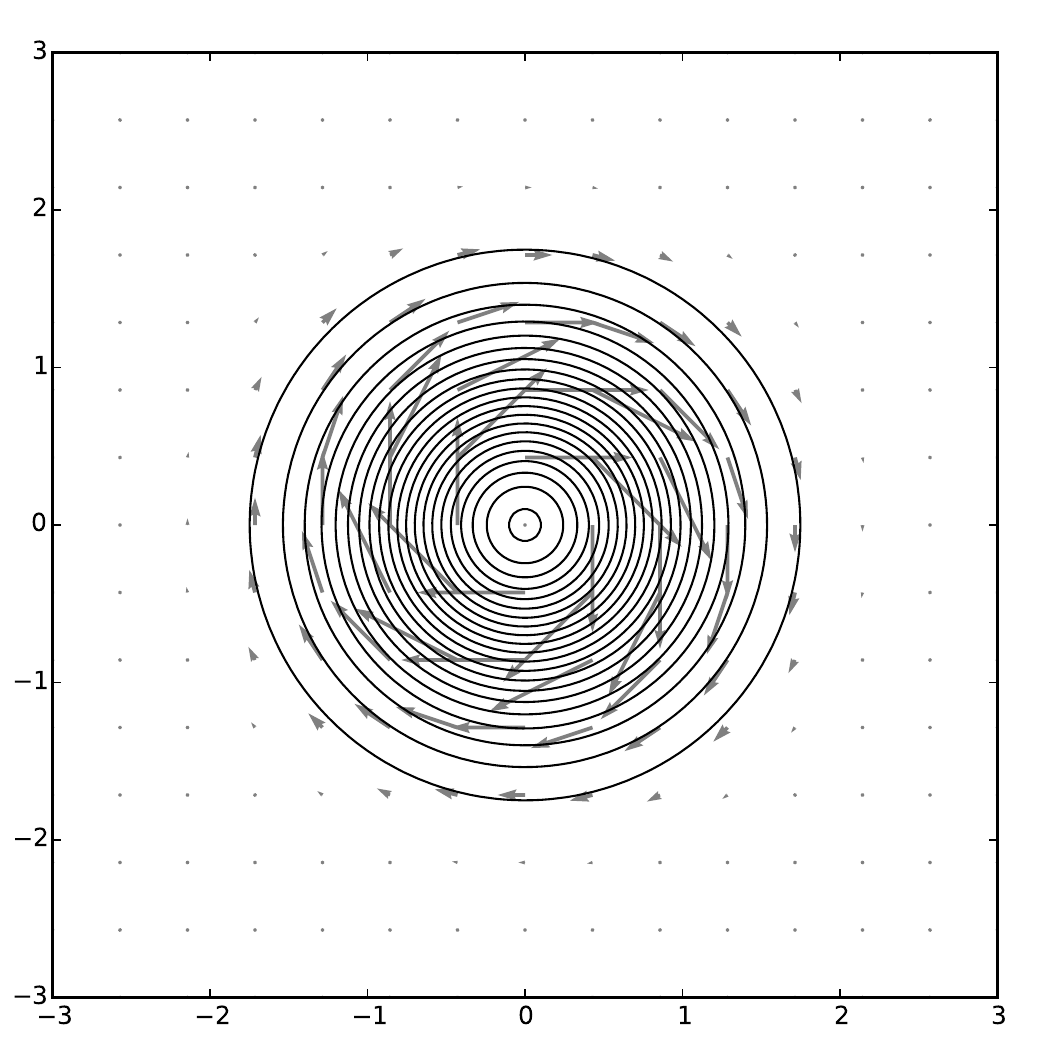} 
	\caption{Various second order MVBs with all $\Gamma$'s set to zero except.
	(left) $\Gamma^{xx} = 1$, (middle) $\Gamma^{xy}=1$, (right) $\Gamma^{xx}=\Gamma^{yy}=1$. }
	\label{fig:second}
\end{figure}

Consider the ansatz for the vorticity,
\begin{align}
  \omega(z,t) = \sum_{i \in S} \sum_{m+n \leq N} \Gamma^{mn}_i(t) \partial_x^m \partial_y^n \delta_{z_i} \,,
  \label{eq:ansatz N}
\end{align}
for spatially constant dynamical variables $\Gamma^{mn}_i(t) \in \R$ for $i \in S$ where $S$ is some countable set.
The stream function is 
\begin{align} \label{eq:stream}
  \psi(z,t) = \sum_{i \in S} \sum_{m+n \leq N} \Gamma^{mn}_i(t) \partial_x^m \partial_y^n G_\delta (z-z_i(t) )
\,.\end{align}
The corresponding velocity field is given by
\begin{align}
\begin{cases}
  u(z,t) = \partial_y \psi(z,t) =  \sum_{i \in S,m+n \leq N} \Gamma^{mn}_i(t) \partial_x^m \partial_y^{n+1} G_\delta (z-z_i(t) )
  \,, \\
  v(z,t) = -\partial_x \psi(z,t) =  - \sum_{i \in S, m+n \leq N} \Gamma^{mn}_i(t) \partial_x^{m+1} \partial_y^n G_\delta (z-z_i(t) )\,.
\end{cases} \label{eq:u N}
\end{align}
Examples of the types of velocity fields produced are depicted in figures \ref{fig:zero} through \ref{fig:second} on page \pageref{fig:zero}.

We seek equations of motion for the $\Gamma^{mn}_i(t)$'s and $z_i(t)$'s
such that the velocity field \eqref{eq:u N} satisfies the vorticity equation \eqref{eq:vorticity}.
In the following calculations, we will not show the explicit time dependence of the dynamical variables.

We now find
\begin{align*}
  \partial_t \omega =	
  \sum_{
  	\substack{
		i \in S \\
		m+n \leq N}}
  	\frac{d \Gamma_i^{mn}}{dt} \partial_x^m \partial_y^n \delta_{z_i} - \Gamma_i^{mn} \frac{dx_i}{dt} \partial_{x}^{m+1} \partial_y^{n} \delta_{z_i}
	- \Gamma_i^{mn} \frac{dy_i}{dt} \partial_{x}^{m} \partial_y^{n+1} \delta_{z_i}\,,
\end{align*}
and
\begin{align*}
  \partial_y \psi \, \partial_x \omega = 
  \sum_{
  	\substack{
		i \in S \\
		m+n \leq N}}
   \partial_y \psi \, \Gamma_i ^{mn}\partial_x^{m+1} \partial_y^n \delta_{z_i}\,.
\end{align*}
By invoking \eqref{eq:func times partial delta} of Appendix
\ref{sec:distributions} we can rearrange the previous equation to
obtain
\begin{align*}
  \partial_y \psi \, \partial_x \omega &=
  \sum_{
  	\substack{
		i \in S \\
		m+n \leq N \\
		\ell,k}}
	\Gamma_i^{mn} (-1)^{m+n+1 + \ell + k} \binom{m+1}{\ell} \binom{n}{k} \partial_x^{\ell} \partial_y^{k+1} \psi(z_i) \partial_x^{m+1-\ell} \partial_y^{n-k} \delta_{z_i}.
\end{align*}
Similarly, we find
\begin{align*}
  \partial_x \psi \, \partial_y \omega &=
  \sum_{
  	\substack{
		i \in S \\
		m+n \leq N\\
		\ell,k}
	}
	\Gamma_i^{mn} (-1)^{m+n+1 + \ell + k} \binom{m}{\ell} \binom{n+1}{k} \partial_x^{\ell+1} \partial_y^k \psi(z_i) \partial_x^{m-\ell} \partial_y^{n+1-k} \delta_{z_i}.
\end{align*}
Substitution of these expressions into \eqref{eq:vorticity} yields
the vanishing of a linear combination of the distributions
$\partial_x^m \partial_y^n \delta_{z_i}$
for $m+n \leq N+1$.
Since each of these distributions is linearly independent
of the others (assuming the $z_i$'s are distinct),
their individual coefficients must each vanish independently.
If we focus on the terms of the sum where $m+n = N$ we obtain coefficients for $\partial_{x} \delta$ and $\partial_{y}\delta$
at the core locations.
The vanishing of these coefficients yields the dynamics for MVB cores
\begin{align}
  \frac{dx_i}{dt} = \partial_y \psi (z_i) \,,\quad  \quad \frac{dy_i}{dt} = - \partial_x \psi(z_i) \label{eq:core dynamics}.
\end{align}
The vanishing of the coefficient of $\delta_{z_i}$ yields
\begin{align*}
	\frac{d\Gamma^{0,0}_i}{dt}  = 0.
\end{align*}

For $\ell+k \leq N$ the vanishing of the coefficient of $\partial_x^{\ell} \partial_{y}^{k} \delta_{z_i}$ yields
\begin{align}
\begin{split}
 &\frac{d\Gamma_i^{\ell k}}{dt} =\\
 &(-1)^{\ell + k}
  \sum_{
    \substack{
      m > \ell \\
      n > k \\
      n+m \leq N}
    }\Gamma_i^{mn} \Bigg[  \binom{n}{k} \binom{m}{\ell-1} +  \binom{n}{k-1} \binom{m}{\ell} \Bigg] \partial_{m-\ell+1,n-k+1} \psi(z_i) \label{eq:Gamma dynamics}
\end{split}
\end{align}
We observe that $d \Gamma_{i}^{\ell k}/dt$ depends on $\psi$ at the vortex cores $z_{i}$, and the vortex core dynamics depend on $\psi$ as well.
Fortunately, we already found that $\psi$ is purely a function of $\Gamma_{i}^{\ell k}$ and $z_{i}$, as stated in \eqref{eq:stream}.
Thus \eqref{eq:Gamma dynamics} and \eqref{eq:core dynamics} (with \eqref{eq:stream}) form a closed finite dimensional system.
Most notably, by construction the vorticity equation \eqref{eq:vorticity} admits the $N$th order MVB ansatz for the vorticity in \eqref{eq:ansatz N} as a solution when the $z_i(t)$'s and the $\Gamma_i(t)$'s satisfy the just derived finite dimensional system.

\begin{rmk}
	In the point vortex method (i.e. the un-regularized case where $L_{\alpha} = 1$), this derivation of the dynamics requires an extra step.  In particular, one must discard the self-interaction term, which we will describe here.
	For point vortices, $\omega = - \Delta \psi$. Substituting the point-vortex ansatz $\omega = \sum_i \Gamma_i \delta(z - z_i(t))$ into the equations of motion \eqref{eq:omega eom} would then
	yield the non-sensical equation
	\begin{align*}
		\dot{z}_i = \nabla^\perp \left( \sum_{j} \Gamma_j \log | z_i - z_j | \right).
	\end{align*}
	We say ``non-sensical'' because the right hand side explodes when you evaluate the $i$th term in the sum, the self-interaction term.
	Historically, it is customary to discard this self-interaction term based on physical and symmetry principles \cite[Chapter 4]{MarchioroPulvirenti1994}.
	In contrast, for blob methods the logarithmic kernel is replaced with a differentiable kernel function, such as a Gaussian.
	This allows one to retain the self-interaction terms.
	In the case of standard vortex blobs (i.e. 0th order MVBs), this distinction makes no difference because the gradient of the kernel vanishes at the origin and the self-interaction term contributes nothing to the dynamics.
	However, the derivatives of the kernel of degree $2$ and higher do not vanish at the origin.
	As a result, the self-interaction terms do contribute to the dynamics for MVBs of order $2$ and higher.	
	The choice to discard the self-interaction terms in \cite{Yanovsky2009}, versus our choice to keep them, explains one of the major discrepancies between our work and \cite{Yanovsky2009}.
	In particular, \cite{Yanovsky2009} was concerned with generalizing the (un-regularized) point vortex method in the same way that we have generalized the vortex-blob method.
	Once the ansatz $\omega = \sum \Gamma^{mn}_i \partial_x^m \partial_y^n \delta( z - z_i )$ was substituted into the equations of motion, they discarded the self-interaction terms in order to handle the singularities in the velocity field. They had no other choice.
	Except for the initial regularization step we took, this discarding of the self-interaction term is the primary place where the derivation of the equations of motion presented here diverges from the derivation in\cite{Yanovsky2009}.
	Discarding the self-interaction term in \cite{Yanovsky2009} lead to contradictory compatibility equations for singularities of degree $2$ and higher.
	This is one regime where the self-interaction terms have an impact on the dynamics in our regularized formulation.
	One of the major findings of \cite{Yanovsky2009} was that one could avoid these contradictory compatibility conditions by limiting one's self to combinations of point-vortices and dipoles.
	Even in this limited scenario, our equations of motion do not match even in a regularized sense, as the vortex cores of the dipoles are not advected by the (singular) velocity field in \cite{Yanovsky2009}.
	Additionally, as the regularization parameter goes to $0$ in our framework, the velocity fields become singular, and the equations of motion for the $\Gamma$'s will explode.
	So we can not expect to observe any form of convergence to the finite valued ODEs of \cite{Yanovsky2009}.
\end{rmk}

\section{Conserved Quantities}
\label{sec:conserved}
  In this section we begin to touch upon some of the symplectic geometry of MVBs.
  To begin, let us consider a general vorticity distribution $\omega$.
  The energy is defined as
  \begin{align*}
  	H(\omega) := \frac{1}{2} \int \omega(z) G_\alpha(z-z') \omega (z') dz\, dz' \equiv \frac{1}{2} \int \psi(z) \omega(z) dz.
  \end{align*}
  where $\psi = G_\alpha * \omega$.
  The vorticity equation, \eqref{eq:vorticity}, can be seen as an instance of Hamilton's equations on a Poisson manifold.
  In this case the Poisson manifold is the space of vorticity distributions, and the Poisson bracket is the vorticity Poisson bracket derived in \cite{MarsdenWeinstein1983}.
  As the Hamiltonian is conserved by Hamilton's equations, we should expect $H(\omega)$ to be constant in time.
  Indeed, we find that if $\omega$ satisfies \eqref{eq:vorticity}, then
  \begin{align*}
  	\frac{dH}{dt}(\omega) &= \frac{d}{dt} \left( \frac{1}{2} \int \omega(z) G_\alpha(z-z') \omega (z') dz\, dz' \right).  \\
	&= \int (G_{\alpha} * \omega (z))\, \partial_t \omega(z) dz  \\
	&= \int \psi(z)\, \partial_t \omega(z) dz \\
	&= \int \psi \left(\partial_x  \omega\,  \partial_y \psi - \partial_y \omega \, \partial_x \psi  \right) dz \\
	&= \int \partial_y \left( \frac{1}{2} \psi^2 \right) \partial_x \omega - \partial_x \left( \frac{1}{2} \psi^2 \right) \partial_y \omega dz \\
  \end{align*}
  By integration by parts, we can remove the partial derivates from the $\omega$'s to find
  \begin{align*}
  	=  \int -\partial_{xy} \left( \frac{1}{2} \psi^2 \right) \omega + \partial_{yx} \left( \frac{1}{2} \psi^2 \right) \omega dz = 0 \\
  \end{align*}
  which vanished by the equivalence of mixed partials.
  
  As \eqref{eq:vorticity} is a Hamiltonian system, we can consider searching for symmetries to find other conserved quantities using Noether's theorem.
  We've relegated the discussion of the relevant symplectic structure to Appendix \ref{sec:symmetries}, where derivations and proofs of the following can be found.
  Here we can summarize the appendix.
  
  It's simple to observe that the Hamiltonian $H$ is translation invariant, and that $H$ is rotationally invariant as long as the kernel $G_{\alpha}$ has rotational symmetry.
  Thus we should expect there to be conserved quantities tied to these symmetries.
  We find that the quantities
  \begin{align*}
	{\bf J}_{\rm lin}(\omega) &= \left( \int -y \, \omega(z) dz , \int x \, \omega(z) dz \right)\\
	{\bf J}_{\rm ang}( \omega) &= \int (x^2 + y^2) \omega(z) dz
  \end{align*}
  are conserved.  The relationship between these quantities and the symmetries of the system is explained in Appendix \ref{sec:symmetries}.
  Alternatively, one can observe the conservation of these quantities by direct calculation in the same way that conservation of energy was verified.
    
  As the MVB ansatz is consistent with \eqref{eq:vorticity} we can substitute the MVB ansatz into the above conserved quantities, to obtain conserved quantities for the MVB evolution, \eqref{eq:core dynamics} and \eqref{eq:Gamma dynamics}.
  We obtain the following conserved quantities:
  \begin{align*}
	  {\bf J}_{\rm ang} &= \sum_i \frac{\Gamma^{0,0}_i}{2} (x_i^2 + y_i^2) - \Gamma_i^{1,0} x_i - \Gamma_i^{0,1} y_i + \Gamma_i^{2,0} + \Gamma_i^{0,2} \,, \\
	  {\bf J}_{\rm lin} &= \sum_i ( \Gamma_i^{0,1} - \Gamma^{0,0}_i y_i , \Gamma^{0,0}_i x_i -\Gamma^{1,0}_i ) \,, \\
	  H &= \sum_{m,n,\ell,k,i} (-1)^{m+n+\ell +k} \Gamma^{mn}_i \Gamma_j^{\ell k} \partial_{m+\ell}^x \partial_{n+k}^y G(z_i-z_j)\,.
  \end{align*}
  Again, the first two quantities, ${\bf J}_{\rm ang}$ and ${\bf J}_{\rm lin}$, are momenta derived from Noether's theorem for the rotational and translational symmetries of the fluid.
  The quantity $H$ is the kinetic energy of the fluid.
  In section \ref{sec:symplectic} we will characterize the MVB dynamics as Hamiltonian systems, with Hamiltonian $H$.

	To each individual MVB there are numerous conserved quantities which can be seen as a manifestation of the conservation of circulation.
	To show this, let $\vec{u} = (u,v) = (\partial_y \psi, - \partial_x \psi)$ be a time-dependent vector field which satisfies \eqref{eq:vorticity}.
	The flow of $\vec{u}$ is the diffeomorphism, $\Phi_t: \mathbb{R}^2 \to \mathbb{R}^2$,
	which sends particle labels at time $0$ to their positions at time $t$.
	If $\omega_t$ is the vorticity at time $t$ then $\omega_t( \Phi_t(z) ) = \omega_0$ is constant in time.
	This conservation law can be seen as a corollary of Kelvin's circulation theorem \cite{ArnoldKhesin1992}.
	As a consequence, the quantity
	\begin{align*}
		J(t) := \int \omega_t( \Phi_t(z) ) f(z) dz
	\end{align*}
	is constant in time for any $f \in C_0^\infty(\R^2)$.
	By applying the change of variables formula and invoking the
	incompressibility condition, $\det(D\Phi) = 1$, we find
	\begin{align*}
		J(t) = \int \omega_t( z) f(\Phi_t^{-1}(z)) dz.
	\end{align*}
	This form of writing $J(t)$ makes sense when $\omega_t$ is a distribution.
	As a result, we find that for a vorticity of the form \eqref{eq:ansatz N} the quantity
	\begin{align}
		J(t) = \sum_{
			\substack{
				i \in S \\
				m+n \leq N
			}
		} \Gamma_i^{mn} (-1)^{m+n} \partial_x^m\partial_y^n( f \circ \Phi_t^{-1}) |_{z = z_i(t)} \label{eq:conserved_circulation}
	\end{align}
	is conserved for any $f \in C_0^\infty(\R^2)$.
	While this conservation law holds for all functions with compact support, $f$, we do not obtain infinitely many conserved
	quantities when $\omega_t$ satisfies the MVB ansatz and $S$ is finite.
	This is because the expression on the right hand side only depends on the $N$th order Taylor expansion of $f$ at $z_i(0) \equiv \Phi_t^{-1}(z_i(t) )$,
	as is illustrated by the Fa\`a di Bruno formula.
	We will not display the Fa\`a di Bruno formula here because it requires nearly a page of notational definitions before to writing it down
	\cite{ConstantineSavits1996}.
	Nonetheless, by computing the cardinality of jet spaces, one would obtain ${\rm card}(S) \frac{ N(N+1)}{2}$ independent conserved quantities
	as a result of \eqref{eq:conserved_circulation}.
	These conserved quantities can be interpreted as a finite dimensional manifestation of the conservation of circulation.

\section{Moments}
\label{sec:moments}
	In this section we present how the moments of the vorticity distribution evolve in time.
	We will find that when the vorticity distribution is that of a MVB, then the moments form a closed dynamical system at finite order.
	
	The $(a,b)^{\rm th}$ moment of the vorticity, $\omega$, centered around the vortex position $(x_i,y_i) \in \R^2$ is given by
	\begin{align*}
		\mu^{ab}_i := \int (x-x_i)^a (y-y_i)^b \omega dxdy\,.
	\end{align*}
	We call the integer $a+b$ the \emph{order} of the moment.
	For a general vorticity, the evolution for the $n$th order moments will depend on the $(n+1)$th and higher order moments
	and so we can not concoct a closed dynamical system for the moments of order $n$ and less.
	However, this is not the case if $\omega$ satisfies the MVB ansatz, and the points $(x_i,y_i)$ are given by the locations of the jet-vortices.
	If $\omega$ satisfies the MVB ansatz \eqref{eq:ansatz N} then
	\begin{align*}
		\mu^{ab}_i = \sum_{
			\substack{
				j \in S \\
				m \leq a ,
				n \leq b
			}
		}
		(-1)^{m+n} \frac{a! b!}{(a-m)!(b-n)!} \Gamma_j^{mn} (x_j - x_i)^{a-m} (y_j - y_i)^{b-n}\,,
	\end{align*}
	for $a+b \leq N$ and $i \in S$.
	Given the points $z_i \in S$, one can write the moments in terms of the circulation
	strengths, the $\Gamma$'s.
	For the moment $\mu_i^{ab}$ with $a+b \leq N$ with $a,b \in \mathbb{N}$
	we may invert this relationship to write $\Gamma_i^{mn} = \Gamma_i^{mn}( \mu)$, i.e. as a function of the moments.
	Invoking the motion equations for the $\Gamma$'s and substituting the relation between the $\Gamma$'s and the $\mu$'s yields a closed dynamical system for the $\mu$'s.

	\begin{rmk}
	This relation between the $\Gamma$'s and the $\mu$'s may also be important in the context of plasma physics, especially when one recalls that \eqref{eq:vorticity} can be interpreted as a one-dimensional plasma model.
	Specifically, phase-space moments of the Vlasov probability distribution form an important  dynamical link between Lagrangian-particle and Eulerian-continuum descriptions.
The phase-space moments of the Vlasov probability distribution provide \emph{collective coordinates} for the Hamiltonian dynamics of ensembles of particles.
	For more explanation of this property of Hamiltonian collectivization of the phase-space moments, see \cite{GuilleminSternberg1990,HolmLysenkoScovel1990,GibbonsHolmTronci2008a,GibbonsHolmTronci2008b}.
	In plasma dynamics, the phase-space moments arise from a Taylor expansion of the Vlasov particle distribution, taken around its centroid in phase space. For planar incompressible flow of an ideal fluid, the phase space comprises the $(x,y)$ Lagrangian coordinates of a fluid particle, and the corresponding moments arise from Taylor expansions around the centroid of the (smooth) vorticity distribution. The duality between the resulting spatial moments of a smooth vorticity distribution and the MVBs corresponding to higher-order singular vorticity distributions also obtained from a Taylor expansion raises the intriguing question of finding a relation between these two types of dynamical description. This question is particularly intriguing because the dynamics of moments beyond quadratic order in general does not close to form a finite-dimensional Hamiltonian system, while the dynamics of MVBs closes at every order. 
	\end{rmk}
	
	\begin{rmk}
	There exist other systems for approximating the dynamics of moments which differ from the one presented here.
	In particular, the equations of motion for the moments here form a closed system at order $N$, whereas other methods for deriving dynamical systems for moments
	\cite{UminskyWayneBarbaro2010, NagemSandriUminskyWayne2009,GibbonsHolmTronci2008a,GibbonsHolmTronci2008b}
	require truncations in order to form a closed system. 	For example, \cite{UminskyWayneBarbaro2010} approximates the stream function as a sum of Hermite functions with evolving centroids and weights.  In order to obtain the evolution for the weights and the centroids they project the viscous vorticity equation onto this space via $L^2$ projection.
	The resulting formulas are explicit and efficient to compute, albeit more complex than the formulas found in this paper.
	The primary source of error for \cite{UminskyWayneBarbaro2010} over long times is the discrepancy between the projected evolution equations and the true evolution equations.
	In contrast, we approximate an Euler fluid with a regularized fluid equation which we solve exactly.
	This is not to say that error is not accumulated in time.
	The primary source of error for our method over long times is the discrepancy between between the regularized fluid equations and the true fluid equations.
	
	Admittedly, the equations of motion for the moments in \cite{UminskyWayneBarbaro2010} bear some resemblance to the equations of motion for the $\Gamma$'s in our method.
	Both are quadratic in their respective variables, with coefficients involving combinatorial functions.  A more precise relationship, if one exists, is difficult to discern.
	Philosophically, the methods share much in common.  However, due to the fundamental approximation technique of projecting the equations of motion versus regularizing them, the methods are indeed distinct.  This difference cascades throughout the study of both methods.  For example, the convergence for \cite{UminskyWayneBarbaro2010} is obtained via the convergence of spectral approximations,
	while the convergence of our method is a corollary of the convergence of a regularized fluid model (see \cite{MumfordMichor2013,FoiasHolmTiti2001} for such convergence proofs).
	\end{rmk}

\section{Numerical Aspects}
\label{sec:numerics}
In this section we discuss various numerical aspects of using MVBs to model fluid dynamics.
We will observe how MVBs can be used to reduce the number of necessary pairwise computations without a drastic compromise in accuracy.
We will also present an algorithm for constructing an initial condition of MVBs from a given stream
function.

\begin{rmk}
We refer to \cite{ChiuNivolaides1988} for a convergence proof and error analysis of the 1st order case.
A convergence proof is beyond the scope of this article.
Suffice it to say, such a proof would likely resemble \cite{ChiuNivolaides1988}.
\end{rmk}

\subsection{Grouping and reduction of pairwise computations}
\label{sec:grouping}

Let us consider the vorticity distribution
\begin{align*}
	\omega = \Gamma_1 \delta_{z_1} + \Gamma_2 \delta_{z_2}\,.
\end{align*}
If $z_1$ and $z_2$ are close, we can define the quantities $\bar{z} = (z_1+z_2)/2$ and $\delta z = z_1 - z_2 $ to obtain the approximation
\begin{align*}
	\int \omega(z) f(z)dz &= \Gamma_1 f(z_1) + \Gamma_2 f(z_2) \\
		&= \Gamma_1 \left( f(\bar{z}) + \partial_x f(\bar{z}) \cdot \frac{\delta x}{2} + \partial_y f(\bar{z}) \cdot \frac{\delta y}{2}  \right)\\
		&\quad + \Gamma_2 \left( f(\bar{z}) - \partial_x f(\bar{z}) \cdot \frac{\delta x}{2} - \partial_y f(\bar{z}) \cdot \frac{\delta y}{2}  \right) + o( h ) 
\end{align*}
where $h = \| \delta z \|$.
Therefore the distribution
\begin{align*}
	\tilde{\omega} = \Gamma \delta_{\bar{z}} + \Gamma^x \partial_x \delta_{\bar{z}} + \Gamma^y \partial_y \delta_{\bar{z}}
\end{align*}
with 
\begin{align*}
	\Gamma = \Gamma_1 + \Gamma_2 \quad,\quad \Gamma^x = \frac{\delta x}{2} (\Gamma_2 -\Gamma_1) \quad,\quad \Gamma^y = \frac{\delta y}{2} (\Gamma_2 -\Gamma_1)
\end{align*}
serves as a $o(h)$ approximation of $\omega$ in the sense of distributions.
Moreover, the stream function $\tilde{\psi} := G_\delta * \tilde{\omega}$ is an $o(h)$ approximation of $\psi := G_\delta * \omega$ in the traditional sense of analysis on functions.

We have just described the first case of grouping two $N$th order MVBs concentrated at $z_1$ and $z_2$ into a single $(N+1)$th order MVB concentrated at 
the average position $\bar{z}$.
More generally, we can consider the ansatz
\begin{align*}
	\omega = \sum_{m+n \leq N} \Gamma_1^{mn} \partial_x^m \partial_y^n \delta_{z_1} + \Gamma_2^{mn} \partial_x^m \partial_y^n \delta_{z_2}
\end{align*}
and observe
\begin{align*}
	&\int \omega(z) f(z)dz = \sum_{m+n \leq N} (-1)^{m+n} \left( \Gamma_1^{mn} \partial_x^m \partial_y^n f(z_1) + \Gamma_2^{mn}  \partial_x^m \partial_y^n f(z_2) \right)\\
		&\quad= \Bigg\{ \sum_{m+n \leq N} (-1)^{m+n} \Gamma_1^{mn} \left( \partial_x^m \partial_y^n  f(\bar{z}) + \partial_x^{m+1} \partial_y^n  f(\bar(z)) \cdot \frac{\delta x}{2} + \partial_x^m \partial_y^{n+1} f(\bar(z)) \cdot \frac{\delta y}{2}  \right)\\
		&\qquad +  (-1)^{m+n} \Gamma_2^{mn} \left( \partial_x^m \partial_y^n  f(\bar{z}) - \partial_x^{m+1} \partial_y^n  f(\bar(z)) \cdot \frac{\delta x}{2} - \partial_x^m \partial_y^{n+1} f(\bar(z)) \cdot \frac{\delta y}{2}  \right) \Bigg\} \\
		&\qquad + o( h ) .
\end{align*}
The above computation implies that the quantity
\begin{align*}
	&\tilde{\omega} :=\\
	 &\sum_{m+n \leq N+1} \left( \Gamma_1^{mn} + \Gamma_2^{mn}
	- \frac{\delta x}{2}( \Gamma_1^{ m-1,n} - \Gamma_2^{m-1,n} ) - \frac{\delta y}{2} ( \Gamma_1^{m,n-1} - \Gamma_2^{m,n-1}) \right) \partial_x^m \partial_y^n \delta_{\bar{z}}
\end{align*}
serves as an $o(h)$ approximation of $\omega$.
Of course, this again implies that the corresponding stream functions are approximated to order $h$ as well.
Note that $\tilde{\omega}$ is concentrated above a single point, $\bar{z}$, while $\omega$ is concentrated above two points.

\begin{rmk}
Such reductions are even more dramatic when considering higher order jets.
In particular, $2^N$ zeroth order MVBs can be approximated with a single $N$th order MVB by applying the above approximations iteratively.

The computation of pairwise interactions in the vortex method was once a major bottleneck in implementing the standard vortex method for real-world applications.
It was not until the invention of the fast multipole method, that it became tractable to compute millions of pairwise interactions by
reducing the complexity from an $\mathcal{O}(n^2)$ calculation to an $\mathcal{O}(n \log (n))$ calculation, where $n$ is the number of vortices \cite{GreengardRokhlin1987}.
However, in the case of viscous fluids with boundaries, vorticity is shed from the boundaries.
As a result, the vortex blob method of \cite{Chorin1973} created new vortices at the boundary by using the Kutta condition as a creation criteria.
For these applications, $n$ will grow in time without bound, and some means of discarding vortices must be invoked.
It is here that the grouping of MVBs  could be useful.
If one merges two $N$th order MVBs to obtained a $(N+1)$th order MVB, the amount of scalars and data typically increases.
So one must still make a tough decision as to what data to discard (e.g. through some tolerance or by simply truncating at level $M$).
Nonetheless, the analysis presented here could shed light on how best to implement this approach.
\end{rmk}

\begin{rmk}
The merging of blobs of vorticity has been studied analytically \cite{MelanderZabuskyMcWilliams1998} and 
numerically \cite{WeissMcWilliams1993,MelanderZabuskyMcWilliams1998,DizesVerga2002},
as well as in the laboratory \cite{FineDriscollMalmbergMitchell1991}.
All of this study has been in the slightly viscous (or nearly inviscid) regime.
The grouping approach discussed here can be used to numerically resolve such collision events.
In theory, there is no issue with collisions because we are considering regularized vortices where
the induced velocity field from a single MVB is always finite.
However, as $\delta$ becomes smaller, the velocity near the vortex core diverges.
This should be of concern as the convergence analysis of the vortex method pre-supposes that $\delta \ll 1$.
Typically such a near collision is handled by using a smaller time-step (as the ODE is quite stiff).
Grouping of MVBs suggests an alternative by avoiding this pair-wise interaction altogether.
Perhaps such an approach could be viewed as a variation of the punctuated dissipation events
described in \cite{WeissMcWilliams1993} where an initial vorticity distribution is found to asymptotically
approach a smoother axisymmetric vortex blob,
and discrete vortex mergers are implemented to model this behavior.
\end{rmk}

\begin{rmk}
There are qualitative questions which arise from mergers.
For example, when two 0th order vortex blobs are near each other, they will typically scatter after some finite time.
Merging these blobs into a single 1st order blob will prohibit this scattering event from ever occurring.
That both the zeroth-order MVB solution and the merged 1st order MVB represent exact solutions of the fluid (after the merger event)
 is attributable to the long term sensitivity to initial conditions near collision events.
 The scattering angle can be virtually anything since zeroth-order MVBs can waltz around each other many times before scattering.
 The amount of time two zeroth order MVBs can spend waltzing around each other, and perhaps the merged solution represent some sort of limiting solution.
 That is to say, the merged solutions can be interpreted as the ``waltzing for eternity'' solution.
 
 The irreversibility of merging is disturbing when one takes it to its extreme, one massive high order MVB.
In order to address this, a means of splitting high order MVBs into lower order ones should be considered.
The primary difficulty here is in determining when to split.
In the case of mergers, we can decide to merge MVBs when they are close.
Such a criterion is not immediately apparent in the case of splitting MVBs.
\end{rmk}

\subsubsection{A numerical experiment with grouping}
\label{sec:numerical_grouping}
\begin{figure}[h!]
	\centering
	\includegraphics[width=0.15\textwidth]{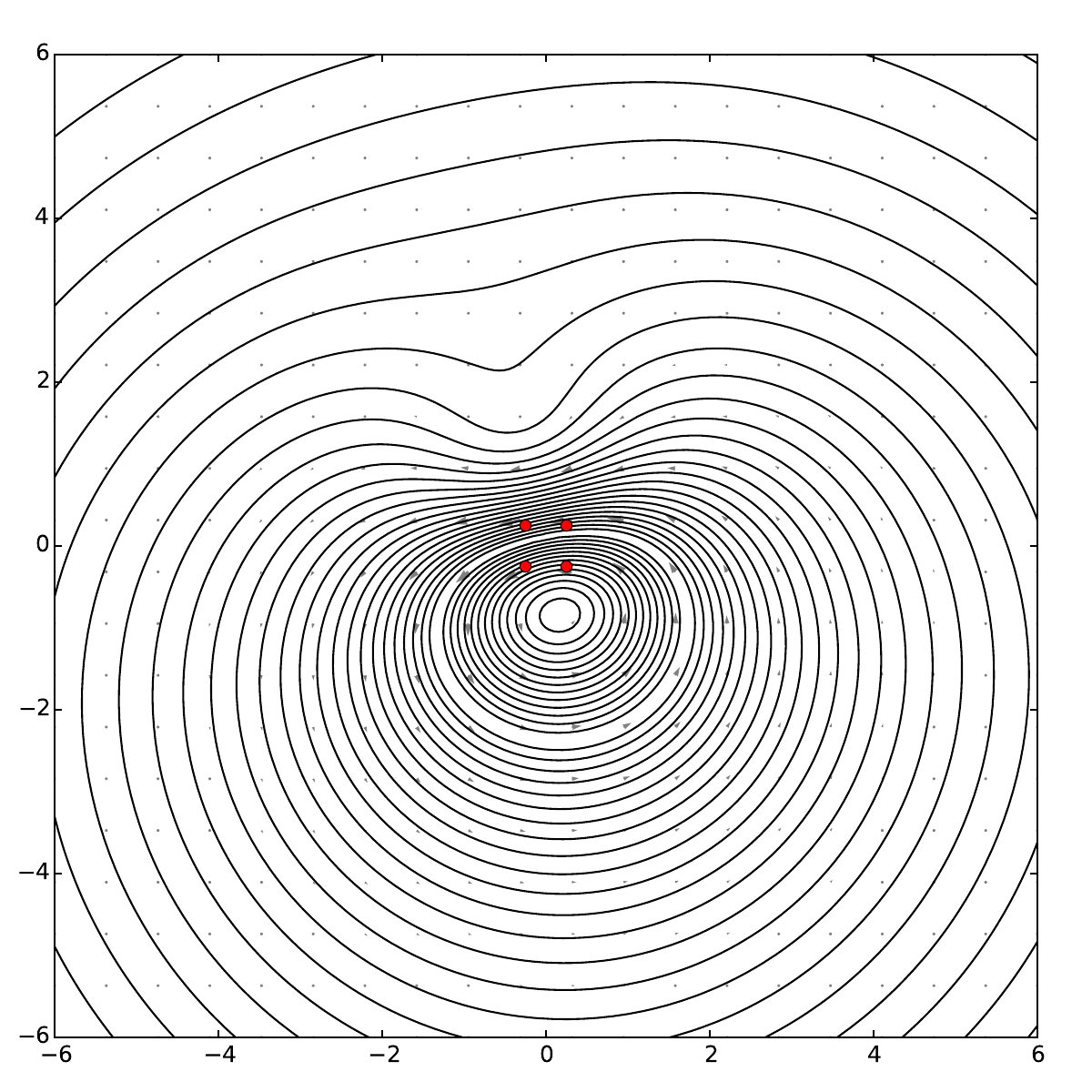}
	\includegraphics[width=0.15\textwidth]{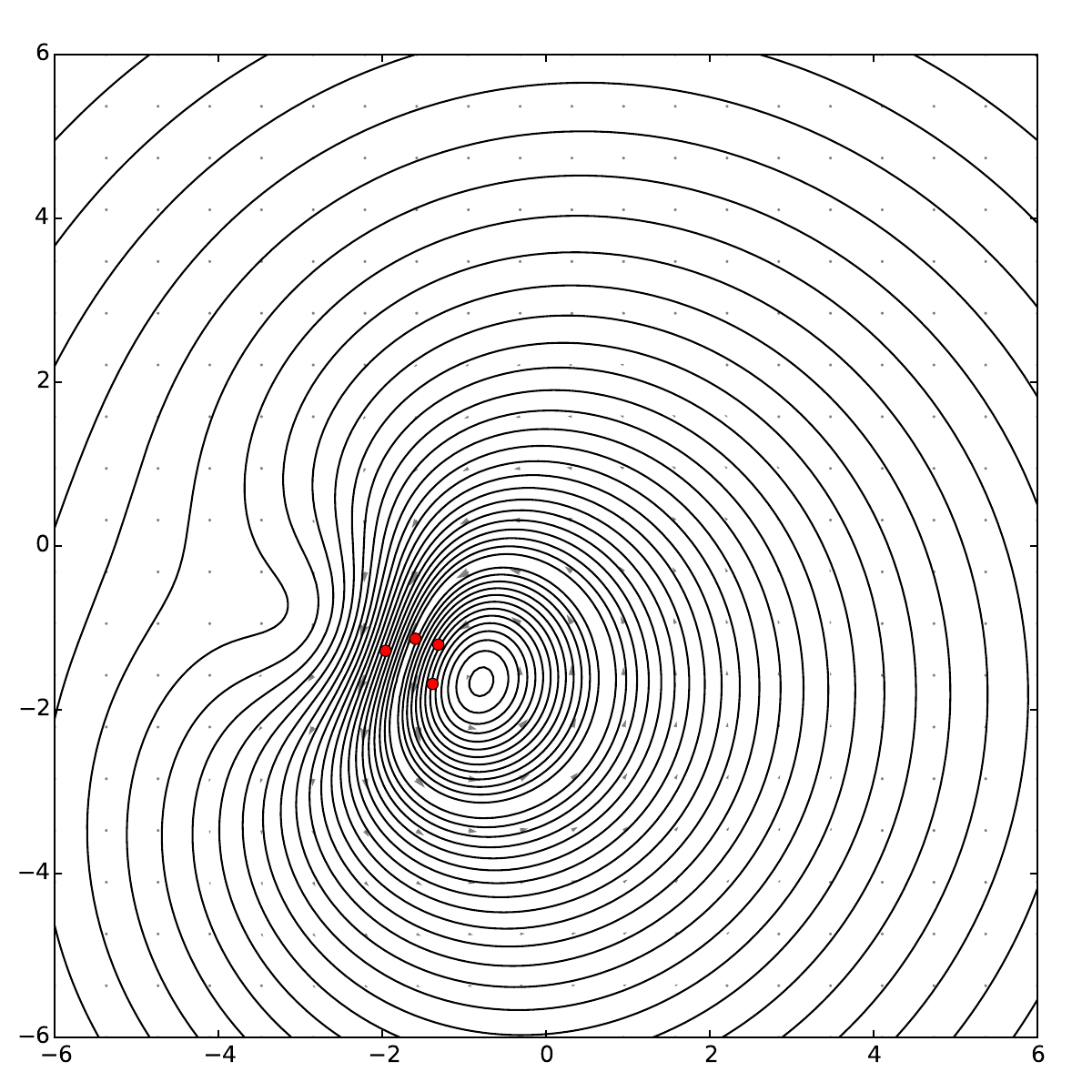}
	\includegraphics[width=0.15\textwidth]{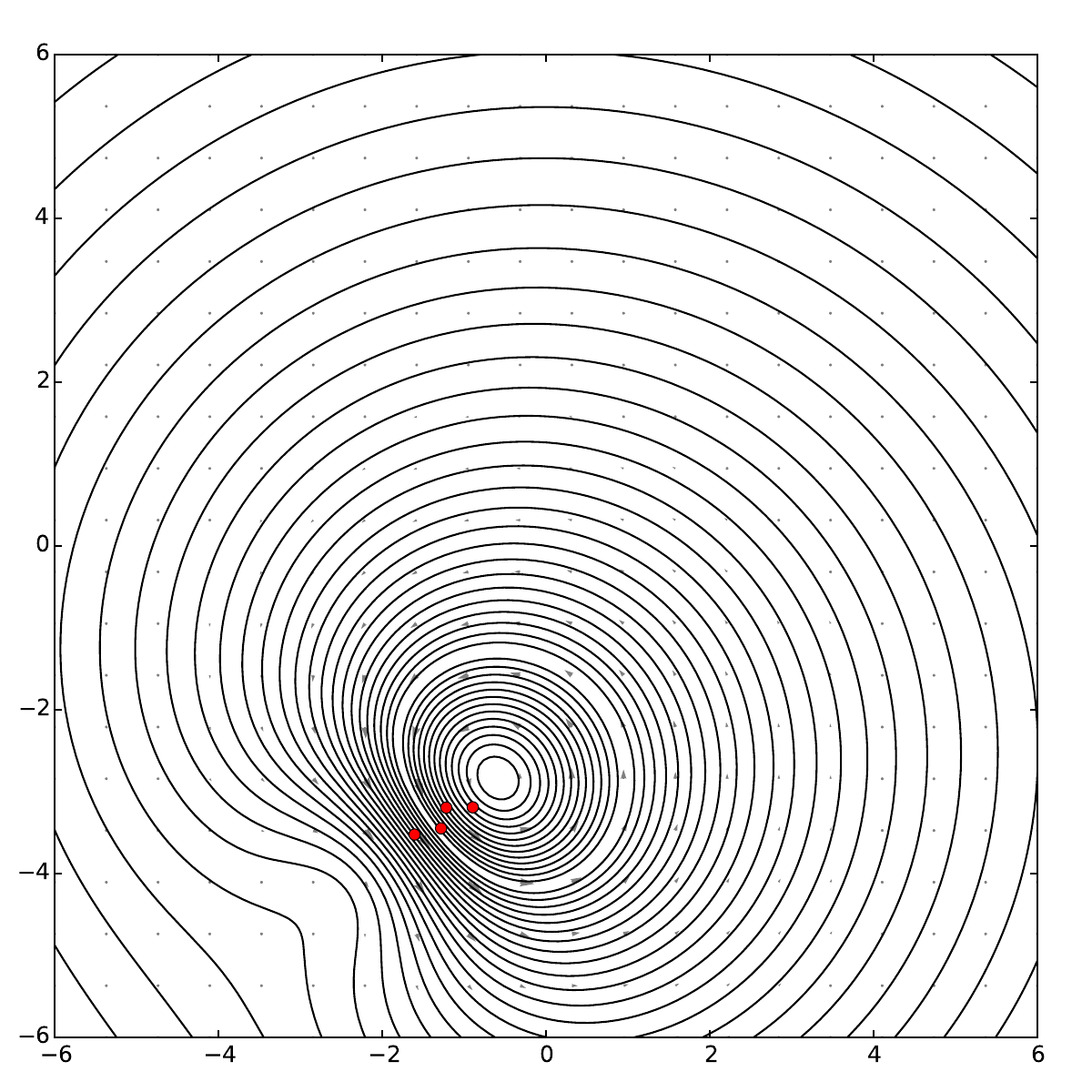}
	\includegraphics[width=0.15\textwidth]{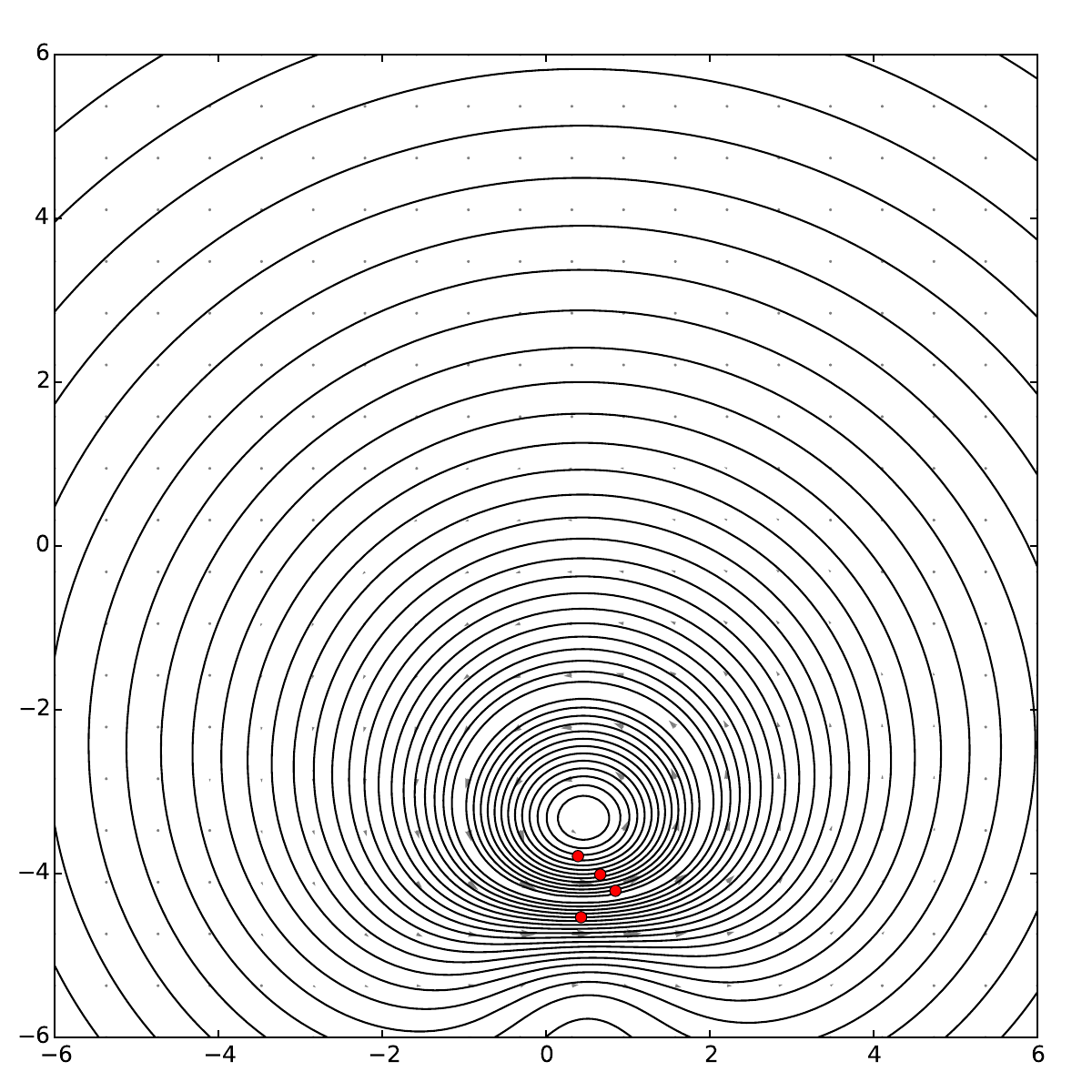}
	\includegraphics[width=0.15\textwidth]{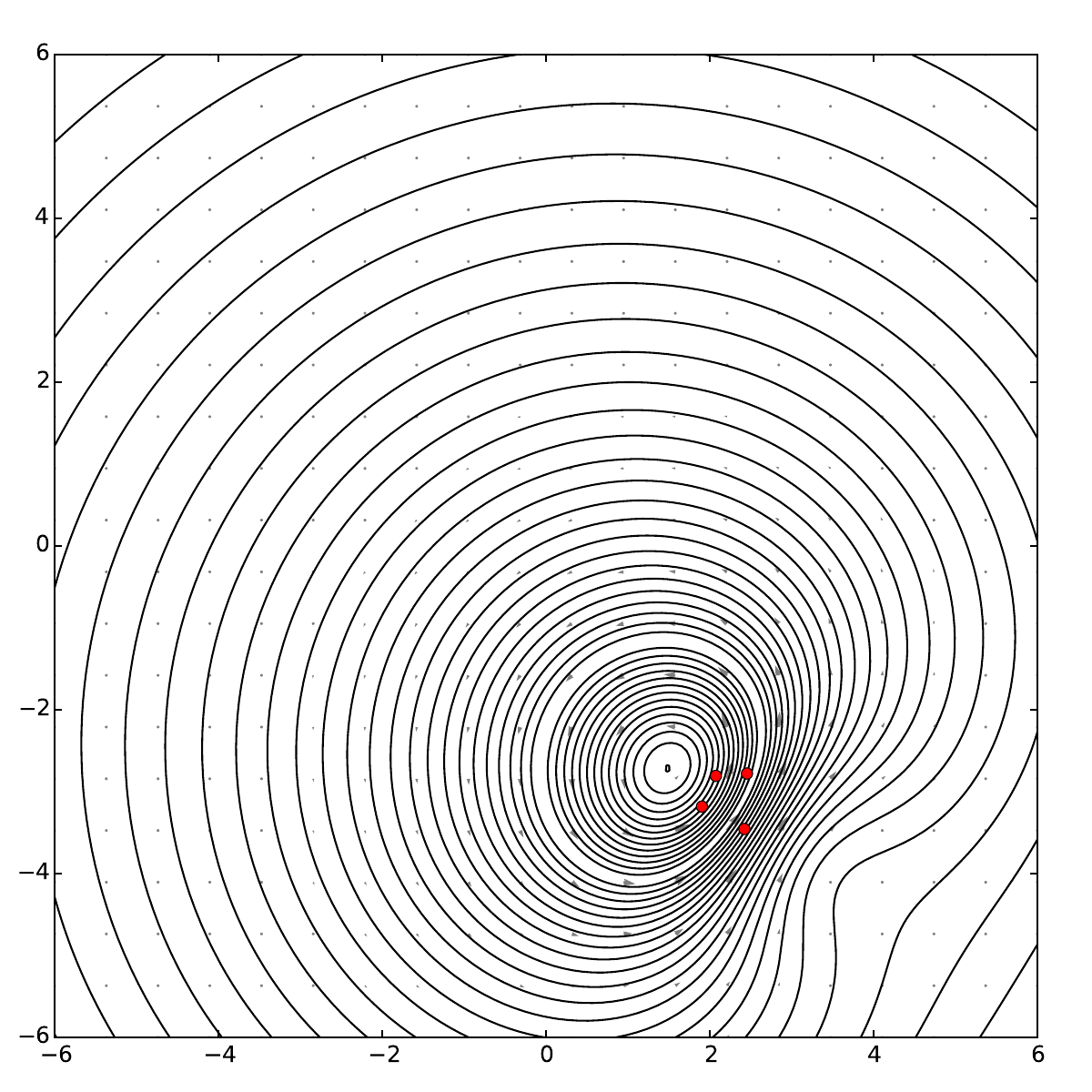}
	\includegraphics[width=0.15\textwidth]{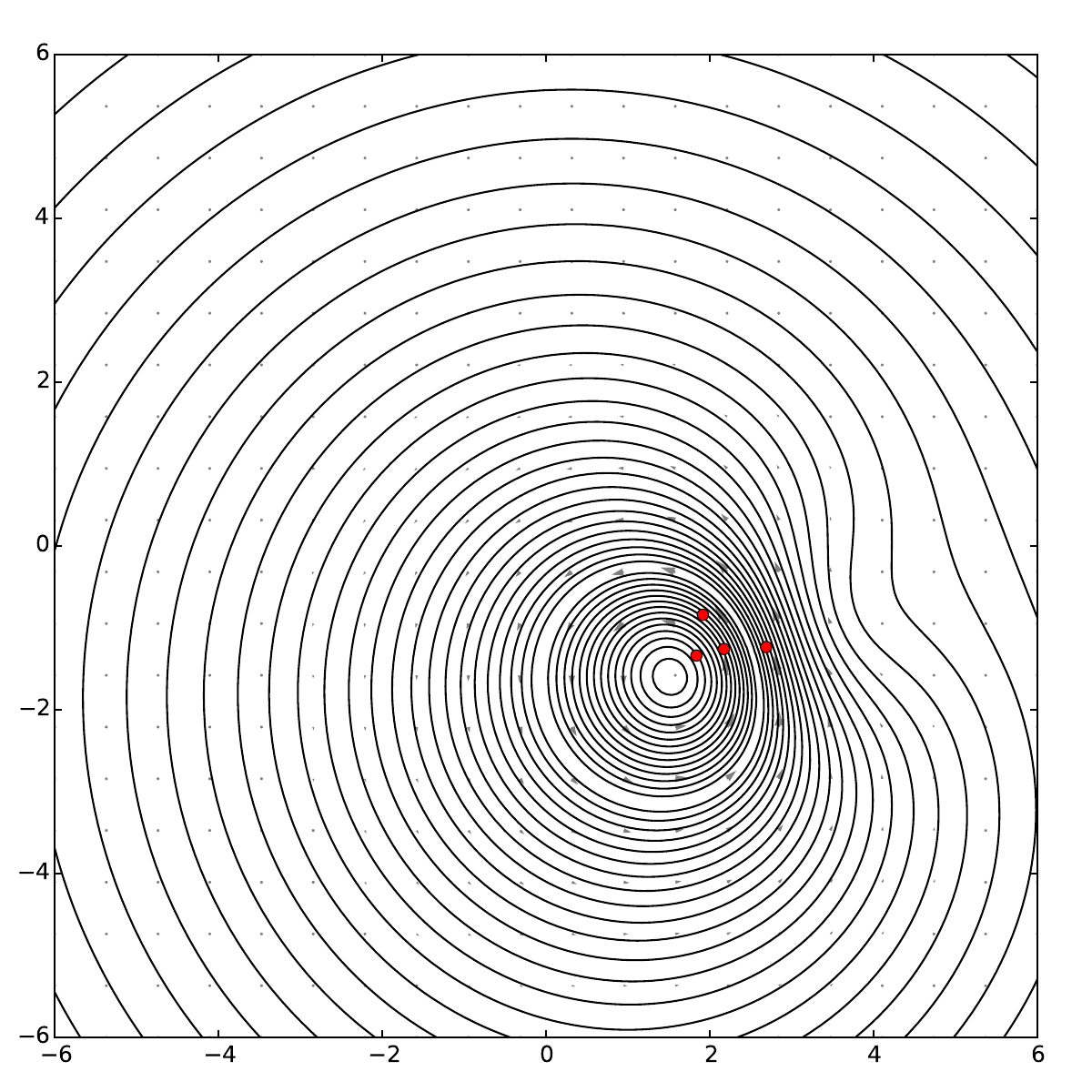}	
	\includegraphics[width=0.15\textwidth]{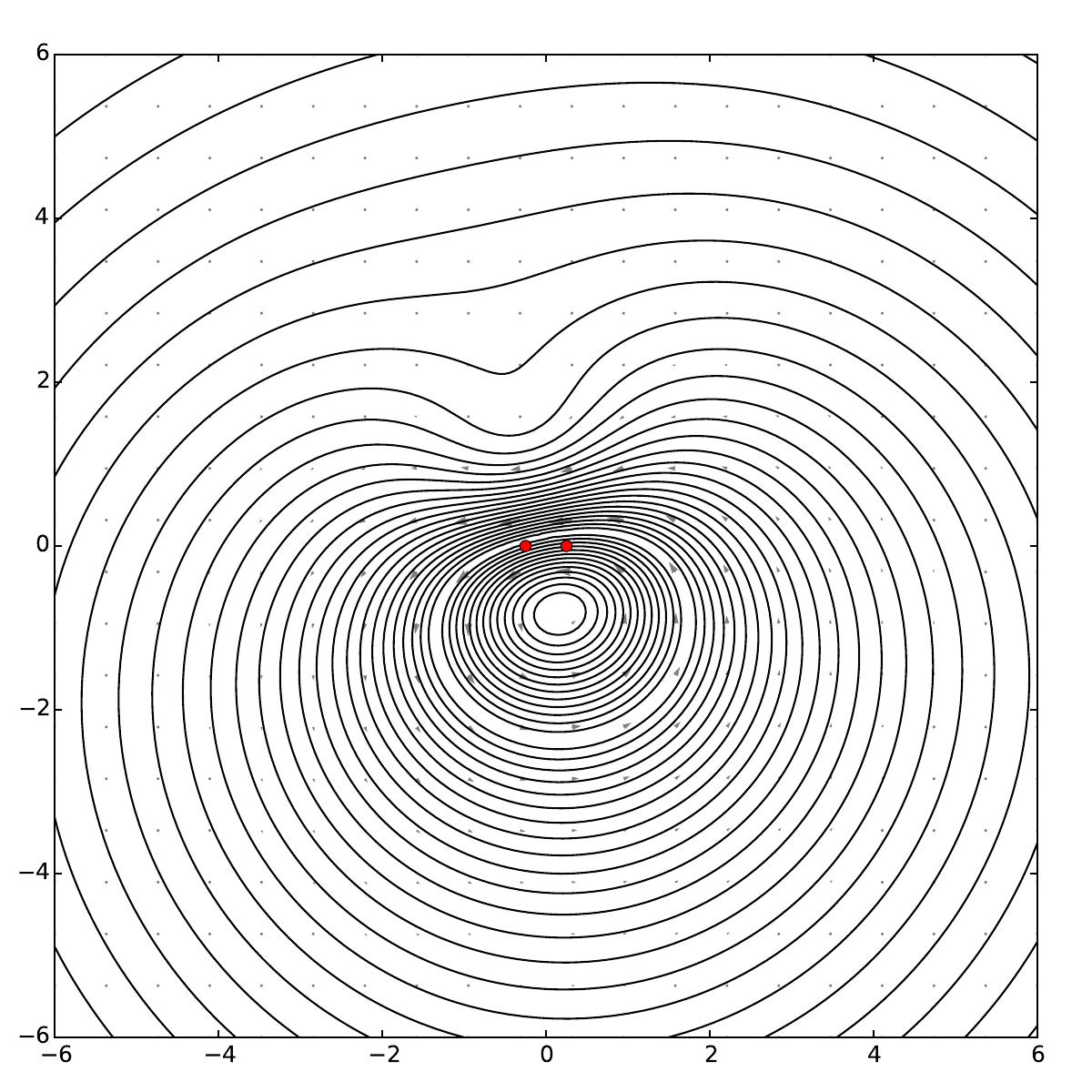}
	\includegraphics[width=0.15\textwidth]{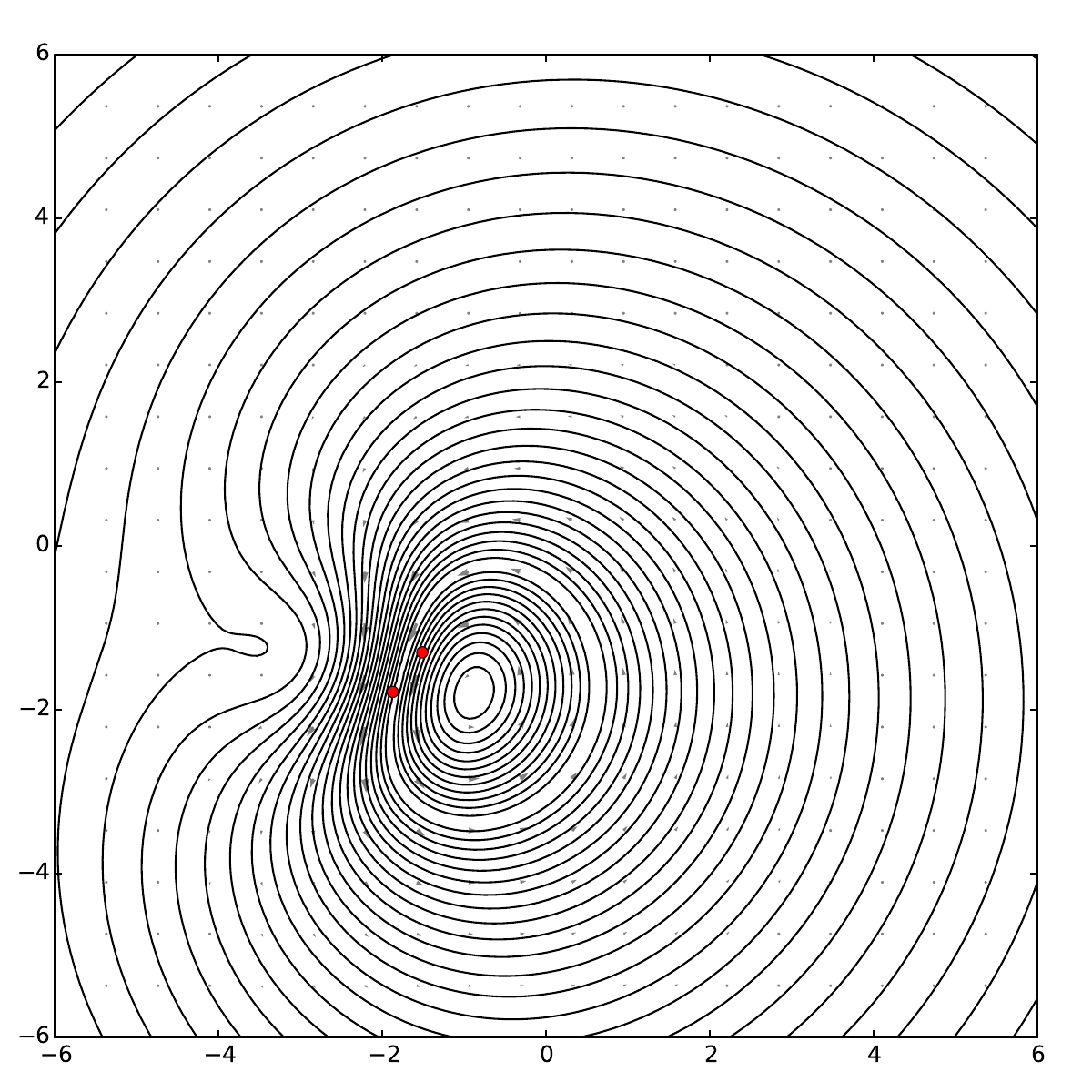}
	\includegraphics[width=0.15\textwidth]{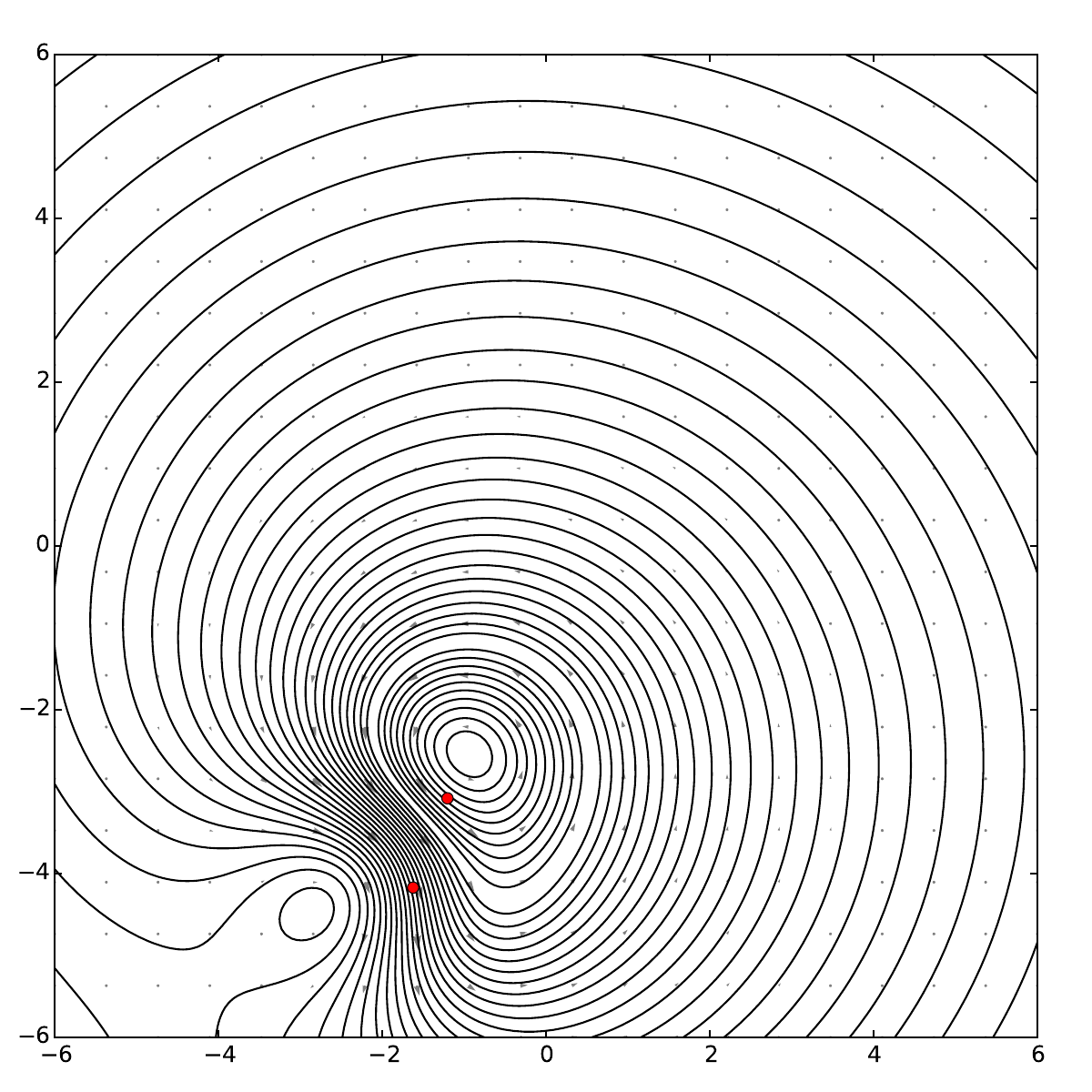}
	\includegraphics[width=0.15\textwidth]{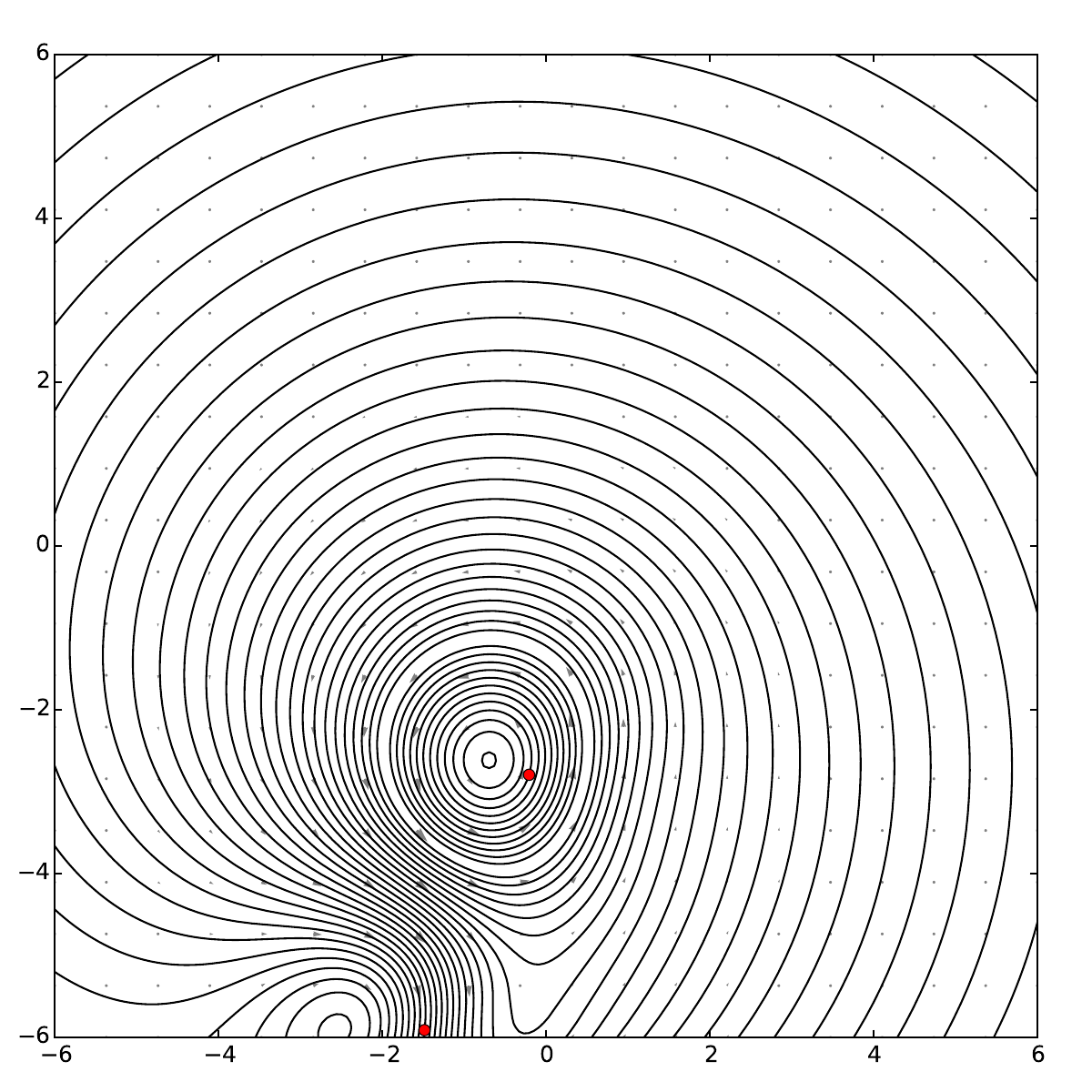}
	\includegraphics[width=0.15\textwidth]{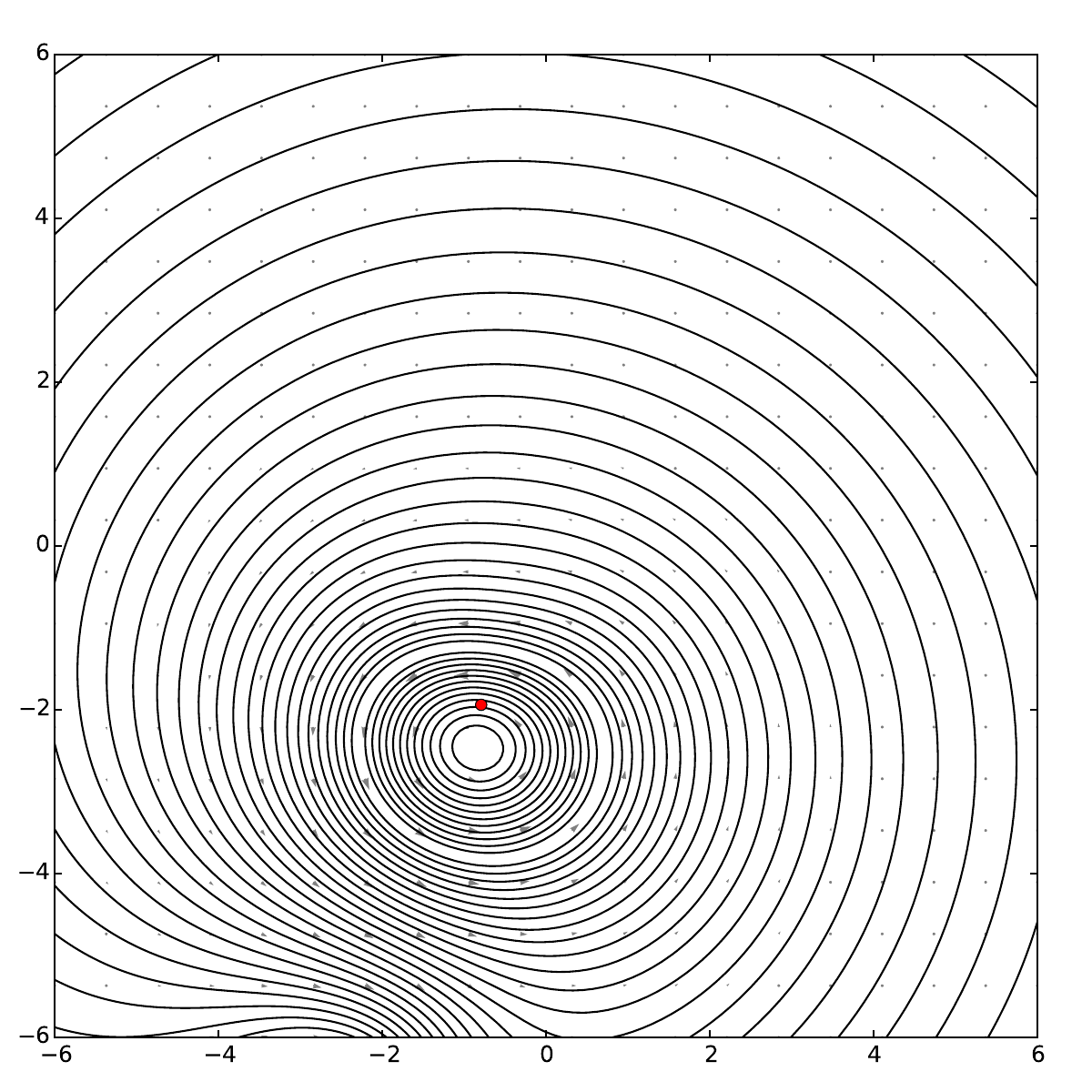}
	\includegraphics[width=0.15\textwidth]{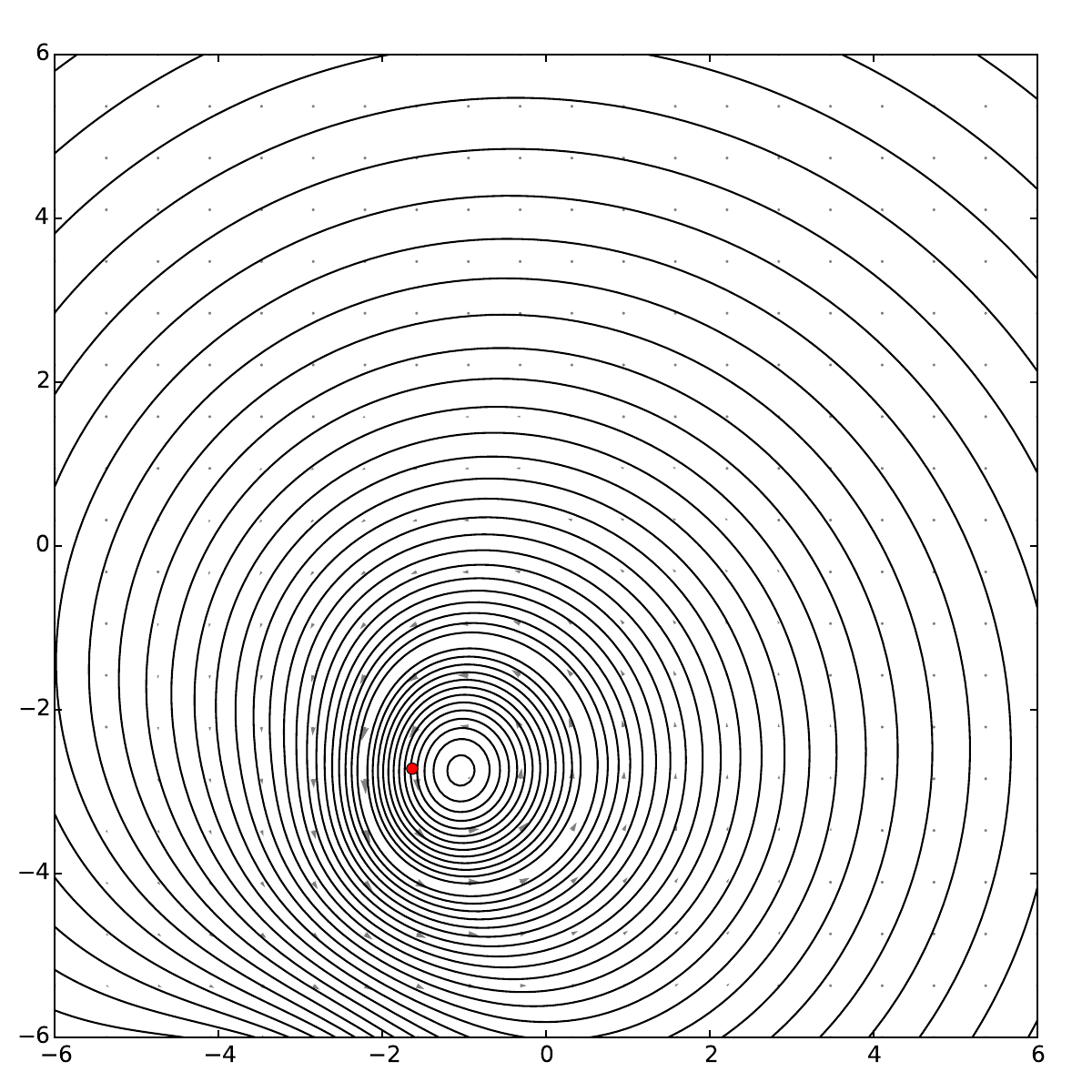}	
	\includegraphics[width=0.15\textwidth]{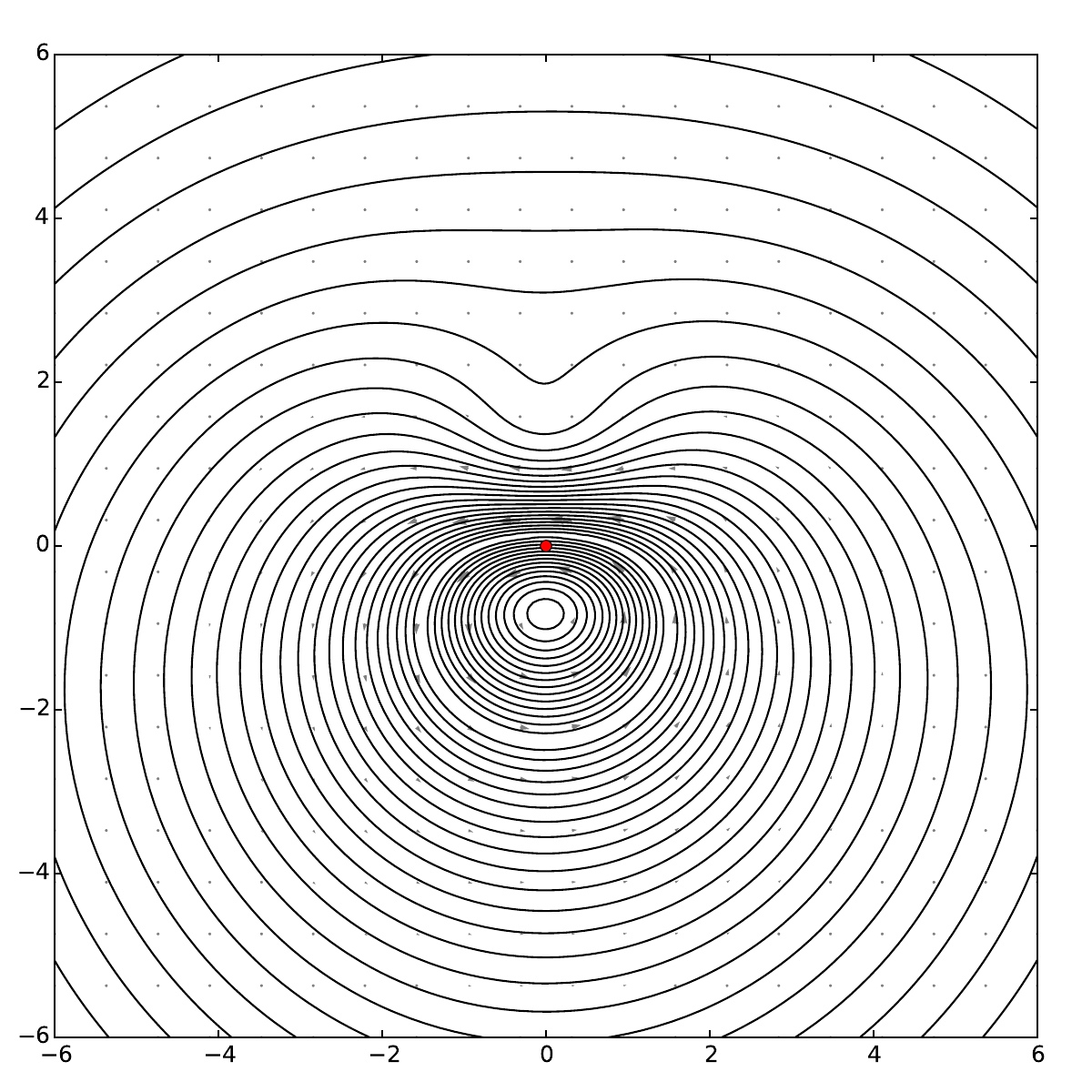}
	\includegraphics[width=0.15\textwidth]{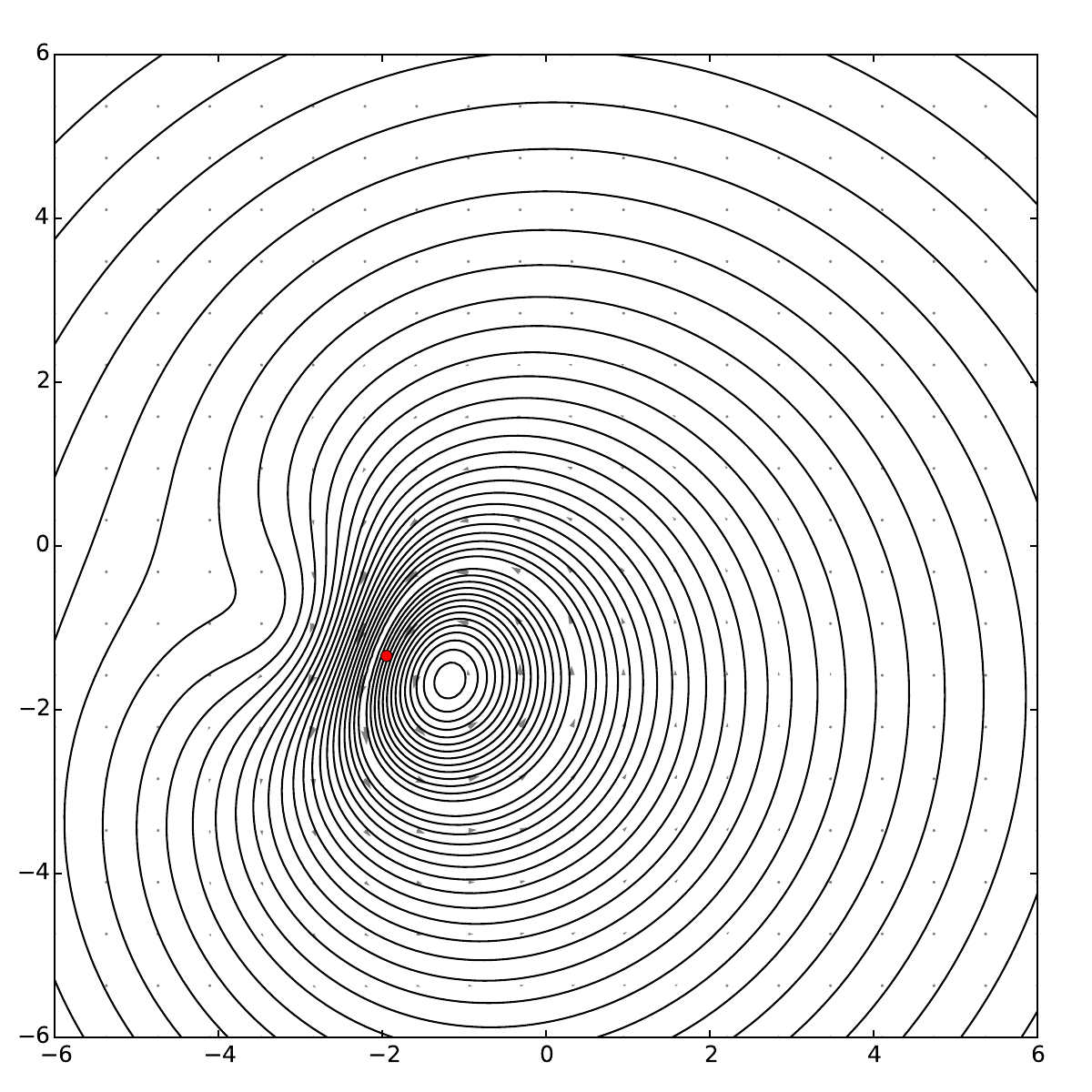}
	\includegraphics[width=0.15\textwidth]{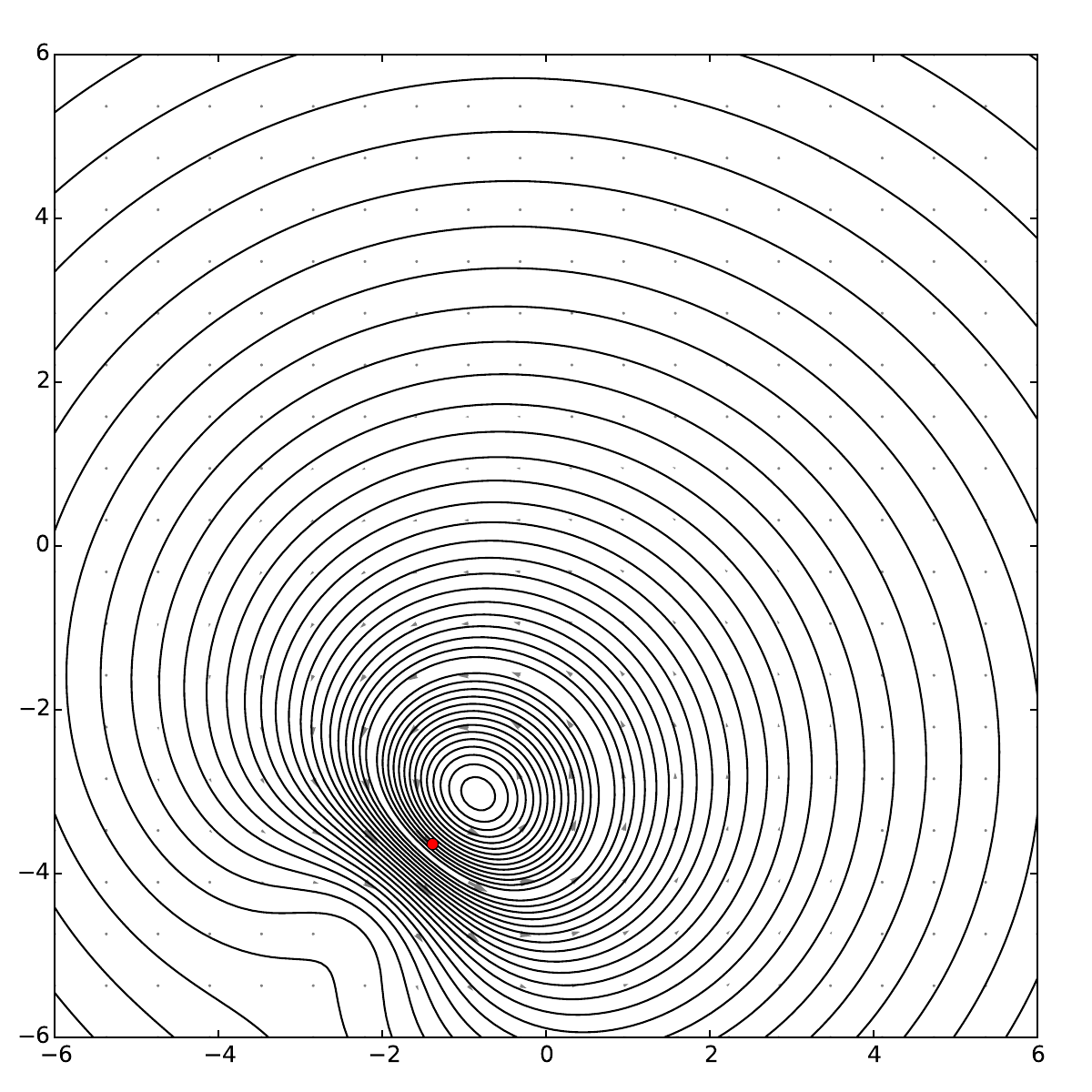}
	\includegraphics[width=0.15\textwidth]{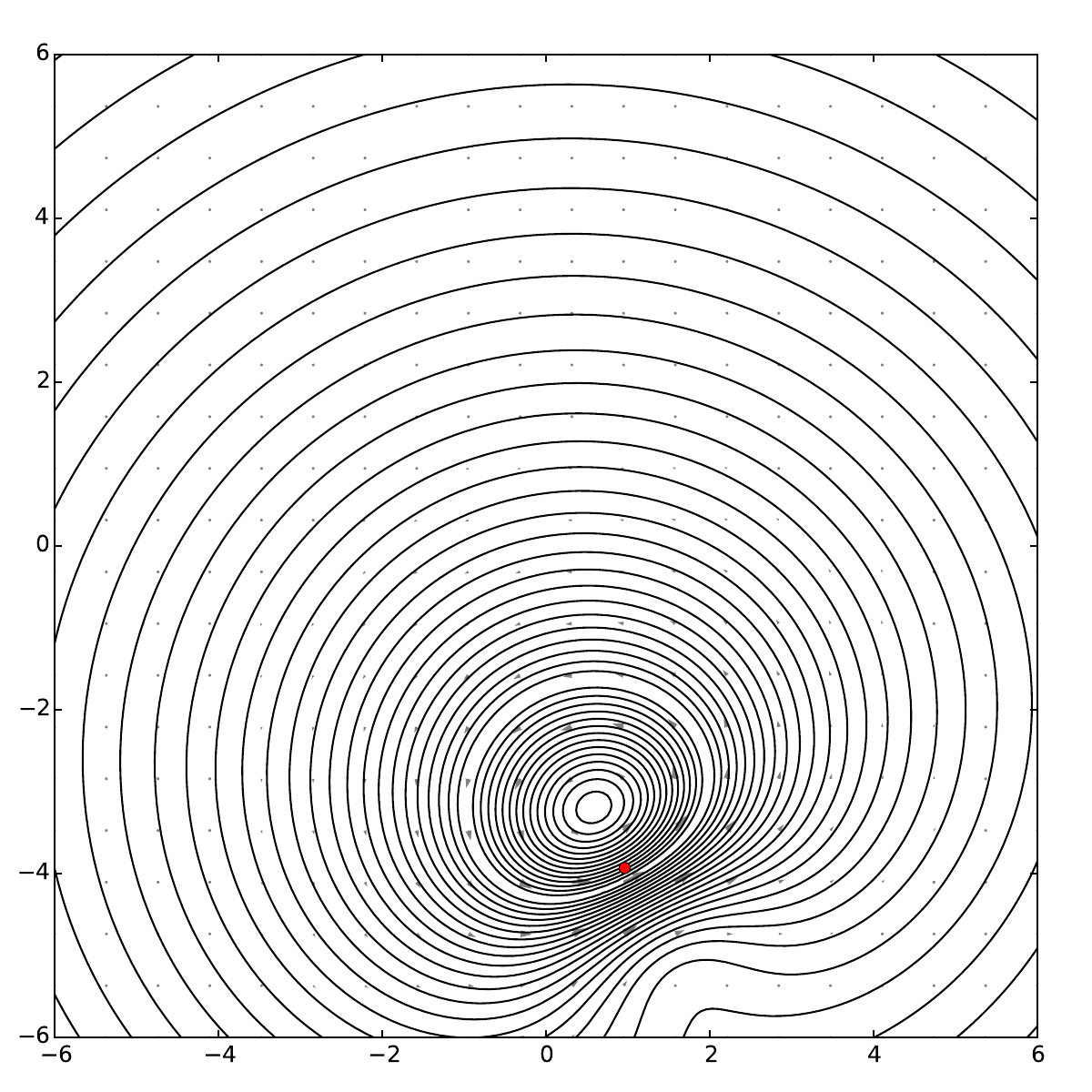}
	\includegraphics[width=0.15\textwidth]{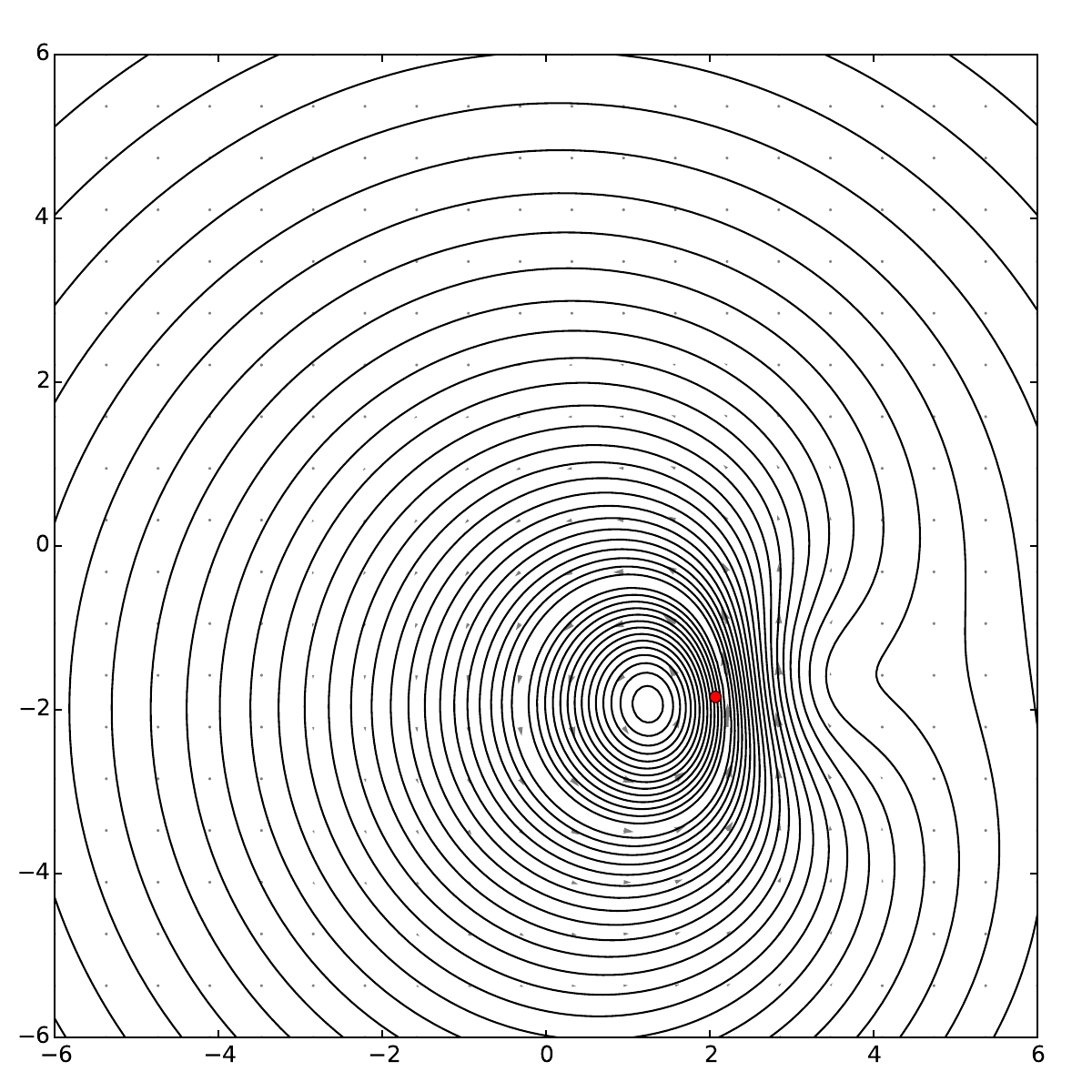}
	\includegraphics[width=0.15\textwidth]{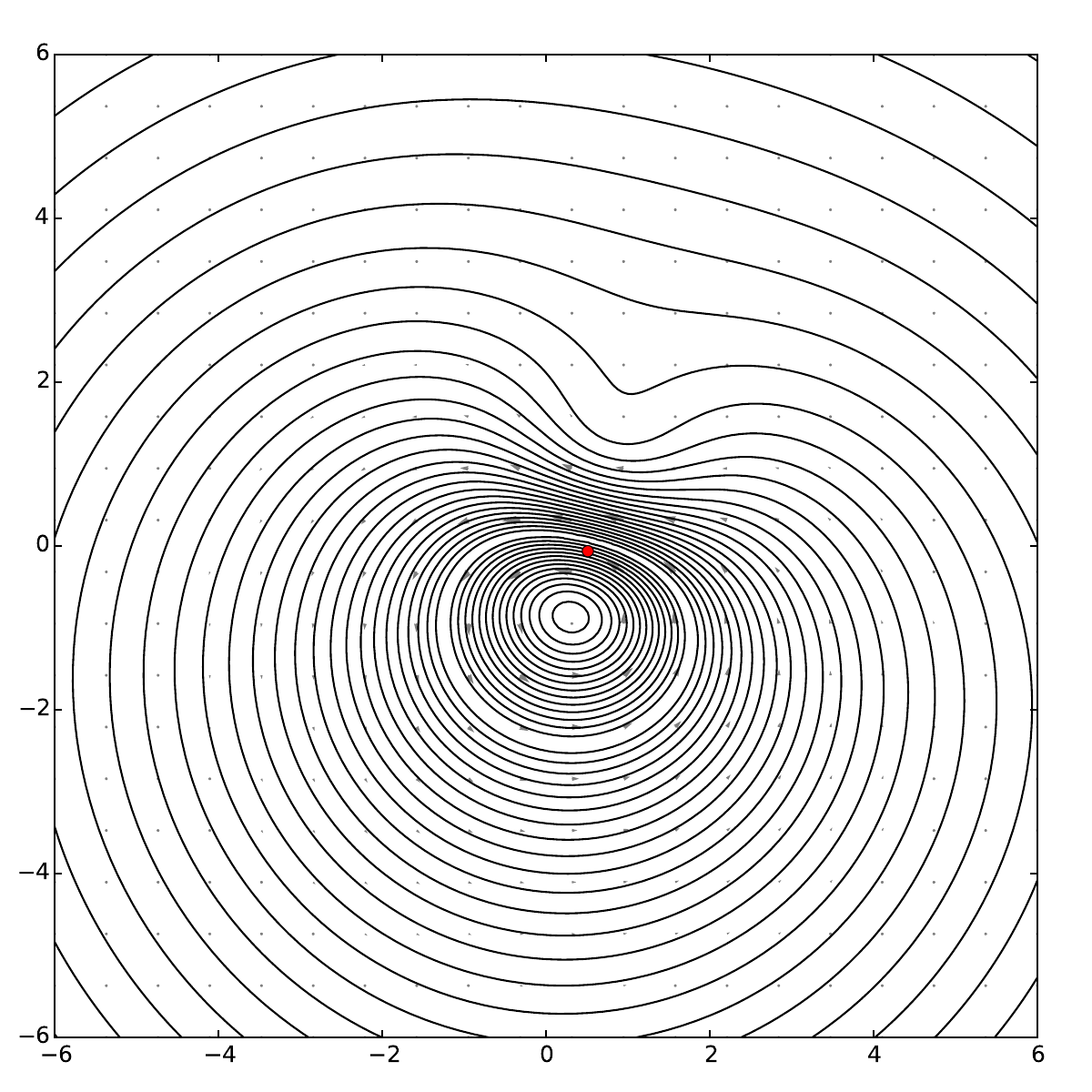}
	\caption{Snapshots of the evolution for various initial conditions at time $t=0,51,101,152,202,253$.
		The top row depicts the evolution of four 0th order MVBs given by the initial condition \eqref{eq:ic_0}.
		The middle row depicts the evolution of two 1st order MVBs obtained by grouping.
		The bottom row depicts the evolution of one 2nd order MVB obtained by grouping.
		}
	\label{fig:grouping}
\end{figure}

For illustrative purposes we can numerically group four 0th order MVBs into two 1st order MVBs, and then one 2nd order MVB.
In particular, we can consider the initial condition
\begin{align}
	\begin{cases}
		z_0 = (-0.25,-0.25) \quad,\quad \Gamma_0 = \phantom{-}0.3 \\
		z_1 = (-0.25,\phantom{-}0.25) \quad,\quad \Gamma_1 = -0.35 \\
		z_2 = (\phantom{-}0.25,\phantom{-}0.25) \quad,\quad \Gamma_2 = -0.2 \\
		z_3 = (\phantom{-}0.25,-0.25) \quad,\quad \Gamma_3 = \phantom{-}0.4 \\		
	\end{cases}
	\label{eq:ic_0}
\end{align}
The corresponding dynamics are depicted in the top row of figure \ref{fig:grouping}.

Next we group $z_1$ with $z_0$ and $z_2$ with $z_3$ in order to obtain two $1$st order MVBs with initial condition
\begin{align}
	\begin{cases}
		z_0 = (-0.25,0.0) \quad,\quad \Gamma_0 = -0.05 \quad,\quad \Gamma^x = 0.0 \quad, \quad \Gamma^y = 0.1625\\
		z_1 = (\phantom{-}0.25,0.0) \quad,\quad \Gamma_1 = \phantom{-}0.20 \quad, \quad \Gamma^x = 0.0 \quad,\quad \Gamma^y = 0.15\phantom{00}\\
	\end{cases}
	\label{eq:ic_1}
\end{align}
The corresponding dynamics are depicted in the middle row of Figure \ref{fig:grouping}.
The dynamics appear qualitatively similar at the beginning of the evolution.
Then the dynamics diverge around time $t=150$ when the two $1$st order MVBs separate from one another, 
in contrast to the dynamics of the $0$th order MVBs.

Finally, we group the two 1st order MVBs to obtain a single 2nd order MVB.
Again, the dynamics appear qualitatively similar at the beginning of the the evolution.
Oddly, the dynamics of the $2$nd order MVB appear qualitatively similar to the $0$th
order case even at $t=253$.
As there is only a single vortex, the separation of vortices mentioned in the $1$st order
MVB experiment is not possible here.  As a result the dynamics of the original $0$th 
order MVB dynamics appears to be approximately recovered.

\subsection{Approximation of initial conditions}
\label{sec:approximation}
In this section we will illustrate how initialize MVBs when given a stream function $\psi$ at time $0$.
We can begin by defining an inner-product on the space of distributions on $\R^2$, given by
\[
	\langle \omega_1 , \omega_2 \rangle_{G_\delta} := \int \omega_1(z) G_\delta (z - \tilde{z}) \omega_2(\tilde{z}) dz d\tilde{z}.
\]
Consequently, the energy of the fluid is given by $H(\omega) = \frac{1}{2} \|\omega\|^2_{G_\delta} = \frac{1}{2} \langle \omega,\omega\rangle_{G_\delta}$.

Let $K$ be a compact set and let $0 < h \ll 1$ be small so that we may define the finite grid $\Lambda_{h} = \{ (ah,bh) \in K \mid (a,b) \in \mathbb{Z}^2 \}$.
\footnote{
	The choice of $K$ should depend on the initial circulation $\omega$, e.g. if the $\omega$ has compact support than any $K$ which contains the support of $\omega$ would be a good candidate.
	Nonetheless, having to choose $K$ is a weakness of the given approximation procedure.}
Given an $\omega \in \mathcal{D}'(\R^2)$, we can attempt to approximate $\omega$ via Dirac-deltas supported on $h \mathbb{Z}^2$.
There is a natural way to do this with respect to the inner product $\langle \cdot , \cdot \rangle_{G_\delta}$.
We could define $\omega_h^{(0)} = \sum_{i \in \mathbb{Z}^2} \Gamma_i \delta_{z_i}$ by requiring the error, $\omega_h^{(0)} - \omega$, to be $\langle \cdot , \cdot \rangle_{G_\delta}$-orthogonal to $\delta_z$ for each $z \in \Lambda_{h}$.
This means that $G_\delta *\omega(z) =  \sum_i \Gamma_i G_\delta (z-z_i)$ for each $z \in \Lambda_{h}$.
Thus $\psi_h^{(0)} = \sum_i \Gamma_i G_\delta (z-z_i)$ can be seen as a $0$th order approximation to $\psi = G_\delta *\omega$
because $\psi_h^{(0)}(z) = \psi(z)$ for all $z \in \Lambda_{h}$.
Therefore, for smooth $\omega$'s, we obtain an error of order $\mathcal{O}( \Delta x)$ for a grid-spacing of $\Delta x$ using $0$th order MVBs.
 
The same reasoning applies if we consider $\omega^{(k)}_h = \sum_{i,m+n \leq N} \Gamma_i^{mn} \partial_x^m \partial_y^n \delta_{z_i}$.
We define the scalars $\Gamma_i^{mn}$ via the equations
\begin{align*}
  \partial_x^\ell \partial_y^k \psi (z_i) = \sum_j (-1)^{m+n} \Gamma_j^{mn} \partial_x^{m+\ell} \partial_y^{n+k} G_\delta(z_i - z_j)
\end{align*}
for $\psi = G_\delta*\omega$, $z_i \in \Lambda_{h}$, and $|\beta| \leq k$.
Then $\psi^{(k)}_h (z)= \sum_{i,\alpha} (-1)^{m+n} \Gamma_k^{mn} \partial_x^m \partial_y^n G_\delta( z - z_i)$
serves as an order $k$ approximation of $\psi$ when $\psi \in C^k$.
In particular, for smooth $\omega$'s, we obtain an error of order $\mathcal{O}( \Delta x^{k+1})$ for a grid-spacing of $\Delta x$ using $k$th order MVBs.

\begin{figure}[h]
        \centering
        \includegraphics[width=0.7\textwidth]{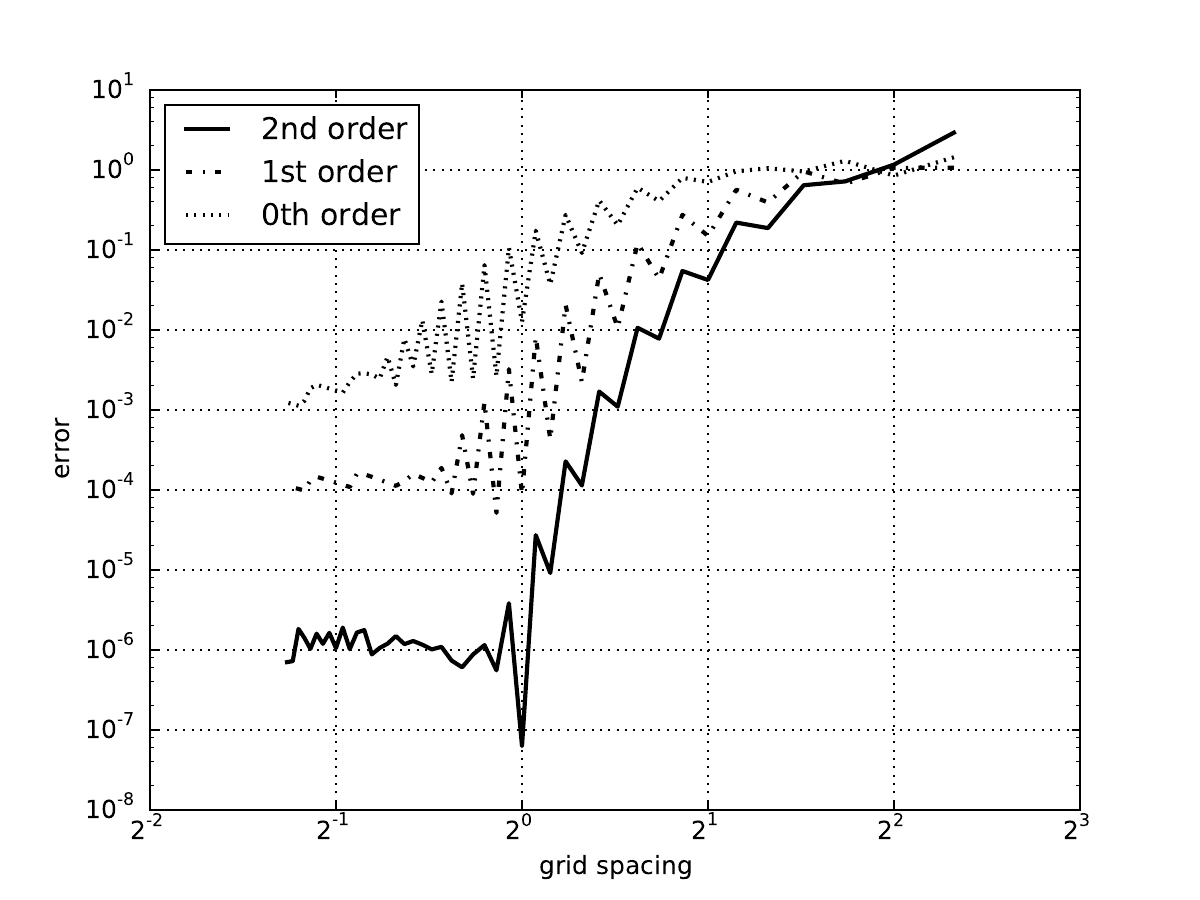}
        \caption{A convergence plot of the error in the sup-norm
        of the reconstructed stream function approximated using MVBs of order $0$, $1$, and $2$.}
        \label{fig:convergence}
\end{figure}

As an example, we numerically compute the corresponding approximations of the stream function
\begin{align}
        \psi(x,y) = \exp( - r^2 ) - \exp( - r^2 / 2 )
        \label{eq:psi_exact}
\end{align}
The results are depicted in Figure \ref{fig:convergence}
where we observe sup-norm convergence on the interior of $K$.
In particular, we measure the sup-norm error
on the subregion ($-3<x,y<3$) with $K = \{ (x,y) \mid -6 \leq x,y \leq 6 \}$.
We observe convergence using MVBs at orders zero, one, and two.
In each case, a grid spacing is reached where the error plateaus (possibly 
due to machine precision).
Nonetheless, higher order MVBs appear to out perform lower order ones for smaller grid spacings.
In particular, we observe slopes in a log-log plot of magnitudes 1,2, and 3, suggesting that 1st, 2nd, and 3rd order convergence rates for 0th,1st, and 2nd order MVBs respectively.

In terms of complexity, in order to achieve a desired error bound, $e_{tol} >0 $, you would need to use a grid with $\mathcal{O}( e_{tol}^{-2/(k+1)} )$ MVBs.
While the number of MVBs drops as $k$ increases, one could object that a high-order MVB is much more complex than a low order one.
However, the number of degrees of freedom for a $k$th order MVB is $2 + \sum_{j=0}^{k} (2^k / k!)$ which monotonically converges to a constant (roughly $9.39$) as $k \to \infty$.
Therefore the number of degrees of freedom is dominated by $\mathcal{O}( e_{tol}^{-2/(k+1)} )$ as well.
In other words, when $\psi$ is highly differentiable we observe benefits in terms of complexity and storage to using a larger $k$ regardless of weather one measures complexity by the number of parameters to keep track of, or the number of MVBs.

\section{Numerical experiments}
\label{sec:numerical_experiments}
In these section we present the results of numerical experiments involving small numbers of vortices, for $N=1,2$, and $3$.

\subsection{Behavior of isolated MVBs}
\label{sec:singles}
Next, we will briefly explore the behavior of a single isolated $k$th order MVB with $\Gamma^{mn} = 0$ with $m+n < k$ for $k=0,1,2$.
This case allows us to investigate the dynamics induced by the higher order circulation variables
in the absence of the lower order ones.

\subsection{Order 0}
The behavior of a single $0$th order MVB is explicitly solvable because the dynamics are stationary.

\subsection{Order 1}  
\begin{figure}[h!]
	\centering
	\includegraphics[width = 0.3\textwidth]{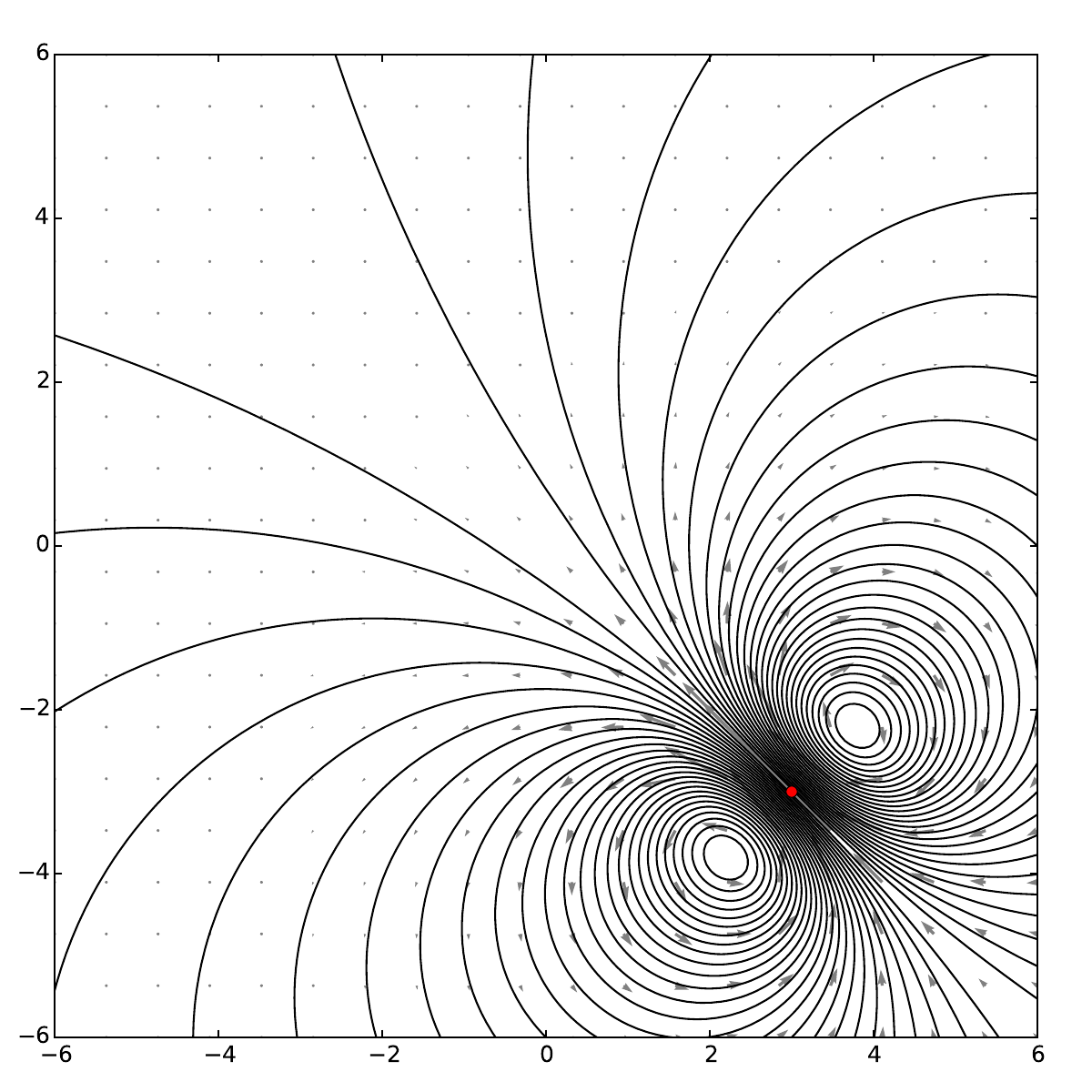}
	\includegraphics[width = 0.3\textwidth]{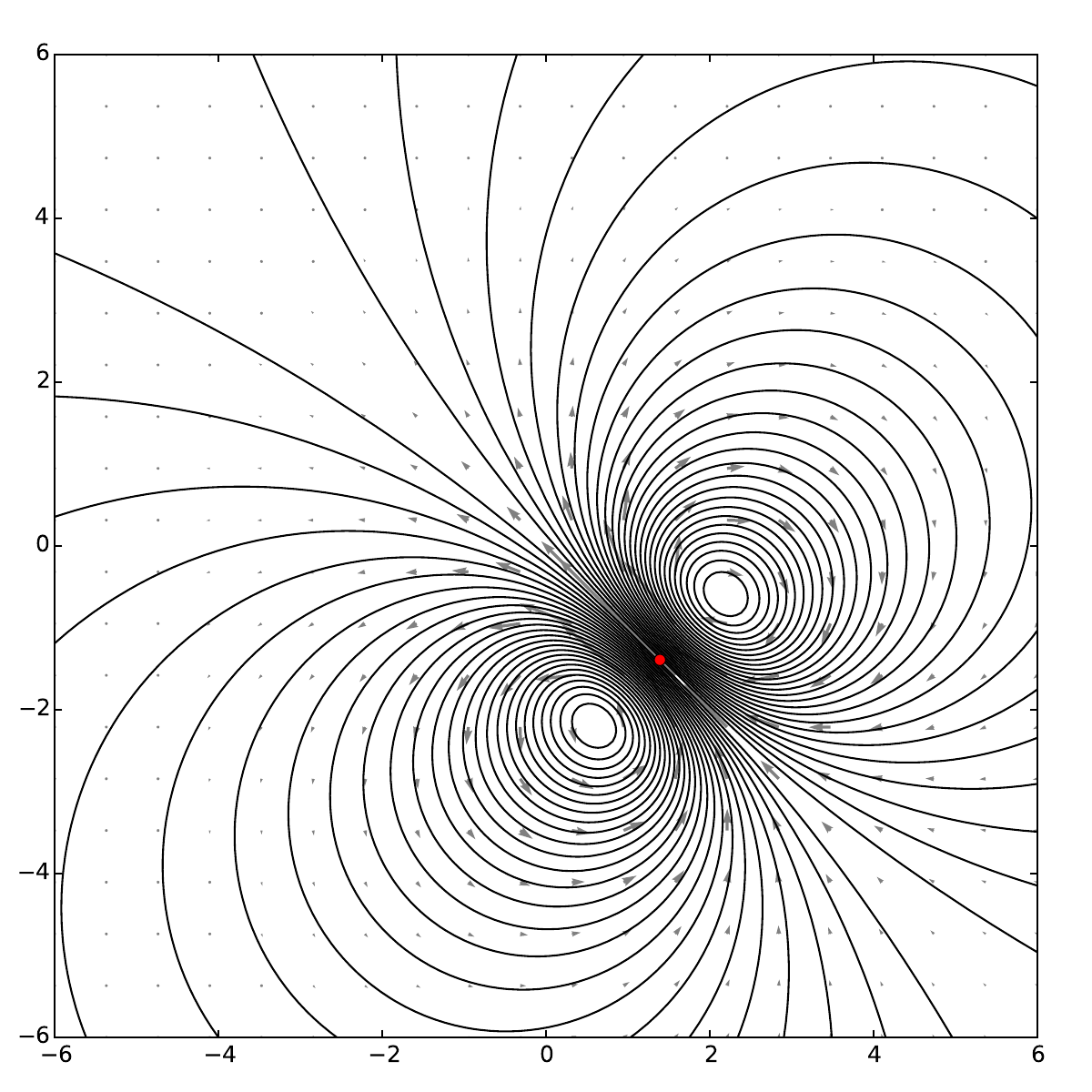}
	\includegraphics[width = 0.3\textwidth]{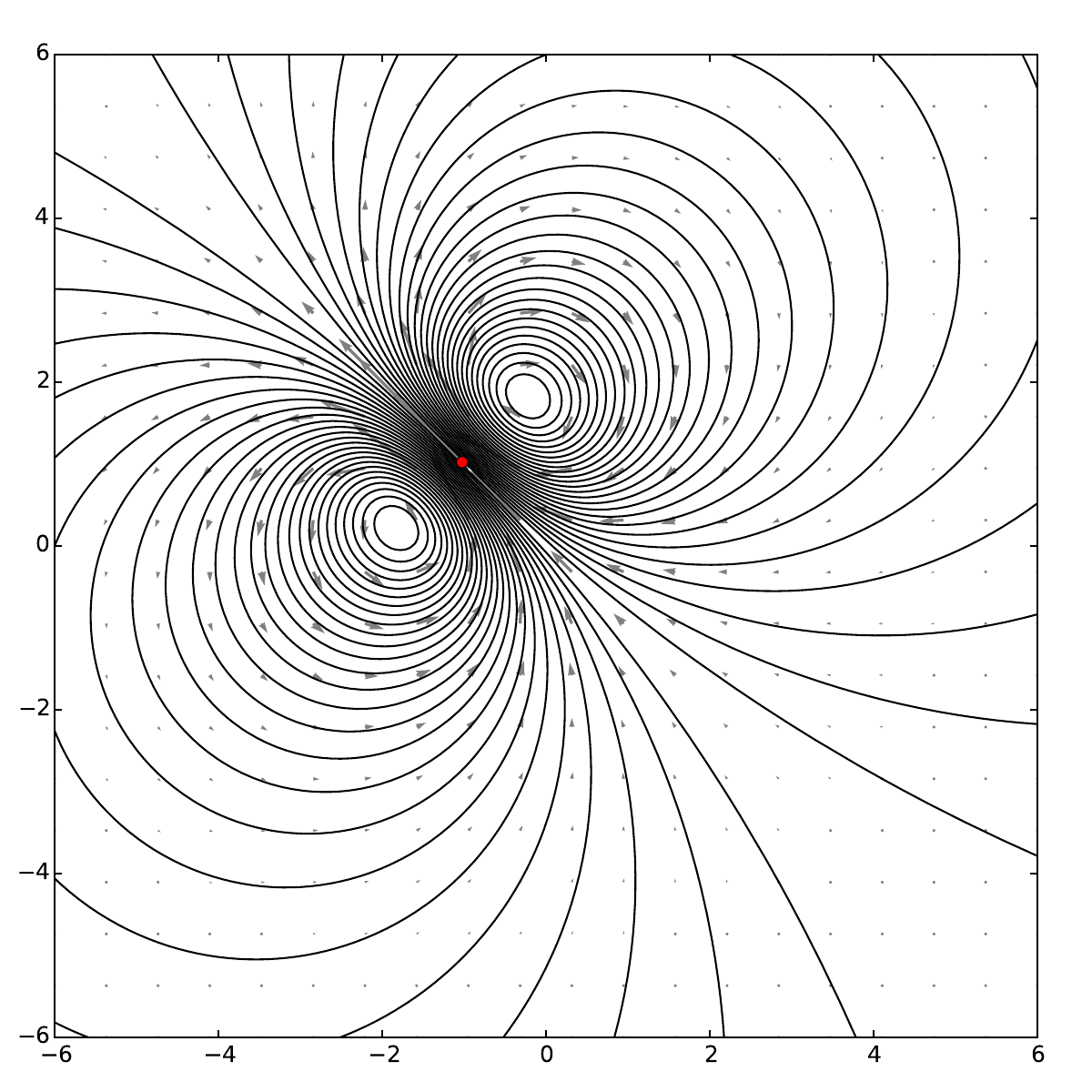}
	\caption{A $1$st order MVB with initial conditions given by \eqref{eq:ic_order1}
		with snapshots taking at $t=0,10,25$.}
	\label{fig:singleton_order1}
\end{figure}

The behavior of a single $1$st order MVB with $\Gamma= 0$ is explicitly solvable.
Given the initial condition $(x(0) , y(0) ,  \Gamma (0) , \Gamma^x(0) , \Gamma^y(0) )$ with $\Gamma(0) = 0$ we find
\begin{align*}
	&x(t) = x(0) + v^x t \,,\quad \quad y(t) = y(0) + v^y t \,,\quad \quad \Gamma(t) = \Gamma(0) \\
	&\Gamma^x(t) = \Gamma^x(0)\,,\quad \quad \Gamma^y(t) = \Gamma^y(0)
\end{align*}
where $v^x = \Gamma^x(0) \partial_{xy}G_\delta(0) + \Gamma^y(0) \partial_{yy}G_\delta(0)$
and $v_y =  -\Gamma^y(0) \partial_{xy}G_\delta(0) - \Gamma^x(0) \partial_{xx}G_\delta(0)$.
In Figure \ref{fig:singleton_order1}
we depict such a trajectory with initial condition
\begin{align}
	x(0) = -3 , y(0) = -3 , \Gamma(0) = 0 , \Gamma^x = 1, \Gamma^y = 1 \label{eq:ic_order1}
\end{align}

\subsection{Order 2}
\begin{figure}[h!]
	\centering
	\includegraphics[width = 0.15\textwidth]{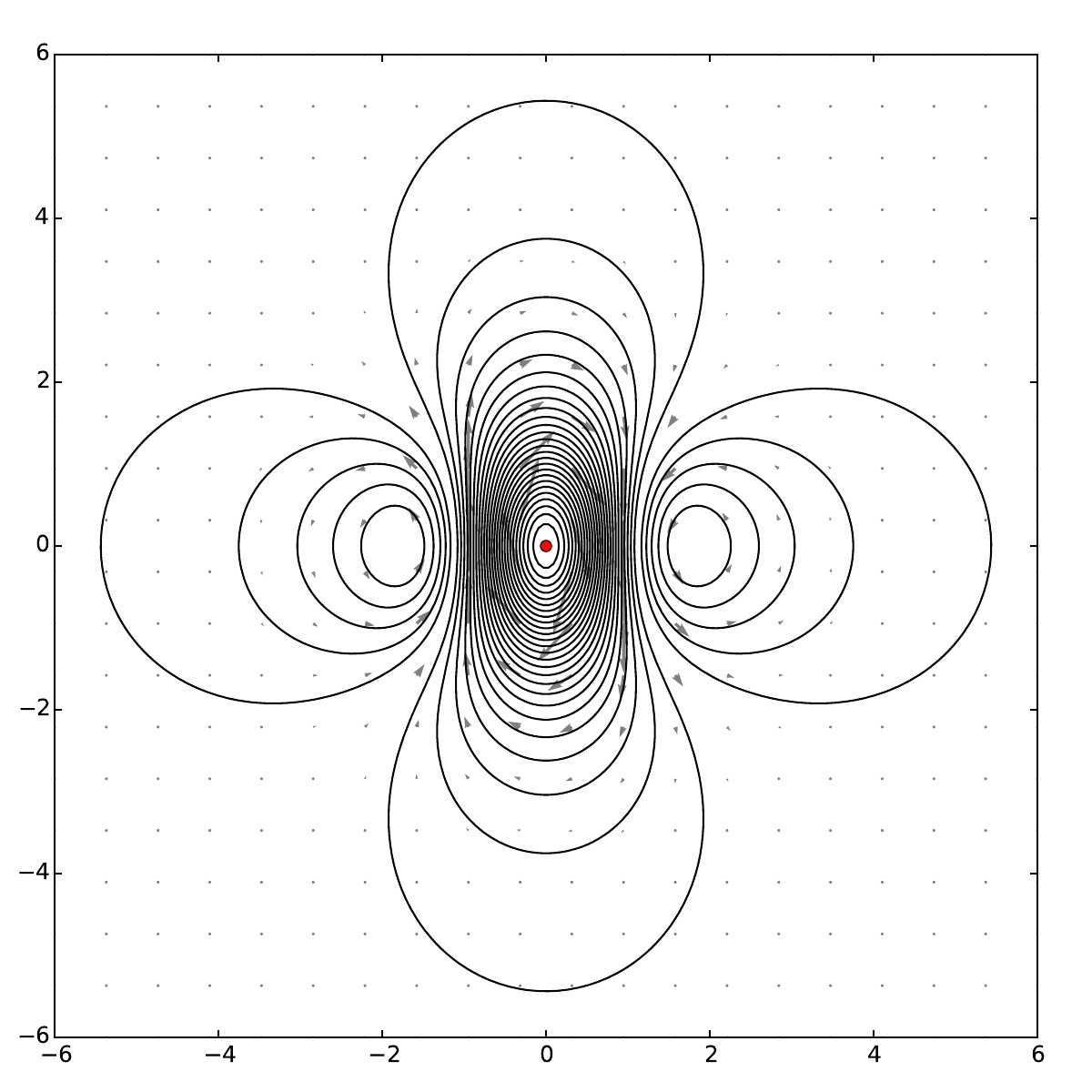}
	\includegraphics[width = 0.15\textwidth]{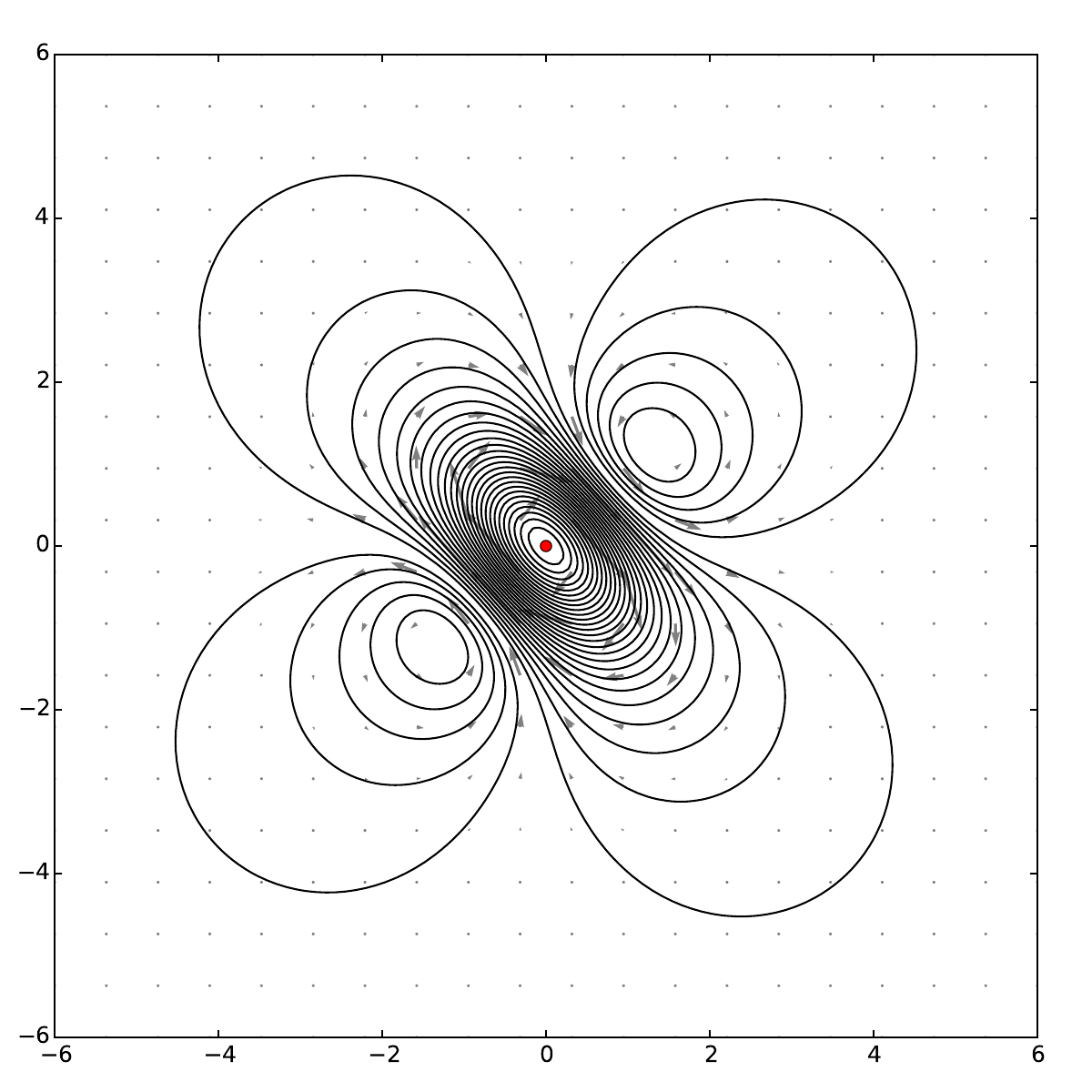}
	\includegraphics[width = 0.15\textwidth]{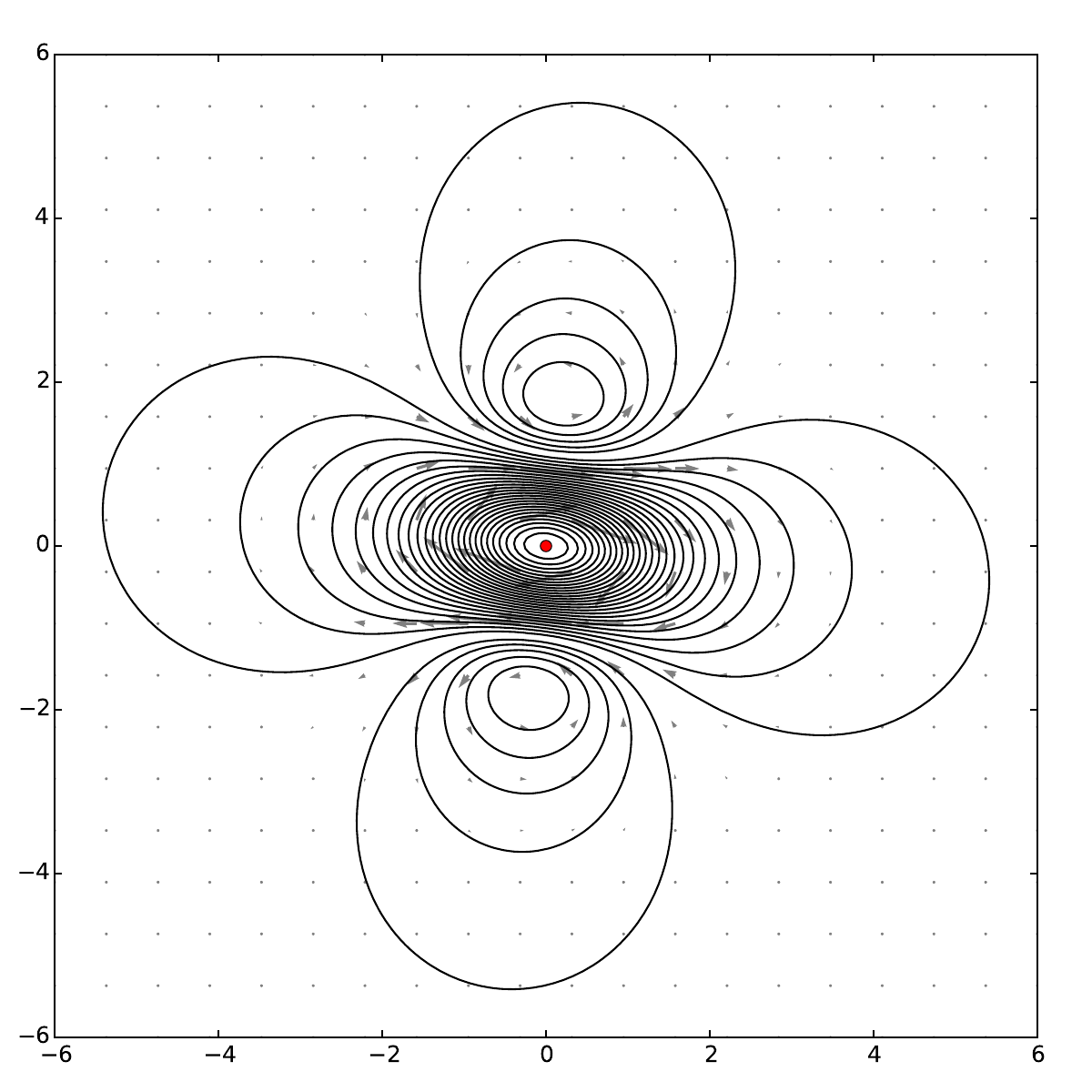}
	\includegraphics[width = 0.15\textwidth]{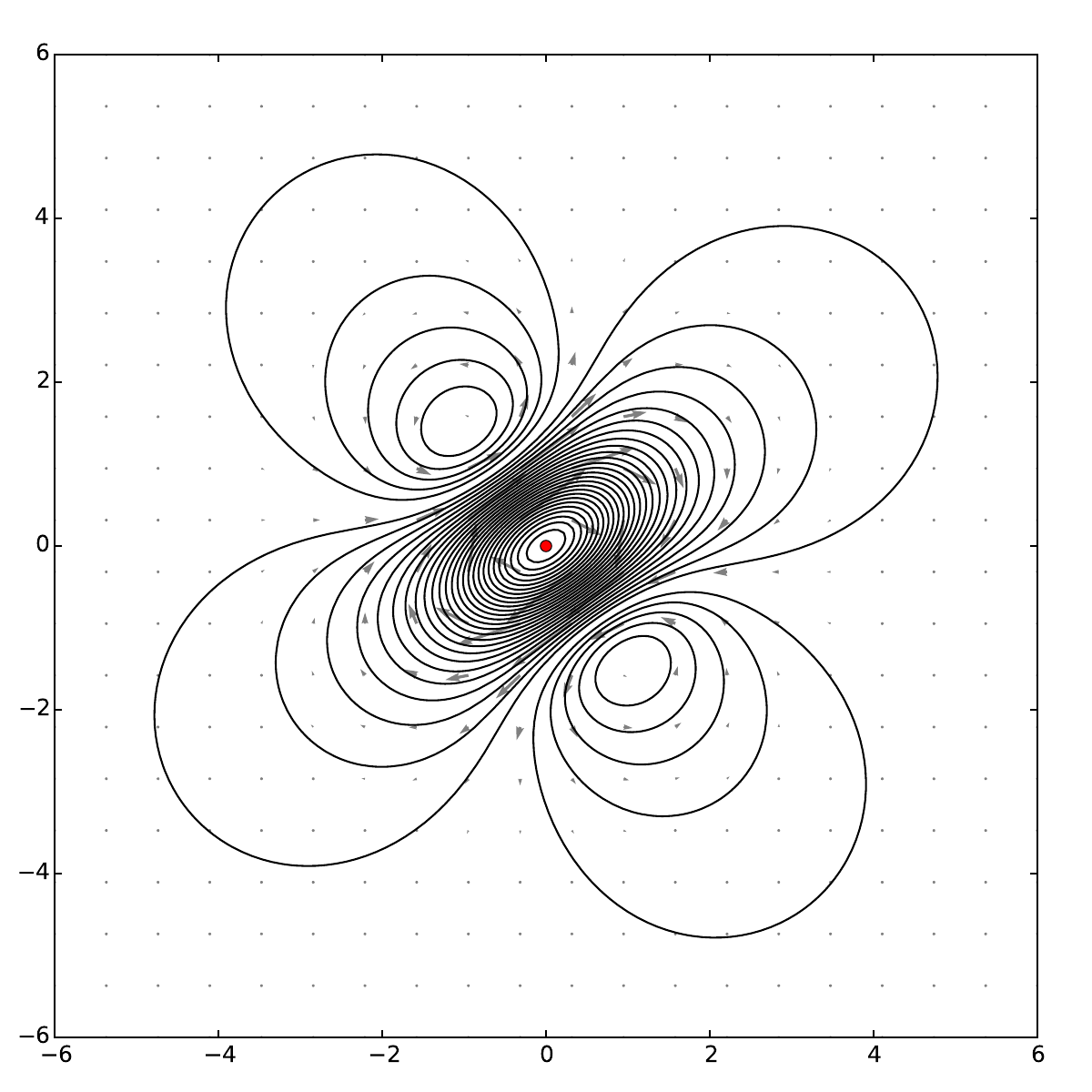}
	\includegraphics[width = 0.15\textwidth]{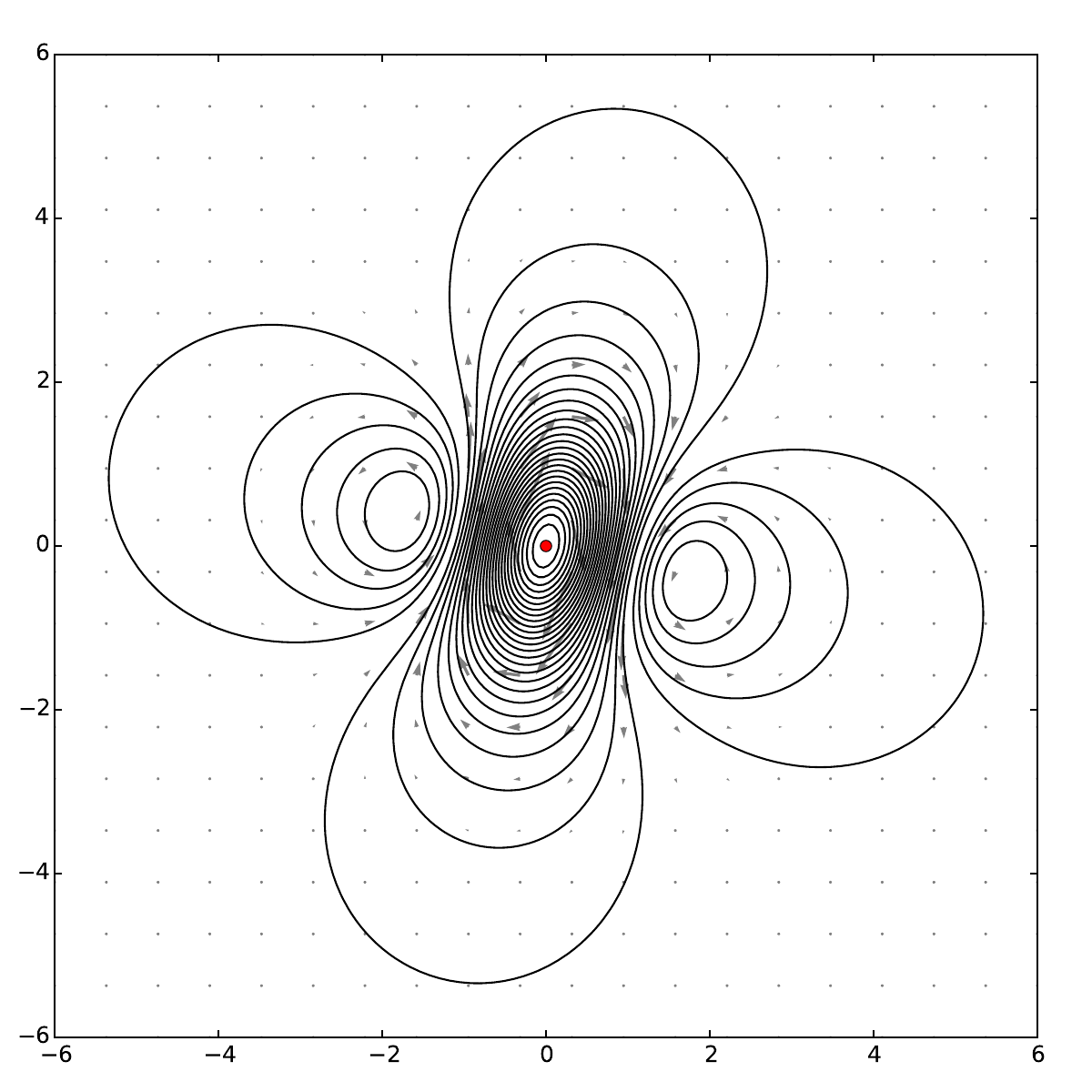}
	\includegraphics[width = 0.15\textwidth]{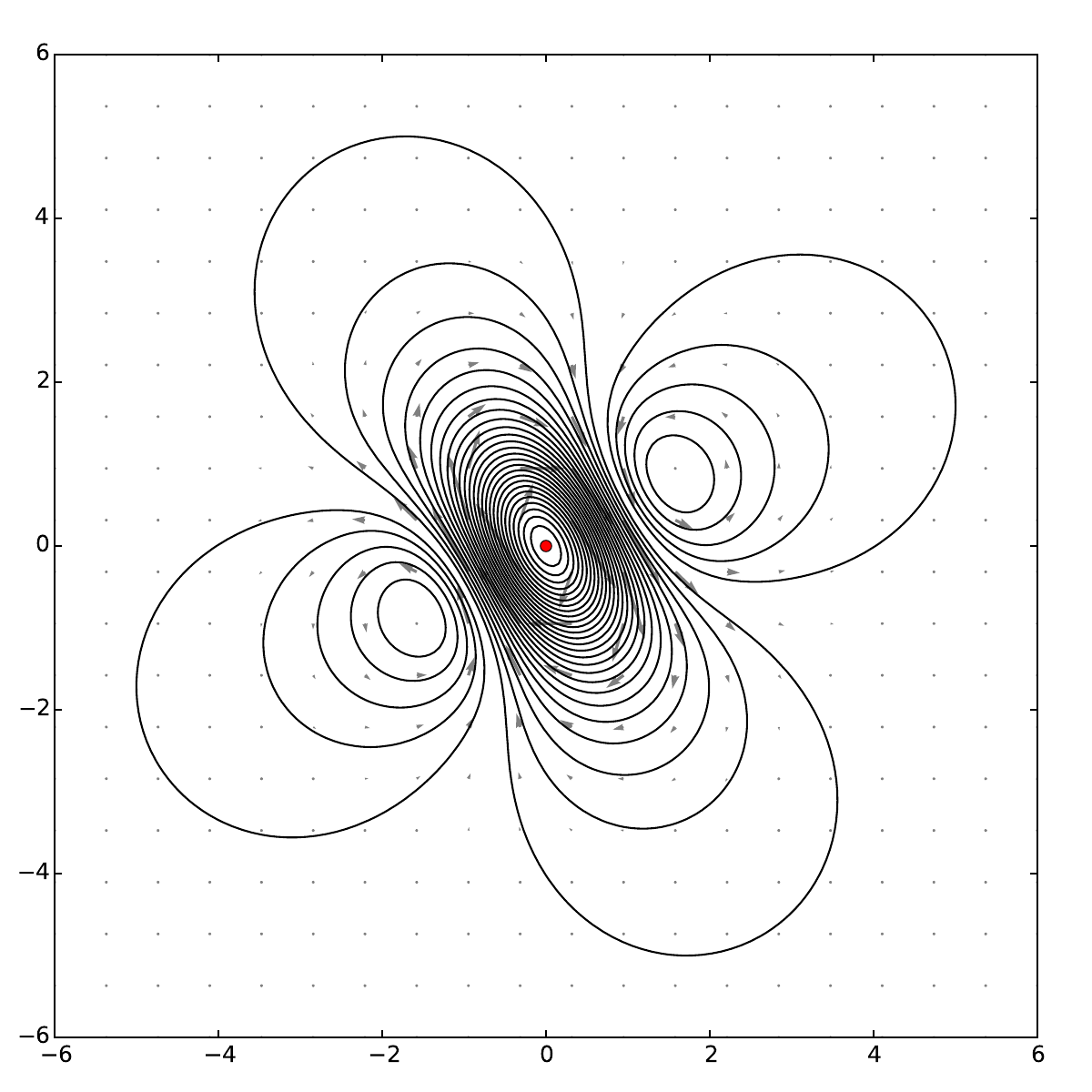}
	\caption{A $1$st order MVB with initial conditions given by \eqref{eq:ic_order2}
		with snapshots taking at $t=0,5,10,15,20,25$.}
	\label{fig:singleton_order2}
\end{figure}

The behavior of a second order vortex does not seem to be explicitly solvable.
Here we consider initial conditions for which 
\begin{align}
	x(0) = 0 \,,\quad\quad y(0) = 0 \,,\quad\quad \Gamma(0)^{xx} = 1 \label{eq:ic_order2}
\end{align}
and all the other circulation variables are initially set to $0$.
The results are depicted in Figure \ref{fig:singleton_order2}.
We observe a structure which rigidly rotates counter-clockwise.

\subsection{A scattering expiriment}
\label{sec:scattering}
\begin{figure}[h!]
	\centering
	\includegraphics[width=0.3\textwidth]{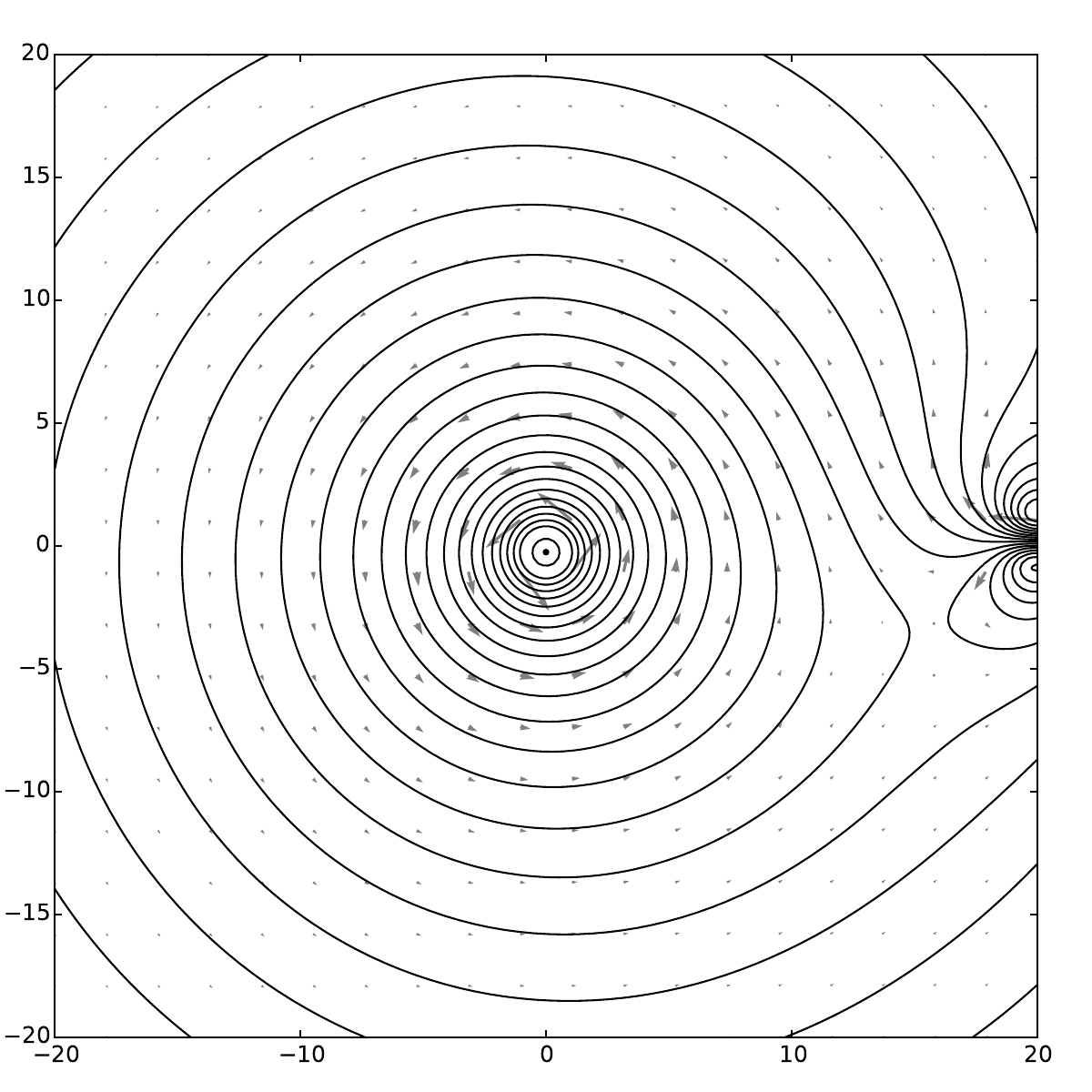}
	\includegraphics[width=0.3\textwidth]{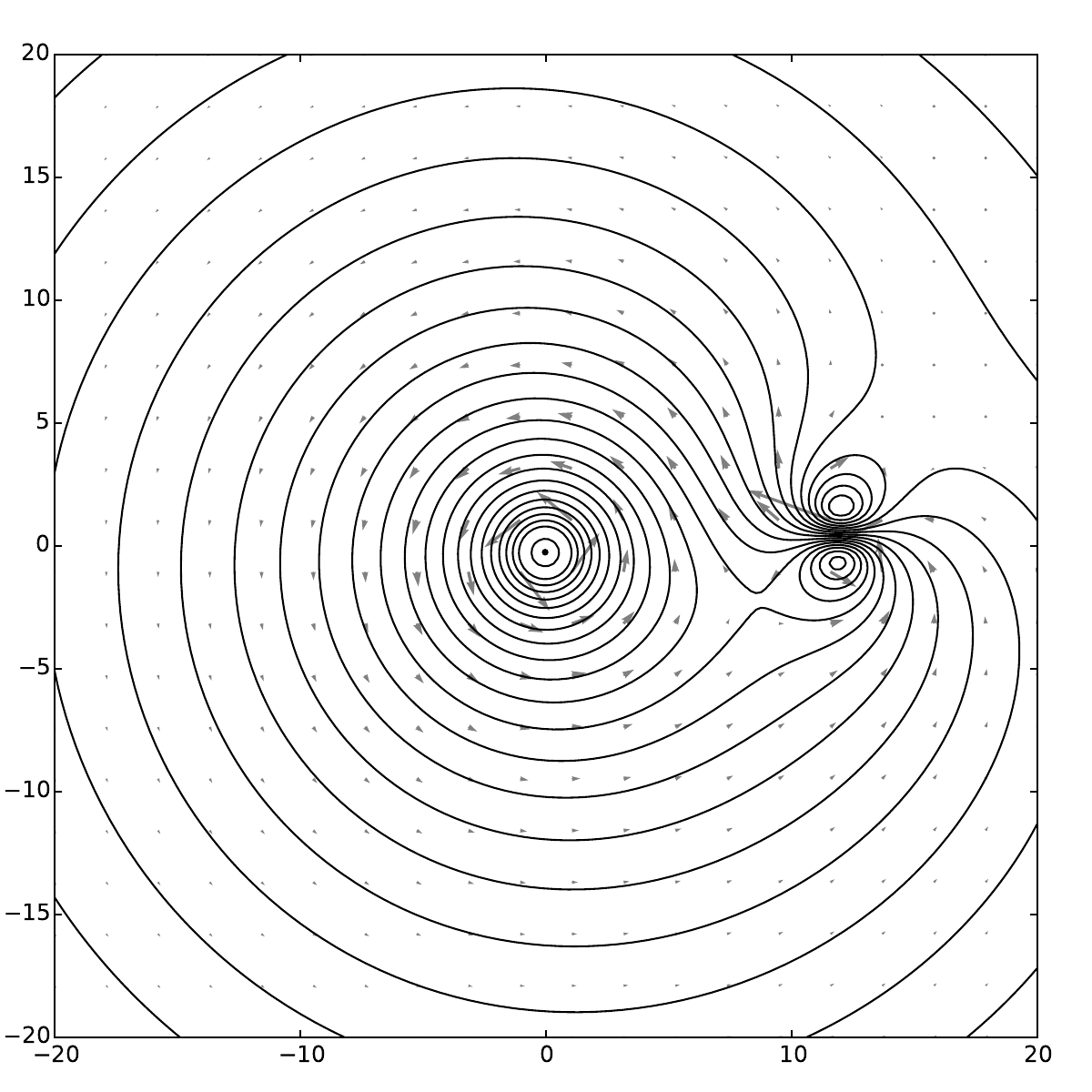}
	\includegraphics[width=0.3\textwidth]{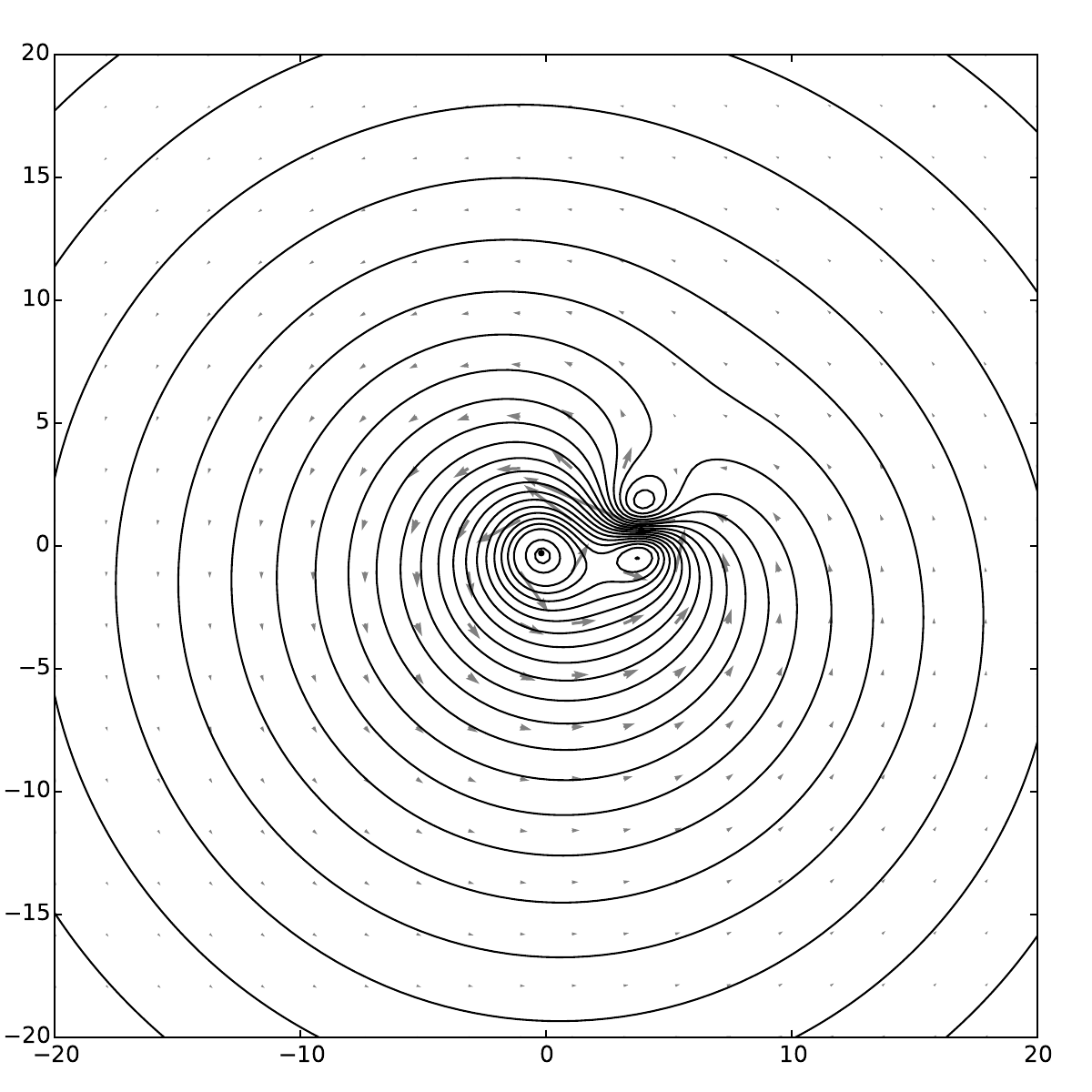}
	\includegraphics[width=0.3\textwidth]{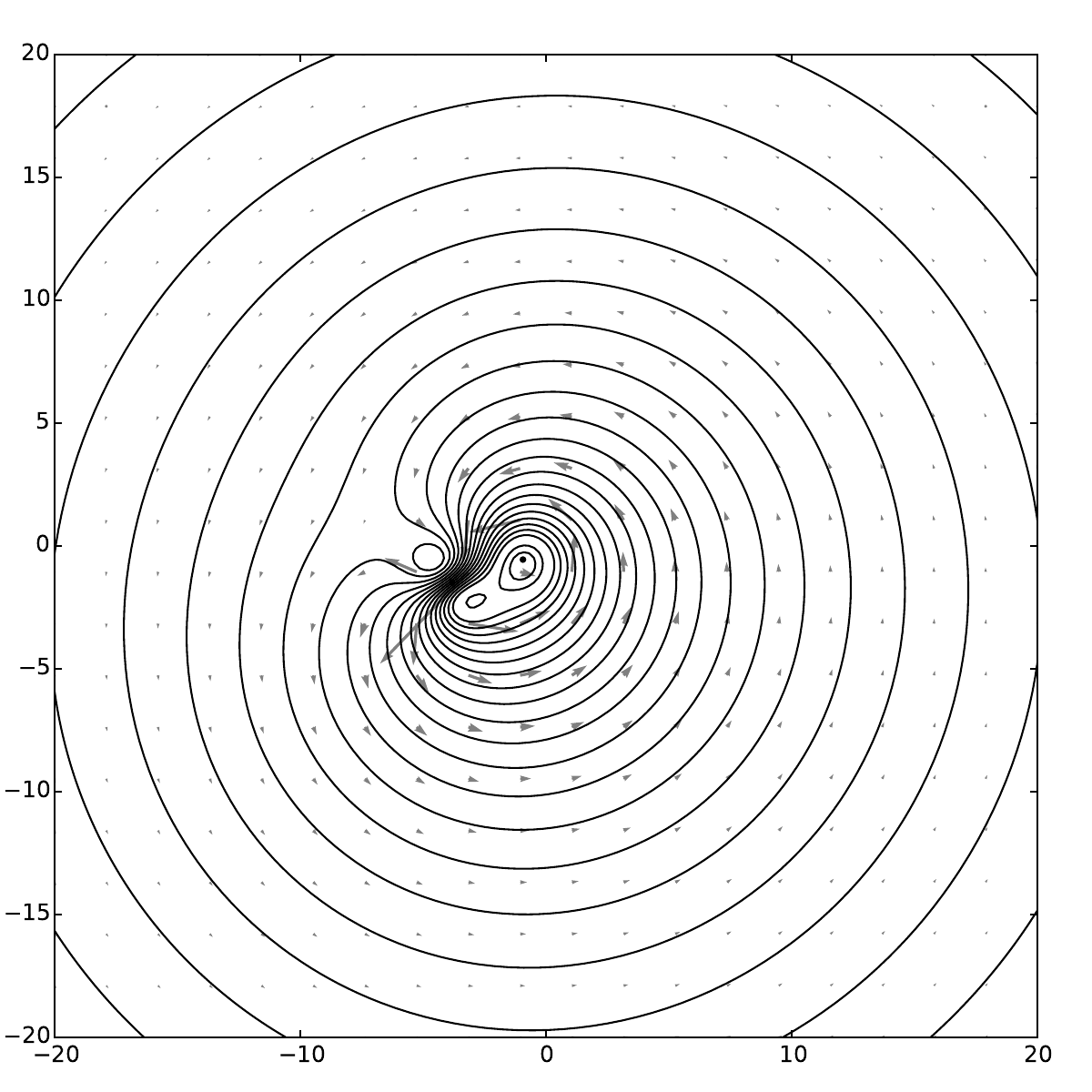}
	\includegraphics[width=0.3\textwidth]{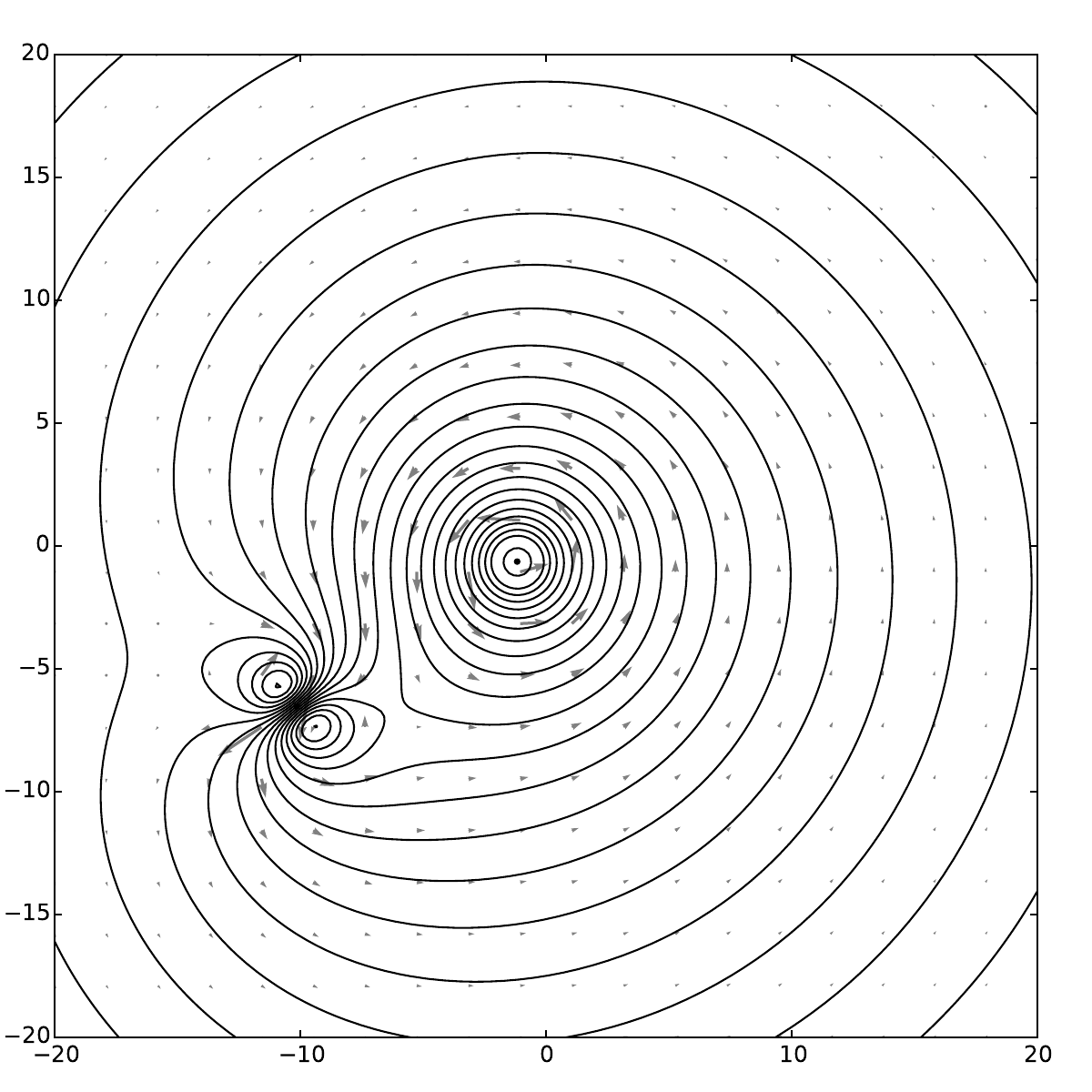}
	\includegraphics[width=0.3\textwidth]{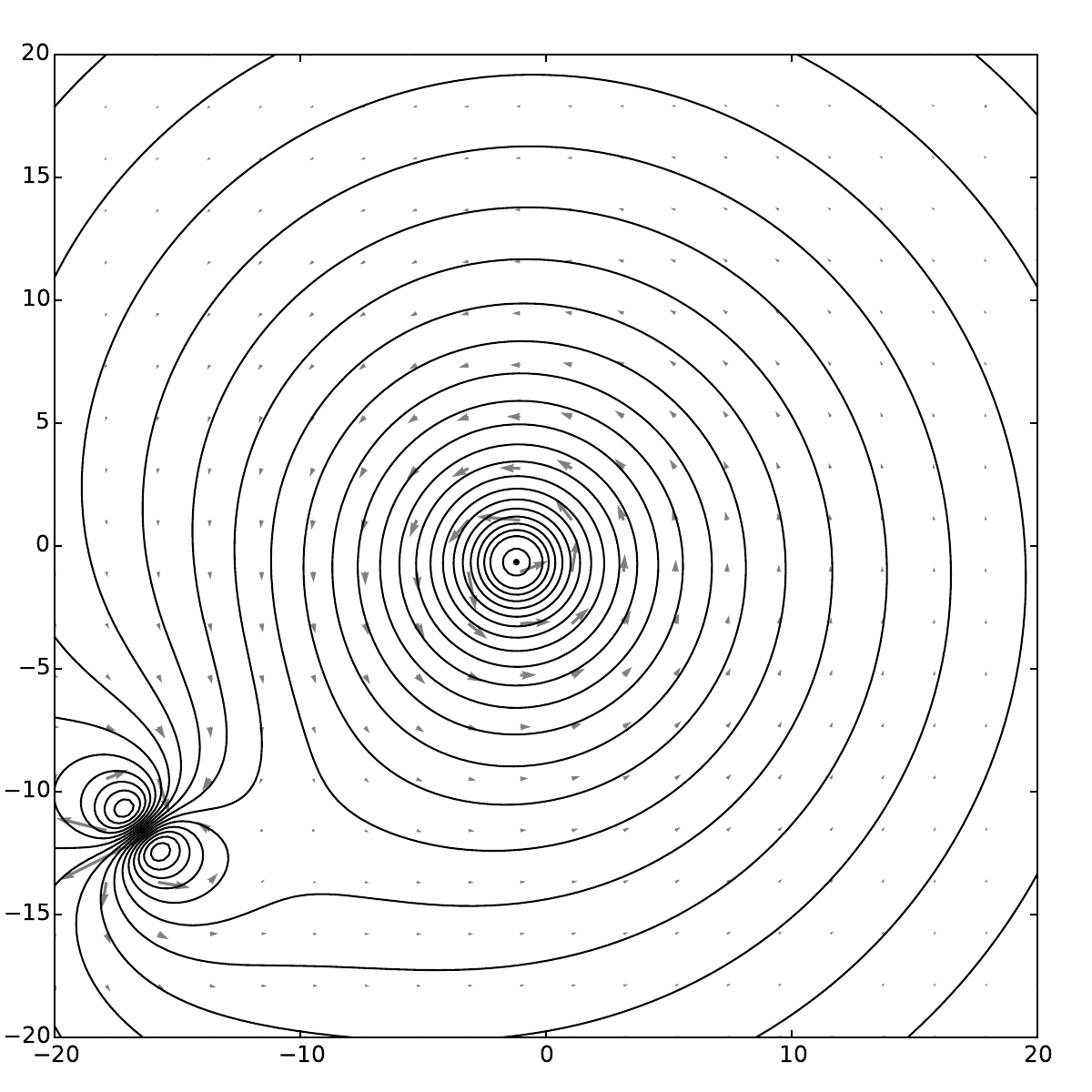}
	\caption{A numerical run is shown with mirror image initial conditions for two 1st order MVBs, as given in \eqref{eq:ic}.
		From left to right and top to bottom these are snapshots at times $t=0,25,50,75,100,125$ respectively.}
	\label{fig:scatter}
\end{figure}

Next we consider two MVBs.  The first is a first order MVB with an initial velocity pointed just slightly above origin.
The second MVB is a standard zeroth order vortex located at the origin.
Specificaly, we consider the initial conditions
\begin{align}
	\begin{cases}
	z_0 = (20.0 , \phantom{-}0.25)\quad, \quad \Gamma_0^{0,0} = 0.0 \quad,\quad \Gamma_0^{1,0} = 0.0 \quad,\quad \Gamma_0^{0,1} = -1.0 \\
	z_1 = (\phantom{2}0.0 , -0.25)\quad, \quad \Gamma_0^{0,0} = 1.0 \quad,\quad \Gamma_0^{1,0} = 0.0 \quad,\quad \Gamma_0^{0,1} = \phantom{-}0.0
	\end{cases}
	\label{eq:ic}
\end{align}
with $\Gamma_i^{mn} = 0$ for $m+n > 1$ and $i=0,1$.
The vortex at the origin appears to remain at the origin throughout the numerical run ($t=0$ to $t=150$).
The first order MVB starts by moving to the left in a straightline until it comes into proximity of the zeroth order vortex.
Then the first order MVB swings around the the zeroth order vortex, traversing an angle of roughly 30 degrees
before zooming off into the lower left quadrant of the plane in a straight line.  These results are depicted in Figure \ref{fig:scatter}

\subsection{The method of images}
\label{sec:method_of_images}
Here we incorporate first order MVBs into the method of images~\cite{Jackson1975,Smith2011}.
We consider the initial conditions consisting of two first order MVBs which are mirror images of each other with respect to the $x$-axis.
By symmetry, the resulting vector-field should be tangential to the $x$-axis, and provides a means of considering a boundary that satisfied the no-penetration condition.
Specifically, we consider the initial condition:
\begin{align}
	\begin{cases}
	z_0 = (1.5 , \phantom{-}1.5)\quad, \quad \Gamma_0^{0,0} = \phantom{-}0.5 \quad,\quad \Gamma_0^{1,0} = \phantom{-}0.5 \quad,\quad \Gamma_0^{0,1} = 1.5 \\
	z_1 = (1.5,-1.5)\quad, \quad \Gamma_0^{0,0} = -0.5 \quad,\quad \Gamma_0^{1,0} = -0.5 \quad,\quad \Gamma_0^{0,1} =1.5
	\end{cases}
	\label{eq:ic moi}
\end{align}
with $\Gamma_i^{mn} = 0$ for $m+n > 1$ and $i=0,1$.

\begin{figure}[h!]
	\centering
	\includegraphics[width=0.3\textwidth]{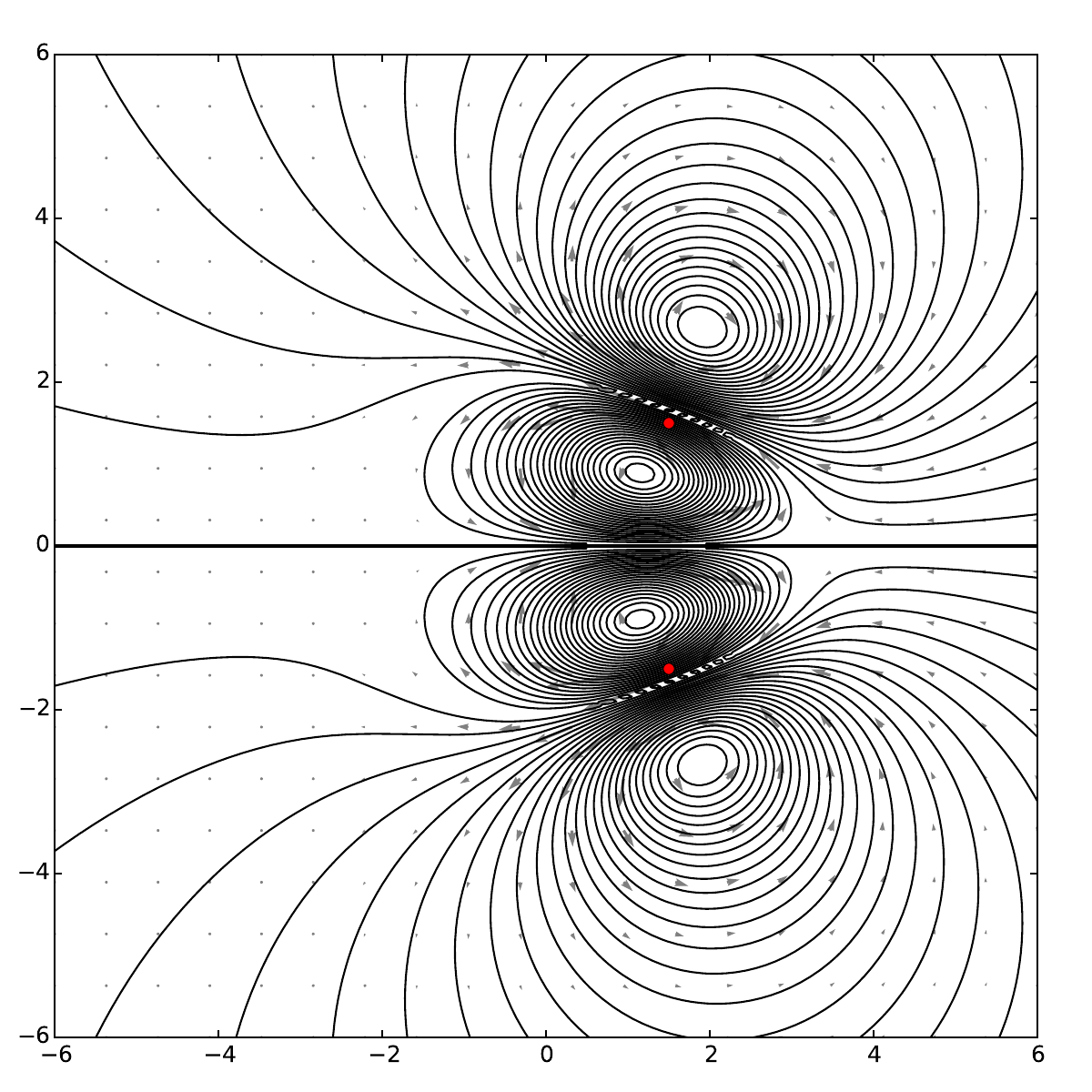}
	\includegraphics[width=0.3\textwidth]{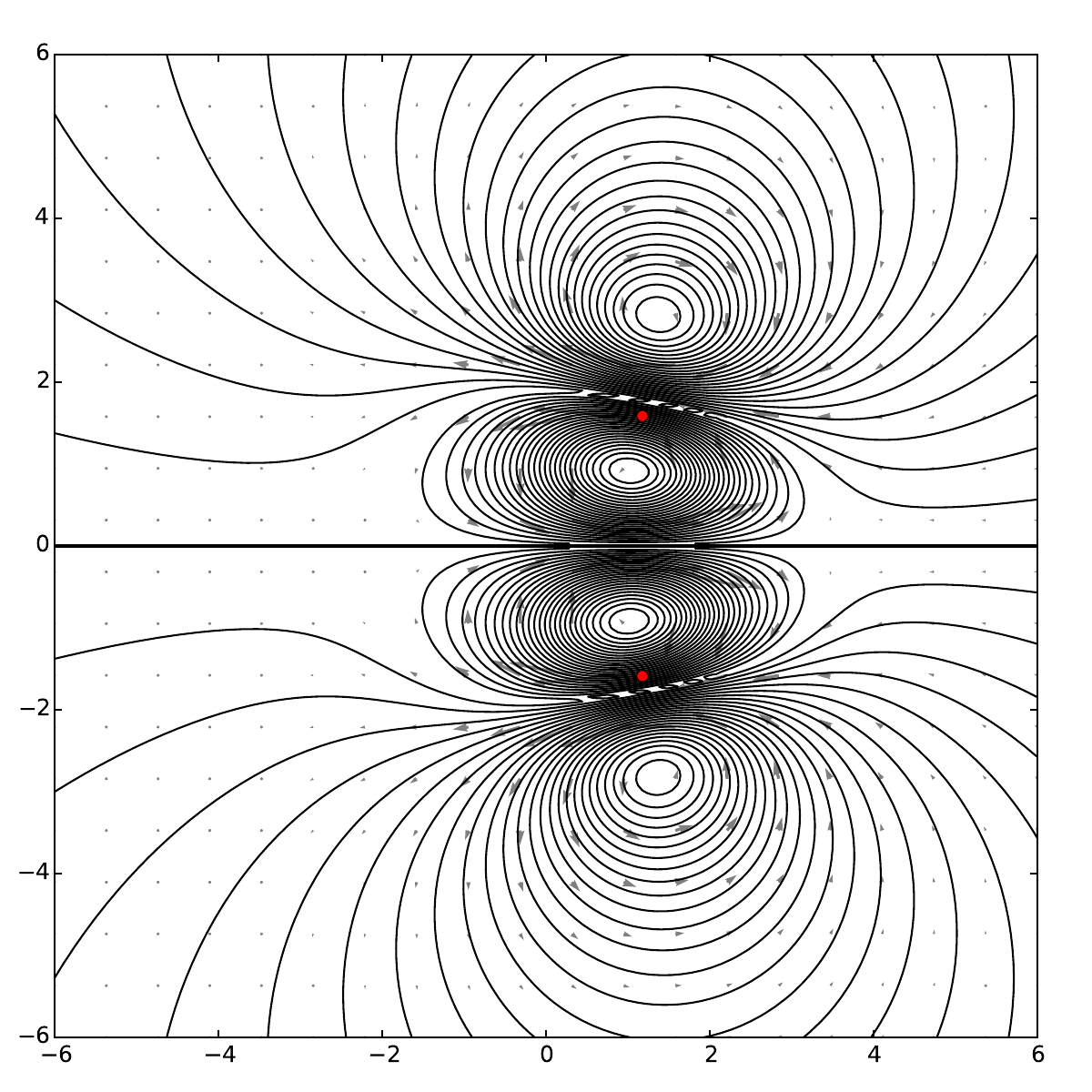}
	\includegraphics[width=0.3\textwidth]{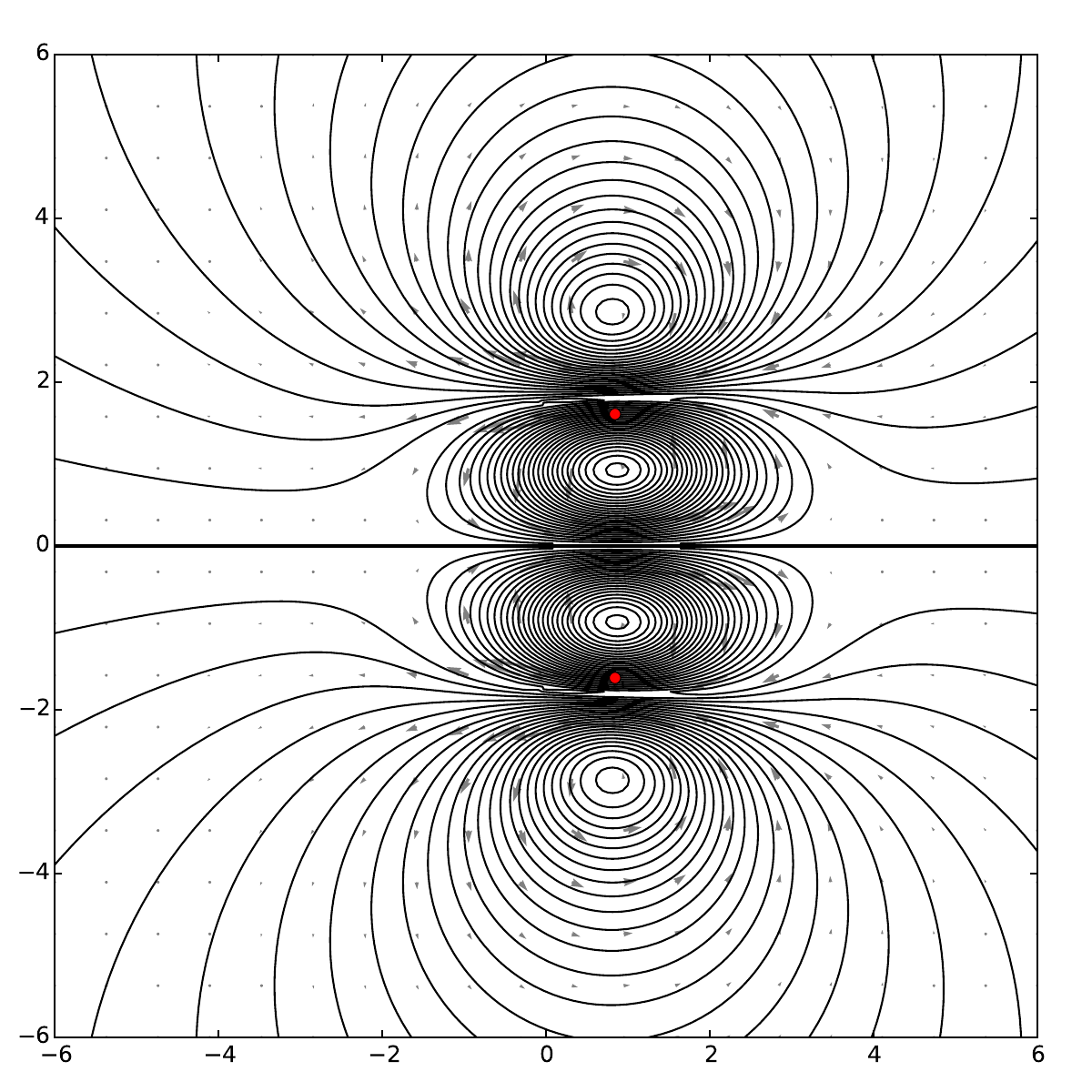}
	\includegraphics[width=0.3\textwidth]{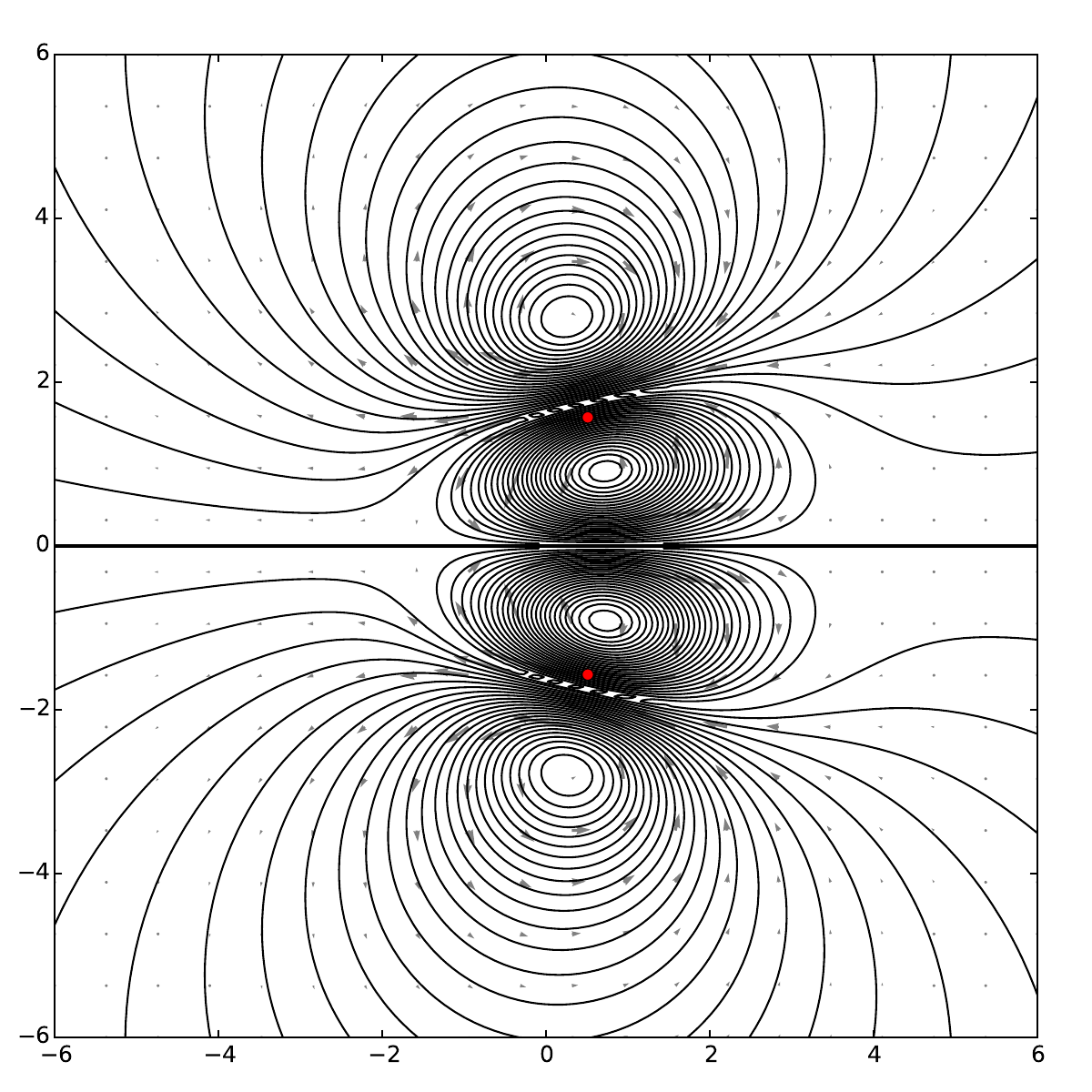}
	\includegraphics[width=0.3\textwidth]{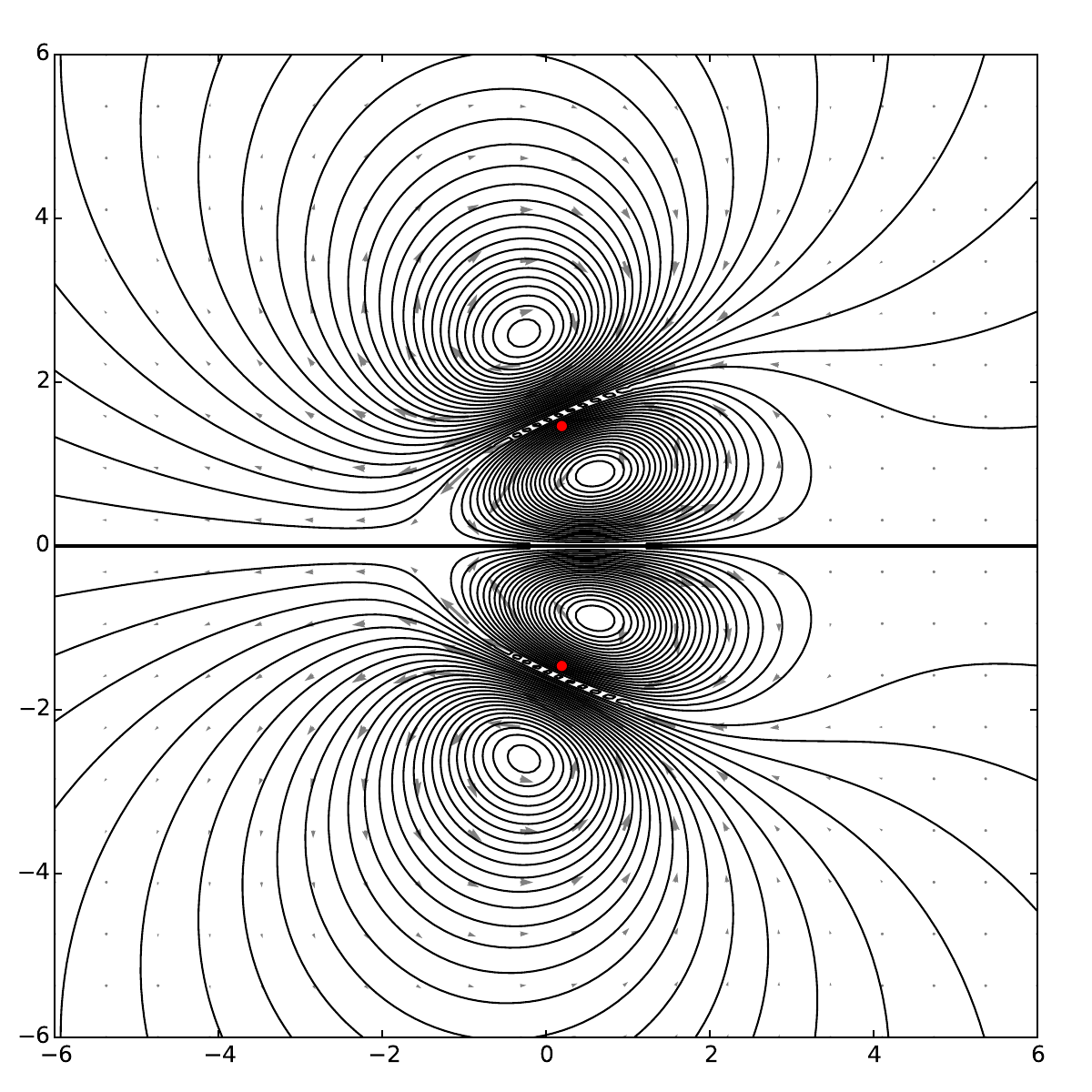}
	\includegraphics[width=0.3\textwidth]{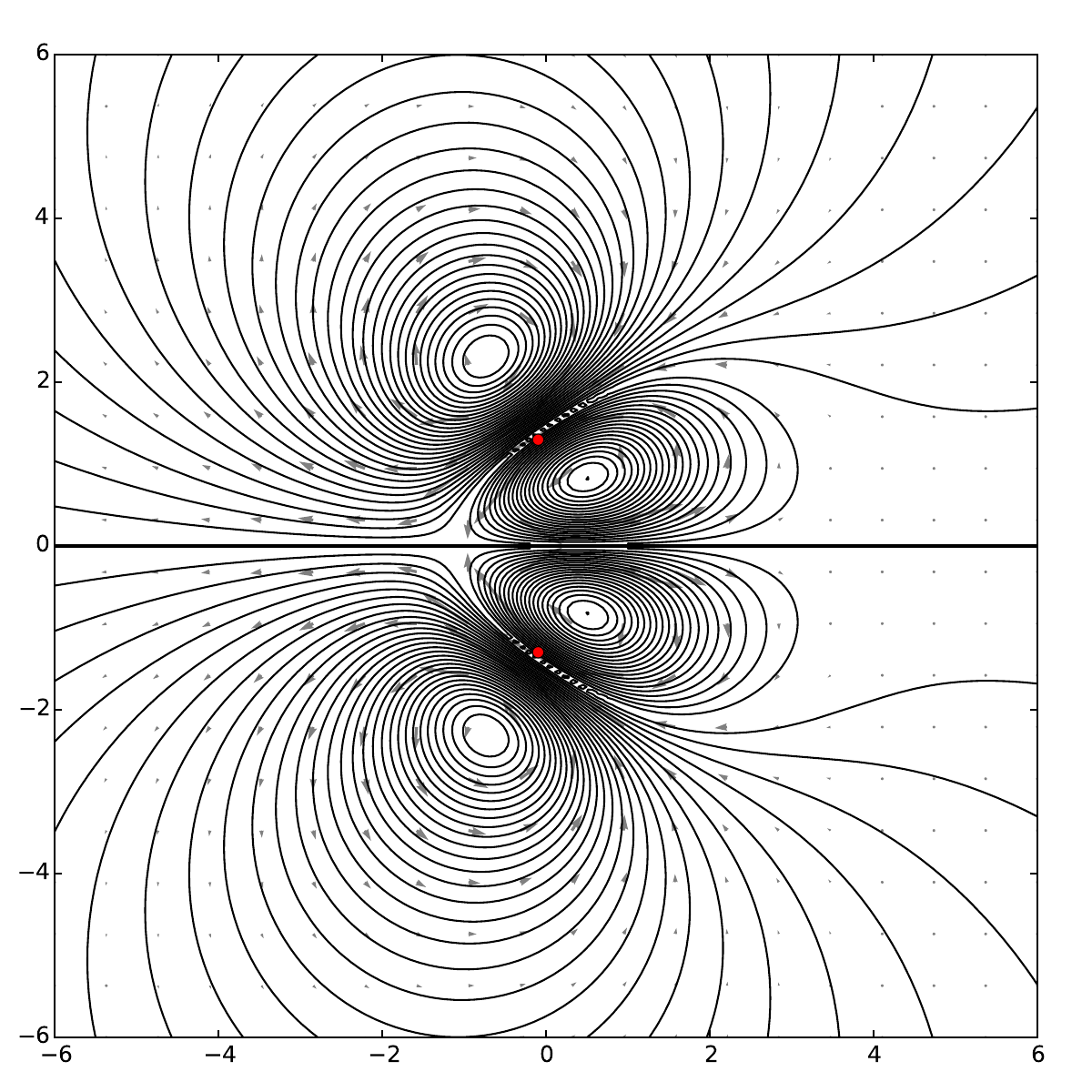}
	\caption{Numerical results are shown for two first order MVBs with mirror-image initial conditions given by \eqref{eq:ic moi}.
		From left to right and top to bottom these are snapshots at times $t=0,1.7,3.4,5,6.7,8.4$ respectively. Apparently, a first order MVB reflects elastically from a fixed boundary, so that its angle of reflection equals its angle of incidence.}
	\label{fig:moi}
\end{figure}

The resulting dynamics depicted in Figure \ref{fig:moi} shows that as a first order MVB approaches a boundary it will turn its motion along the boundary and then move away so that its angle of reflection equals its angle of incidence. 

\section{Hamiltonians and symplectic structures}
\label{sec:symplectic}
In modern Hamiltonian mechanics, as described in~\cite{FOM,Arnold2000},
the Hamiltonian is a function on a symplectic manifold, which produces equations of motion.
An important instance of a symplectic manifold is a coadjoint orbit (defined below).
In this section we compute the coadjoint orbit of a MVB
as well as the associated symplectic structure.
The coadjoint orbit of an initial vorticity distribution $\omega_0$ comprises the set
\begin{align*}
  \Orb(\omega_0) := \{ \omega_0 \circ \varphi^{-1} \mid \varphi \in \SDiff(\R^2) \}.
\end{align*}
In fact $\Orb(\omega_0)$ inherits the structure of a smooth manifold,
and a tangent vector on $\Orb(\omega_0)$ at the point $\tilde{\omega} \in \Orb(\omega)$ is given by 
a distribution of the form $\pounds_{\vec{u}}[\tilde{\omega}] := u \partial_x \tilde{\omega} + v \partial_y \tilde{\omega}$ for some (non-unique) divergence free vector field
$\vec{u} = (u,v) \in \mathfrak{X}_{\rm div}(\R^2)$.
The symplectic structure is nothing more than a special case of the one derived via the Kirillov-Kostant-Souriau theorem \cite[see the boxed formula on p.303]{FOM}.
In particular, the symplectic structure on $\Orb(\omega)$ is given by
\begin{align}
  \Omega_\omega( \pounds_{\vec{u}_1}[\omega] , \pounds_{\vec{u}_2}[\omega] ) = \int \omega(z)  ( u_1(z) v_2(z) - v_1(z) u_2(z) ) dz. \label{eq:symplectic}
\end{align}
When $\omega$ is a smooth distribution, the symplectic structure may be identified with a differential $2$-form and this formula matches the symplectic form derived on page 313 of \cite{MarsdenWeinstein1983}.
In the case that $\omega$ satisfies the ansatz \eqref{eq:ansatz N}, we find that given any $\varphi \in \SDiff(\R^2)$
	that
	\begin{align*}
		\int \omega_0(\varphi^{-1}(z)) f(z) dz &= \int \omega_0(z) , f (\varphi(z)) dz \\
                &= \gamma_i^\alpha \partial_{\alpha}|_{z=Z_i} (f \circ\varphi)(z).
	\end{align*}
	Here we have used the change of variables formula and the fact that $\det( D\varphi) = 1$.
	By the multivariate Fa\'a di Bruno formula, the expression $\partial_{\alpha}|_{z=Z_i} (f \circ\varphi)(z)$
	is a sum of the partial derivatives of $f$ at the points $\varphi(Z_i)$
	of order less than that of the multi-index $\alpha$ \cite{ConstantineSavits1996}.
        Thus $ \omega_0 \circ \varphi^{-1}$ is contained in the finitely parametrized subset $M^{(k)} := \{ \sum_{|\alpha| \leq k} \Gamma_i^\alpha \partial_{\alpha} \delta_{Z_i} \}$
        for any $\varphi \in \SDiff( \R^2)$.
        Therefore $\Orb(\omega_0)$ is a finite-dimensional manifold when $\omega_0$ satisfies the jet-vortex ansatz.

Having identified a symplectic manifold, $\Orb(\omega_0)$, we can then ask the question \emph{``are the dynamics Hamiltonian on $\Orb(\omega_0)$?"}
Of course, the answer is \emph{``yes"}.  This is the primary content of \cite{MarsdenWeinstein1983}.
We provide our own explanation here for convenience.

For a general vorticity distribution $\omega$, we may consider the kinetic energy Hamiltonian
\begin{align}
  H(\omega) = \frac{1}{2} \int \omega(z) G_\delta(z-\tilde{z}) \omega(\tilde{z}) dz d\tilde{z}. \label{eq:Hamiltonian}
\end{align}
Where $\omega$ may be of the form \eqref{eq:ansatz N}.
  In order to find Hamilton's equations on $\Orb(\omega_0)$ choose some $\omega \in \Orb(\omega_0)$ and calculate the vector $X_H(\omega)$ tangent to $\Orb(\omega_0)$
  given by Hamilton's equations.
  It must be the case that $X_H(\omega) =  \pounds_{\vec{u}}[\omega]$ for some (non-unique) vector-field $\vec{u} = (u,v) \in \mathfrak{X}_{\rm div}(\R^2)$.
  Our goal is to solve for $\vec{u}$.
  By the definition of the Hamiltonian vector field $X_H$ we see that for any
  $\vec{u}' = (u',v') \in \mathfrak{X}_{\rm div}(\R^2)$
  \begin{align*}
    \int \omega(z) \left( u(z) v'(z)  - v(z) u'(z) \right) dz &=
    \Omega_{\omega}( \pounds_{\vec{u}}[\omega] , \pounds_{\vec{u}'}[\omega] ) = - \int \frac{\delta H}{\delta \omega} (z) \left( \pounds_{\vec{u}'}[\omega] \right)(z) dz  \\
    &= - \int G_\delta(z-\tilde{z}) \omega (\tilde{z}) \left( \pounds_{\vec{u}'}[\omega] \right)(z) d\tilde{z} dz .
  \end{align*}
  If we let $\psi := G_\delta*\omega = \int G_\delta(\cdot -\tilde{z}) \omega (\tilde{z}) d \tilde{z}$ then integration by parts implies
  \begin{align*}
     \int \omega(z) \pounds_{\vec{u}'}[\psi](z) 
    =  \int \omega(z)  \left( u'(z) \partial_x \psi (z)+ v'(z) \partial_y \psi(z) \right) dz.
  \end{align*}
  We see that $\vec{u} = (-\partial_y \psi,\partial_x \psi)$ is one possible solution.
  As $\Omega$ is non-degenerate on the tangent spaces of $\Orb(\omega)$, this is the unique solution.
  As a result, the evolution prescribed by $X_H$ is precisely \eqref{eq:vorticity}.
  This proves that \eqref{eq:vorticity} can be seen as a Hamiltonian equation on $\Orb(\omega_0)$ with respect to the symplectic structure \eqref{eq:symplectic} and the Hamiltonian \eqref{eq:Hamiltonian}.

\subsection{The first order case}
Let us illustrate these Hamiltonian results for the case of the first order MVB.
Let $z_1,\dots,z_n \in \R^2$ be distinct
and define the initial vorticity distribution
\begin{align*}
  \omega_0 = \sum_{i=1}^N \gamma_i \delta_{z_i} + \gamma_i^x \partial_x \delta_{z_i} + \gamma_i^y \partial_{y} \delta_{z_i}.
\end{align*}
We desire the to determine the coadjoint orbit, $\Orb(\omega_0)$, and the symplectic structure.

  Indeed, we find that for any function $f$
  \begin{align*}
    \int \omega \left(\varphi^{-1}(z)\right) f(z)dz &:= \int \omega (z)f\left(\varphi(z)\right) dz \\
    &= \gamma_i f(\varphi(Z_i)) \\
    &\quad - \gamma_i^x \partial_x \varphi^x|_{z=Z_i} \partial_x f |_{z=\varphi(Z_i)}
    - \gamma_i^x \partial_x \varphi^y|_{z=Z_i} \partial_y f |_{z=\varphi(Z_i)} \\
    &\quad - \gamma_i^y \partial_y \varphi^x|_{z=Z_i} \partial_x f |_{z=\varphi(Z_i)}
    - \gamma_i^y \partial_y \varphi^y|_{z=Z_i} \partial_y f |_{z=\varphi(Z_i)}
  \end{align*}
  Collecting like terms we find
  \begin{align*}
    \omega \circ \varphi^{-1} = \gamma_i \delta_{\varphi(Z_i)} + \Gamma_i^x \partial_x \delta_{\varphi(Z_i)} + \Gamma_i^y \partial_y \delta_{\varphi(Z_i)}
  \end{align*}
  where
  \begin{align*}
    \Gamma = 
    \begin{bmatrix}
      \Gamma_i^x \\ \Gamma_i^y 
    \end{bmatrix}
    =
    D\varphi(Z_i) \cdot
    \begin{bmatrix}
      \gamma_i^x \\ \gamma_i^y
    \end{bmatrix}
  \end{align*}
  By varying $\varphi$ we can obtain 
  any collection of distinct points $z_1,\dots,z_n \in \R^2$
  and any collection of non-zero vectors $\Gamma_1,\dots,\Gamma_n \in \R^2 \backslash \{0\}$.
  This proves
  \begin{align*}
  \Orb(\omega_0) &= \left\{ \sum_{i=1}^n \gamma_i \delta_{z_i} + \Gamma_i^x \partial_x \delta_{\tilde{z}_i} + \Gamma_i^y \partial_{y} \delta_{z_i}
  \mid z_i \in \R^2, (\Gamma_i^x,\Gamma_i^y) \in \R^2 \backslash \{0\} \right\} \\
  &\cong  \{ (z_1,\dots,z_n,\Gamma_1,\dots,\Gamma_n) \mid z_i \in \R^2, \Gamma_i \in \R^2 \backslash \{0\} , ( i \neq j \implies z_i \neq z_j ) \}.
  \end{align*}
  
  To derive the symplectic structure recall the symplectic structure for a general vorticity \eqref{eq:symplectic}.
  Now let $\omega = \gamma_i \delta_{z_i} + \Gamma_i^x \partial_x\delta_{z_i} + \Gamma_i^y \partial_y \delta_{z_i}$.
  In this case the left hand side of \eqref{eq:symplectic} can be computed with respect to 
  divergence free vector field $\vec{u} = (u,v)$ and $\vec{u}' = (u',v')$ as
  \begin{align*}
    \int \omega(z) \left( u v' - v u' \right) (z) dz &= \gamma_i ( u(z_i)v'(z_i) - v(z_i) u'(z_i) ) \\
    &\quad + \Gamma_i^x ( u_{,x} v' + u {v'}_{,x} - {u'}_{,x}v - u' v_{,x})|_{z = z_i} \\
    &\quad + \Gamma_i^y ( u_{,y} v' + u {v'}_{,y} - {u'}_{,y}v - u' v_{,y})|_{z = z_i}
  \end{align*}
  Note that this is written entirely in terms of the 1st order Taylor expansion of $\vec{u}$ and $\vec{u}'$ evaluated 
  at $z_i$.
  Moreover, $\pounds_{\vec{u}}[\omega] = \gamma_0 u(z_i) \partial_x \delta_{z_i} + \dots$
  also has the property that it only depends on
  the first order Taylor expansion of $u$ and $v$ at the points $z_1,\dots,z_n$.
  Therefore, both sides of \eqref{eq:symplectic}
  can be written as a function of the finite collection of 
  numbers $u(z_i), Du(z_i), v(z_i),Dv(z_i)$.
  The result then follows by identifying the scalars
  \begin{align*}
    u(z_i) \mapsto u_{z_i}\\
    u^x_{,x}(z_i) \Gamma^x_i + u^x_{,y}(z_i) \Gamma^y_i \mapsto \dot{\Gamma}_i^x \\
    u^y_{,x}(z_i) \Gamma^x_i + u^y_{,y}(z_i) \Gamma^y_i \mapsto \dot{\Gamma}_i^y.
    \end{align*}
 This proves that the symplectic structure on $\Orb(\omega_0)$ is more concretely written as
  \begin{align}
  \begin{split}
    \Omega( (\dot{z},\dot{\Gamma}), (\delta z,\delta \Gamma) ) &=
    \gamma_i ( \dot{x}_i \cdot \delta y_i - \delta x_i \cdot \dot{y}_i ) \\
    &\quad + \dot{\Gamma}_i^x \cdot \delta y_i
    - \dot{\Gamma}_i^y \cdot \delta x_i
    + \delta \Gamma_i^y \cdot \dot{x}_i 
    - \delta \Gamma_i^x \cdot \dot{y}_i
    \end{split} \label{eq:first_order_symplectic}
  \end{align}

In essence, we have determined a finite-dimensional Hamiltonian system whos solutions solve \eqref{eq:vorticity} when $\psi$ is related to $\omega$ via an appropriate regularization.

\begin{rmk}
  The use of this symplectic structure shows that the map $(z_i, \Gamma_i,\Gamma_i^x,\Gamma_i^y) \mapsto \omega \in \Orb(\omega)$ is a symplectic momentum map.
\end{rmk}

\begin{rmk}
The corresponding Poisson bracket can be represented in tabular form by:
\begin{center}
\begin{tabular}{|c|c|c|c|c|c|}
\hline
	$\{ \cdot , \cdot \}$ & $x$ & $y$ & $\Gamma$ & $\Gamma^x$  & $\Gamma^y$ \\ \hline
	$x$ & 0 & 1 & 0 & 0 & 1 \\ \hline
	$y$ & -1 & 0 & 0 & -1 & 0 \\ \hline 
	$\Gamma$ & 0 & 0 & 0 & 0 & 0 \\ \hline
	$\Gamma^x$ & 0 & -1 & 0 & 0 & 1 \\ \hline
	$\Gamma^y$ & 1 & 0 & 0 & -1 & 0  \\ \hline
\end{tabular}
\end{center}

The way to use this table is as follows.  Let $H = H(\xi)$ be our Hamiltonian where
$\xi = (x_1,\dots,x_n,y_1,\dots,y_n,\Gamma_1,\dots,\Gamma_n,\Gamma_1^x,\dots,\Gamma_n^x,\Gamma_1^y,\dots,\Gamma_n^y)$.
Hamilton's equations are then given by
given by
\begin{align*}
	\frac{df}{dt} = \sum_{i,j} B^{ij} \pder{f}{\xi^i} \pder{H}{\xi^j},
\end{align*}
for any function $f$, where $B^{ij}$ denotes the corresponding entry of the table.
In particular, when $f=\xi^i$, one recovers the equations of motion for the dynamics
of the positions and strengths for a set of $n$ 1st order MVBs. 
Poisson geometers call $B^{ij}$ a \emph{Poisson tensor} \cite{FOM}.
\end{rmk}


\section{Conclusion}
\label{sec:Conclusion}
In this paper we have considered a generalization of the standard vortex blob method, obtained 
by augmenting the vortices with higher order circulation variables
and dubbing them \emph{multipole vortex blobs} (MVBs).
By viewing the vorticity equation as an advection equation, we have obtained equations of motion for these MVBs.

The extra degrees of freedom of MVBs resulted in richer dynamics near the vortex core.
Moreover, these new vorticity carrying elements exhibited a variety of novel types of solution behavior.
We also observed faster convergence rates in space using higher order MVBs.
Moreover, we proposed a scheme to decrease the number of pairwise interactions, by grouping
MVBs of lower order into a smaller number of MVBs of higher order.
Lastly, the implications of Kelvin's circulation theorem were substantially richer in the case of MVBs than they were for the standard vortex blob method.

We have demonstrated the behavior of the MVBs with a sequence of simple numerical experiments consisting of small numbers of MVBs of various degrees.
We found that 1st order MVBs correspond to sums of vortex blobs and regularized dipoles which simply propagate themselves forward, while the 2nd order circulation variables activate richer (non-propagating) dynamics near the vortex core.

Finally, we derived the symplectic structure of MVBs using methods from  \cite{MarsdenWeinstein1983}.
The resulting structure turned out to be a direct generalization of the standard
symplectic structure for vortex blobs.

The multiscale nature of ideal fluids is the principal obstacle to obtaining accurate models~\cite[Ch. 3]{Chorin1994}.
The use of MVBs augments the standard vortex blob method by allowing for singular vorticity distributions which model dynamics below the regularization length scale (i.e. at order $\delta^k$ with $\delta\ll1$ for a $k$th order jet-vortex). As the dynamics of MVBs are relatively easy to derive, and their analysis is tractable, we believe that MVBs will be of considerable value in understanding the place of regularized fluid models within the computational fluids community at large
and they should provide renewed interest in the vortex blob method.

Future avenues of inquiry could include:
\begin{itemize}
	\item MVBs on manifolds, such as the sphere
	\item The convergence properties of the MVB method
	\item How does one choose the regularization length-scale in relation to the grid resolution.  This relationship is addressed quite well for zeroth order MVBs in~\cite{BealeMajda1982}.  It is not clear if higher order MVBs change those results.
	\item An investigation of the kinetic theory of MVBs.
\end{itemize}

\section{Acknowledgements}
Both authors gratefully acknowledge partial support by the European Research
Council Advanced Grant 267382 FCCA to DDH.
We thank Anatoly Tur and Vladimir Yanovsky for helping us navigate the literature and relate our paper to earlier work.
We also thank Stefan Llewellyn Smith for his helpful comments before our initial submission.

\appendix

\section{Distributions}
\label{sec:distributions}
The vorticity, $\omega$, should be viewed as a distribution
and the term ``$\partial_x \omega$'' should be viewed
as a distributional derivative.
When $\omega \in \mathcal{D}'(\R^2)$ is a smooth distribution there is little harm in naively interpreting $\omega$ as a smooth function on $\R^2$.
However, when $\omega$ is not smooth (e.g. a Dirac delta distribution),
then one needs to invoke the mathematics of distributions
as distinct from that of real valued functions.
Therefore, we have included this appendix to remind the reader of the basic theory of distributions.
The main reference for this section is \cite{Hormander2003}.

The space of distributions $\mathcal{D}'(\R^2)$ is the dual-vector space to the space of smooth functions with compact supper $C^\infty_0(\R^2)$.
Therefore a distribution is defined by how it maps functions to real numbers.

The distributional derivative of a given distribution $\omega \in \mathcal{D}'(\R^2)$ in the $i$th coordinate direction may be defined as the distribution $\partial_i \omega$ obtained by
\begin{align*}
	\int \partial_i \omega (z) f(z) dz = - \int \omega(z) \partial_i f (z) dz.
\end{align*}
For example, the Dirac-delta distribution, $\delta_0$, is defined as the unique distribution such that
\begin{align*}
	\int \delta_0(z) , f (z) dz = f(0) \,,\quad \quad \forall f \in C_0^\infty(\R^2).
\end{align*}
The distributional derivative, $\partial_i \delta_0$, is given by
\begin{align*}
	\int \partial_i \delta_0 (z) f(z) dz = -\partial_i f(0) \,,\quad \quad \forall f \in C_0^\infty(\R^2).
\end{align*}
Given a distribution $\omega \in D(\R^2)$ and a function $g \in C_0^\infty(\R^2)$ one can define the
distribution $g \, \omega$ as
\begin{align*}
	\int (g \, \omega)(z)  f(z) dz = \int \omega(z) g(z)f(z) dz \,,\quad \quad \forall f \in C_0^\infty(\R^2).
\end{align*}
For example, we find that $g \, \delta _0 = g(0) \delta_0$.
A slightly more involved, but standard,  example is given by the computation of $g\, \partial_i \delta_0$.
We find
\begin{align*}
	\int (g \, \partial_i \delta_0) (z)  f(z) dz = \int \partial_i \delta_0(z) g(z)f(z) dz =  - g(0) \partial_i f(0) - \partial_ig(0) f(0).
\end{align*}
Therefore
\begin{align*}
	g \, \partial_i \delta_0 = g(0) \partial_i \delta_0 - \partial_i g(0) \delta_0.
\end{align*}
On the left hand side, note that $g(0)$ and $\partial_i g(0)$ are merely real numbers,
which are multiplying the distributions $\partial_i \delta_0$ and $\delta_0$.
More generally, we find
\begin{align*}
	\int (g\, \partial_x^m \partial_y^n \delta_0)(z) f(z) dz &= (-1)^{m+n} \partial_x^m \partial_y^n (fg)(0) \\
		&= (-1)^{m+n} \sum_{\ell,k=0}^{m,n} \binom{m}{\ell} \binom{n}{k}
		\left(\partial_{x}^{\ell} \partial_y^k f(0) \right) 
		\left(\partial_{x}^{m-\ell} \partial_y^{n-k} g(0) \right) 
\end{align*}
which means
\begin{align}
	g \, \partial_x^m \partial_y^n \delta_0 =
		(-1)^{m+n} \sum_{\ell,k=0}^{m,n} (-1)^{\ell + k}
		\binom{m}{\ell} \binom{n}{k}
		\left(\partial_{x}^{m-\ell} \partial_y^{n-k} g(0) \right) 
		\partial_{x}^{\ell} \partial_y^k \delta_0.
                \label{eq:func times partial delta}
\end{align}

\section{Symmetries and conservation laws}
\label{sec:symmetries}
The main reference for the material presented in this section is \cite{FOM}.
Let $G$ be a Lie group with Lie algebra $\mathfrak{g}$.
We will denote the dual of $\mathfrak{g}$ by $\mathfrak{g}^*$.
A (left) group action of $G$ on a manifold $S$ is a map $\varrho: G \times S \to S$
such that $\varrho( g h , x) = \varrho(g , \varrho(h , x))$ for all $g,h, \in G$ and $x \in S$.
\begin{rmk}
The group action $\varrho$ is not to be confused with the fluid density, often denoted as $\rho$ in fluid mechanics.
This appendix relates to more general mathematical constructions which are useful, but not necessarily in the usual purview of fluid mechanics.
In particular, the symbol $\varrho$ is the Greek letter `r', which refers to the word ``representation''
as in ``representation theory''.
\end{rmk}
One can also construct a group action, $D\varrho: G \times TS \to TS$,  given by
$D\varrho( g , u ) := \left.\frac{d}{dt}\right|_{t=0} \varrho(g, x(t) )$ for $u = \frac{dx}{dt} \in T_x S$ and $g \in G$.
There is also a natural Lie-algebra action, which one could also denote by $\varrho: \mathfrak{g} \times S \to TS$
defined by $\varrho( \xi , x) := \left. \frac{d}{d \epsilon} \right|_{\epsilon =0} \varrho( g_\epsilon , x)$.
In particular, the map $\varrho( \xi , \cdot ) : S \to TS$ is a vector field on $S$ which we call the
\emph{infinitesimal generator} of $\xi$.
When no confusion arises, it is typical to use the notation $g \cdot x$, $g \cdot u$, and $\xi \cdot x$ to denote $\varrho(g,x)$,$D\varrho(g,u)$,  and $\varrho(\xi,x)$, respectively.
Finally, if $(S,\Omega)$ is a symplectic manifold, then we say that $G$ acts \emph{symplectically} (or \emph{canonically})
if $\Omega_{g \cdot x}( g \cdot u , g \cdot v) = \Omega_x(u,v)$ for all $g \in G$, $u,v \in T_x S$ and $x \in S$.

Let us now recall the notion of a momentum map \cite[Definition 4.2.1]{FOM}.
Given a symplectic manifold $(S,\Omega)$ and a Lie group $G$ which acts on $(S,\Omega)$ symplectically, a momentum map is a map ${\bf J}:S \to \mathfrak{g}^*$ such that
\begin{align*}
  d \langle {\bf J} , \xi \rangle = i_{\xi_S} \Omega\,,
\end{align*}
where $\langle {\bf J} , \xi \rangle$ denotes the real-valued function on $S$ obtained by pairing ${\bf J}$ with an arbitrary
 $\xi \in \mathfrak{g}$, and where $\xi_S \in \mathfrak{X}(S)$ is the infinitesimal generator of $\xi$ on $S$.
Equivalently, we could express the previous condition as
\begin{align}
  d \langle {\bf J},\xi \rangle(x) \cdot v_x = \Omega( \xi\cdot x , v_x ) \label{eq:momap}
\end{align}
for all $\xi \in \mathfrak{g}$, $x \in S$ and $v_x \in T_xS$.
Momentum maps are significant for a number of reasons.
In particular, given a Hamiltonian on $S$ with $G$-symmetry, the momentum map ${\bf J}$ will be conserved under the evolution of
Hamilton's equations \cite[Theorem 4.2.2]{FOM}.
This is the Hamiltonian version of Noether's theorem.

In our case $S = {\rm Orb}(\omega_0) = \{ \omega_0 \circ \varphi^{-1} \mid \varphi \in \SDiff(\mathbb{R}^2) \}$ is a coadjoint orbit of some vorticity distribution on $\mathbb{R}^2$.
Tangent vectors on $S$ are of the form $$\pounds_{\vec{u}} [\omega] := u \partial_x \omega + v \partial_y \omega$$ for a (perhaps non-unique) divergence free vector-field $\vec{u} = (u,v)$.
Under this identification, the symplectic form at some $\omega \in S$ is given by an application of Kostant's formula \cite{FOM}.
This is derived in section \ref{sec:symplectic} and found to be
\begin{align*}
  \Omega_\omega( \pounds_{\vec{u}_1}[\omega] , \pounds_{\vec{u}_2} [\omega] ) :=
  \int \omega(z) ( \vec{u}_1(z) \times \vec{u}_2(z) ) dz.
\end{align*}
where $\vec{u}_1$ and $\vec{u}_2$ are divergence free vector-fields and $\times$ denotes the planar cross product.
Here we interpret the planar cross product as taking values in the space of real-numbers so that $\vec{u}_1 \times \vec{u}_2$ is
merely a smooth function.

We now will translate formula \eqref{eq:momap} to this more specific scenario.
Assume $G$ acts upon $\mathbb{R}^2$, then $G$ also acts upon distributions and upon $S$ by symplectic group actions.
In this context, a momentum map ${\bf J}$ associated to a $G$-action is defined by the equation
\begin{align}
  \int \frac{\delta  \langle {\bf J} , \xi \rangle }{\delta \omega} (z)  \left( \pounds_{\vec{u}}[\omega] \right)(z) dz = \int \omega(z)  \left( (\xi \cdot z) \times \vec{u}(z) \right)dz \label{eq:momap2}
\end{align}
Where $\xi \cdot z$ denotes the action of $\xi \in \mathfrak{g}$ on $z \in \R^2$
and ``$\times$'' denotes the planar cross-product.

\subsection{Translational symmetry and ${\bf J}_{\rm lin}$}

The group $\mathbb{R}^2$ acts upon $\mathbb{R}^2$ by translation.
That is to say, by sending any $\tilde{z} \in \R^2$ to $z + \tilde{z} \in \R^2$ for any $\tilde{z}  \in \R^2$.
This fact induces an action on smooth functions.
In particular, there is a natural (right) action on $C^{\infty}(\mathbb{R}^2)$ sending the function $\phi(z)$ to the function  $(\tilde{z})^*\phi(z) := \phi(z-\tilde{z})$.
We denote the inverse operation by $(\tilde{z})_* \phi(z) := \phi( z + \tilde{z})$.
This induces a (left) action on distributions which sends $\omega \in \mathcal{D}'(\R^2)$ to the distribution $(\tilde{z})_*\omega(z) := \omega( z +\tilde{z})$.
As a translation of $\mathbb{R}^2$ by $\tilde{z}$ is a volume-preserving diffeomorphism, we see that the coadjoint orbit $S \subset \mathcal{D}(\mathbb{R}^2)$ is invariant under this action.
Moreover, we observe the action, restricted to $S$, is symplectic because
\begin{align*}
  \Omega_{ \tilde{z}_* \omega}( \tilde{z}_*\pounds_{u}[ \omega] , \tilde{z}_*\pounds_v[\omega] ) &= \Omega_{\tilde{z}_* \omega} ( \pounds_{\tilde{z}_*u}[ \tilde{z}_*\omega] ,
  \pounds_{\tilde{z}_*v}[\tilde{z}_*\omega] ) \\
  &= \int \tilde{z}_* \omega(z)  \left( \tilde{z}_*(u \times v) \right)(z) dz \\
  &= \int \omega( z+\tilde{z}) \left(u \times v \right)(z + \tilde{z}) dz \\ 
  &= \int \omega(z) \left( u \times v \right)(z) dz  \\
  &= \Omega_{\omega}( \pounds_u[\omega], \pounds_v[\omega]).
\end{align*}

Given this symplectic action, we can seek a momentum map, ${\bf J}_{\rm lin}:S \to (\mathbb{R}^2)^*$.  Consider an arbitrary element of the Lie-algebra $\delta \tilde{z}  = (\delta \tilde{x}, \delta \tilde{y}) \in \mathbb{R}^2$ and use \eqref{eq:momap2} to obtain
\begin{align*}
  \int \frac{\delta \langle {\bf J}_{\rm lin} , \delta \tilde{z} \rangle}{\delta \omega}(z)   \pounds_{\vec{u}}[\omega] dz
  = \int \omega(z) \left( \delta \tilde{x} \, v(z) - \delta \tilde{y} \, u(z) \right) dz
\end{align*}
We can re-write the right hand side as
\begin{align*}
  = \int \omega(z)  \pounds_{\vec{u}} [ \delta \tilde{x} \, y - \delta \tilde{y} \, x] dz
\end{align*}
and upon integrating by parts this is equivalent to
\begin{align*}
  = - \int \pounds_{\vec{u}}[\omega](z)  \left( \delta \tilde{x}\, y - \delta \tilde{y}\, x \right) dz
\end{align*}
Therefore, ``cancelling'' the arbitrary vector $\pounds_v[\omega]$ from both sides we find
\begin{align*}
  \frac{\delta  \langle {\bf J}_{\rm lin} , \delta \tilde{z} \rangle}{\delta \omega} = \delta \tilde{y} x - \delta \tilde{x} y
\end{align*}
Integrating by $\omega$ we find
\begin{align*}
    {\bf J}_{\rm lin}(\omega) = \left( -\int \omega(z) y dz \, , \, \int \omega(z) x dz \right) \in (\mathbb{R}^2)^* \equiv \mathbb{R}^2
\end{align*}
If $\omega$ satisfies the MVB ansatz
\begin{align*}
  \omega = \Gamma_i \delta_{z_i} + \Gamma_i^x \partial_x \delta_{z_i} + \Gamma_i^y \partial_y \delta_{z_i} + \dots
\end{align*}
then
\begin{align*}
  {\bf J}_{\rm lin}(\omega) = \sum_i ( \Gamma_i^{y} - \Gamma_i y_i , \Gamma_i x_i -\Gamma_i^{x} ).
\end{align*}
The terms of the MVBs beyond the first order do not influence ${\bf J}_{\rm lin}$.

\subsection{Rotational symmetry and ${\bf J}_{\rm ang}$}
The group $\SO(2)$ acts upon $\mathbb{R}^2$ by rotations about the origin sending $z \in \R^2$ to $R_\theta \cdot z := ( \cos(\theta)x-\sin(\theta)y,\sin(\theta)x+\cos(\theta)y )$.
For any $\theta \in \SO(2)$, there is a natural action on $C^{\infty}(\mathbb{R}^2)$ sending the function $\phi \in C^{\infty}(\mathbb{R}^2)$ to the function $\theta^*\phi(x,y) := \phi(\cos(\theta)x-\sin(\theta)y,\sin(\theta)x+\cos(\theta)y)$.  The corresponding action on vector-fields and all other objects on $\mathbb{R}^2$ follows naturally.  
 In particular, the left-action on distributions sends $\omega \in \mathcal{D}'(\mathbb{R}^2)$ to the distribution $\theta_* \omega(z) := \omega( R_\theta \cdot z)$.
Again, we can verify that the coadjoint orbit $S$ is invariant under this action, and that $\SO(2)$ acts symplectically upon $S$
through computations which are analogous to those performed in the previous subsection.
Given this symplectic action, we can seek a momentum map, ${\bf J}_{\rm ang}:S \to \mathfrak{so}(2)^* \equiv \mathbb{R}$.  Consider an arbitrary element of the Lie-algebra $\xi \in \mathfrak{so}(2) \equiv \mathbb{R}$ and use \eqref{eq:momap2} to obtain
\begin{align*}
  \int  \frac{\delta  \langle {\bf J}_{\rm ang} (z) , \xi \rangle }{\delta \omega} (z) \, \pounds_{\vec{u}}[\omega] (z) dz
  = \int \omega(z) \left( -y\xi v - x\xi u \right) dz
\end{align*}
We can re-write the right hand side and integrate by parts to find
\begin{align*}
  = -\xi \int \omega(z) \pounds_{\vec{u}} \left[\frac{x^2+y^2}{2}\right] dz
  = \xi \int \pounds_{\vec{u}}[\omega](z) \left( \frac{x^2+y^2}{2} \right) dz.
\end{align*}
As the vector $\pounds_{\vec{u}}[\omega] \in T_\omega S$ is arbitrary we find
\begin{align*}
  \frac{\delta \langle {\bf J}_{\rm ang} , \xi \rangle}{\delta \omega} = \xi \frac{x^2+y^2}{2}
\end{align*}
Integrating by $\omega$ we find
\begin{align*}
    {\bf J}_{\rm ang}(\omega) = \frac{1}{2}\int \omega(z) ( x^2 + y^2) dz \in \mathbb{R} \equiv \mathfrak{so}(2)^*
\end{align*}
If $\omega$ satisfies the MVB ansatz
\begin{align*}
  \omega = \Gamma_i \delta_{z_i} + \Gamma_i^x \partial_x \delta_{z_i} + \Gamma_i^y \partial_y \delta_{z_i} + \Gamma_i^{xx} \partial_{xx} \delta_{z_i} + \dots
\end{align*}
then
\begin{align*}
  {\bf J}_{\rm ang}(\omega) = \frac{\Gamma_i}{2} (x_i^2 + y_i^2) - \Gamma_i^x x_i - \Gamma_i^y y_i + \Gamma_i^{xx} + \Gamma_i^{yy} .
\end{align*}
The terms of the MVBs beyond the second order do not influence the angular momentum ${\bf J}_{\rm ang}$.


\providecommand{\bysame}{\leavevmode\hbox to3em{\hrulefill}\thinspace}
\providecommand{\MR}{\relax\ifhmode\unskip\space\fi MR }
\providecommand{\MRhref}[2]{%
  \href{http://www.ams.org/mathscinet-getitem?mr=#1}{#2}
}
\providecommand{\href}[2]{#2}

\end{document}